\documentclass[%
reprint,
 aps,
nofootinbib,
 amsmath,amssymb,
 rmp,
 floatfix,
 showpacs,
longbibliography,
eqsecnum,
footmisc,
graphicx
]{revtex4-2}

\usepackage[colorlinks,citecolor=blue,linkcolor=blue,urlcolor=blue]{hyperref}
\usepackage{graphicx}
\usepackage{dcolumn}
\usepackage{bm}
\usepackage{color}
\usepackage{amsfonts}
\usepackage{txfonts}
\usepackage{multirow}
\usepackage{soul}
\usepackage[english]{babel}
\usepackage{titlesec}
\usepackage{upgreek}
\begin{document}
\title{Measuring interfacial Dzyaloshinskii-Moriya interaction in ultra-thin magnetic films}

\author{M. Kuepferling}
\author{A. Casiraghi}
\author{G. Soares}
\author{G. Durin}
\affiliation{Istituto Nazionale di Ricerca Metrologica, Strada delle Cacce 91, 10135, Torino, Italy}
\author{F. Garcia-Sanchez}
\affiliation{Dep. of Applied Physics, University of Salamanca, Plaza de los Caidos s/n 37008, Salamanca, Spain}
\author{L.~Chen}
\author{C.~H.~Back}
\affiliation{Technical University Munich, James-Frank-Str.~1, 85748 Garching, Germany}
\author{C.~H.~Marrows}
\affiliation{School of Physics and Astronomy, University of Leeds, Leeds LS2 9JT, United Kingdom}
\author{S. Tacchi}
\affiliation{CNR, Istituto Officina dei Materiali - Perugia, c/o Dipartimento di Fisica e Geologia\\Università Perugia, Via A. Pascoli, 06123 Perugia, Italy}
\author{G. Carlotti}
\affiliation{Dipartimento di Fisica e Geologia, Università di Perugia, Via Pascoli, 06123 Perugia, Italy}

\date{\today}

\begin{abstract}
The Dzyaloshinskii-Moriya interaction (DMI), being one of the origins of chiral magnetism, is currently attracting considerable attention in the research community focusing on applied magnetism and spintronics. For future applications, an accurate measurement of its strength is indispensable. Here we present a review of the state-of-the-art of measuring the coefficient of the Dzyaloshinskii-Moriya interaction, the DMI constant $D$, focusing on systems where the interaction arises from the interface between two materials (i.e. interfacial DMI). We give an overview of the experimental techniques as well as their theoretical background and models for the quantification of the DMI constant. The measurement techniques are divided into three categories: a) domain wall-based measurements, b) spin wave-based measurements and c) spin-orbit torque-based measurements. We analyze the advantages and disadvantages of each method and compare $D$ values at different interfaces. The review aims to obtain a better understanding of the applicability of the different techniques to various stacks and of the origin of apparent disagreements among literature values.
\end{abstract}

\maketitle

\tableofcontents

\section{Introduction}\label{sec:Introduction}
Topological spin structures such as chiral domain walls (DWs) and skyrmions are emerging as promising information carriers for future spintronics technologies \cite{DIE-20}. 
Electric currents can drive such spin structures with an unprecedented level of efficiency, which makes them particularly attractive for innovative storage devices, including the racetrack memory \cite{PAR-15}. The crucial ingredient needed for stabilizing these chiral spin textures is the Dzyaloshinskii-Moriya Interaction (DMI). 

The DMI is an anisotropic exchange interaction favoring a canted spin arrangement and has a net contribution only in systems without a center of inversion. It originates from the spin-orbit coupling (SOC) which acts as a perturbation on localized spin states. Given two neighboring spins $\vec{S}_ i $ and $\vec{S}_{j}$, the DMI contributes to the Hamiltonian locally with a first bilinear energy term described by the following expression: 
\begin{equation}
\mathcal{H}_{DMI}=\vec{D}_{ij}\cdot (\vec{S}_i \times \vec{S}_{j})
\label{eq:hamiltonian}
\end{equation}
where $\vec{D}_{ij}=D_{ij}(\hat r_{ij} \times \hat{n})$ is the local DMI vector which can be described by its magnitude $D_{ij}$, with $\hat r_{ij}$ being the unit vector linking the two neighboring spins and $\hat{n}$ being along the direction of symmetry breaking. This expression is part of a generalized exchange interaction, which relates $\vec{D}_{ij}$ to the isotropic exchange $-J_{ij} \vec{S}_i \cdot \vec{S}_{j}$ with exchange constant $J_{ij}$, i.e. Heisenberg exchange in the case of localized electrons with a direct orbital overlap, or any other direct or indirect exchange described by the above Hamiltonian. Contrary to the Heisenberg type exchange, which favors collinear alignment, the DMI promotes an orthogonal arrangement between $\vec{S}_ i $ and $\vec{S}_j$, with a chirality imposed by the direction of $\vec{D}_{ij}$. 

The DMI was first proposed in the 1950s for antiferromagnets such as $\alpha$-Fe$_2$O$_3$ to account for the existence of a weak ferromagnetism \cite{DZY-57,MOR-60}. In the following decades, several magnetic materials, such as spin glasses, orthoferrites, manganites or superconducting cuprates \cite{FER-80,  LEV-81, LUO-99,BOG-02,COF-91, FER-91}, were investigated for their non-collinear or helical magnetism and for the influence of DMI on the magnetic state. An important step towards multilayers, on which this review focuses, was done by \textcite{FER-91}, where anisotropic pair interactions on surfaces or interfaces were considered.

From an experimental point of view, the development of sophisticated imaging techniques to visualize the magnetic state, such as spin-polarized scanning tunnel microscopy (SPSTM) or spin-polarized low energy electron microscopy (SPLEEM), was crucial for triggering a renewed interest in the topic at the beginning of the century, highlighted in the important works \textcite{BOD-07, HEI-00, HEI-11}. In non-centrosymmetric single crystals with FeSi structure\footnote{Often called B20 according to classification of the journal ``Strukturbericht''}, such as the materials MnSi, FeGe, etc., it was shown that DMI gives rise to skyrmion lattices and other exotic spin textures at low temperature, which could be directly visualized \cite{MUE-09, ROS-06, YU-10}.

 %Theoretical models in non-collinear magnetism are often based on ab-initio calculations and density functional theory \cite{SAN-98}.

\begin{figure}
	\centering
	\includegraphics[width=0.90\linewidth]{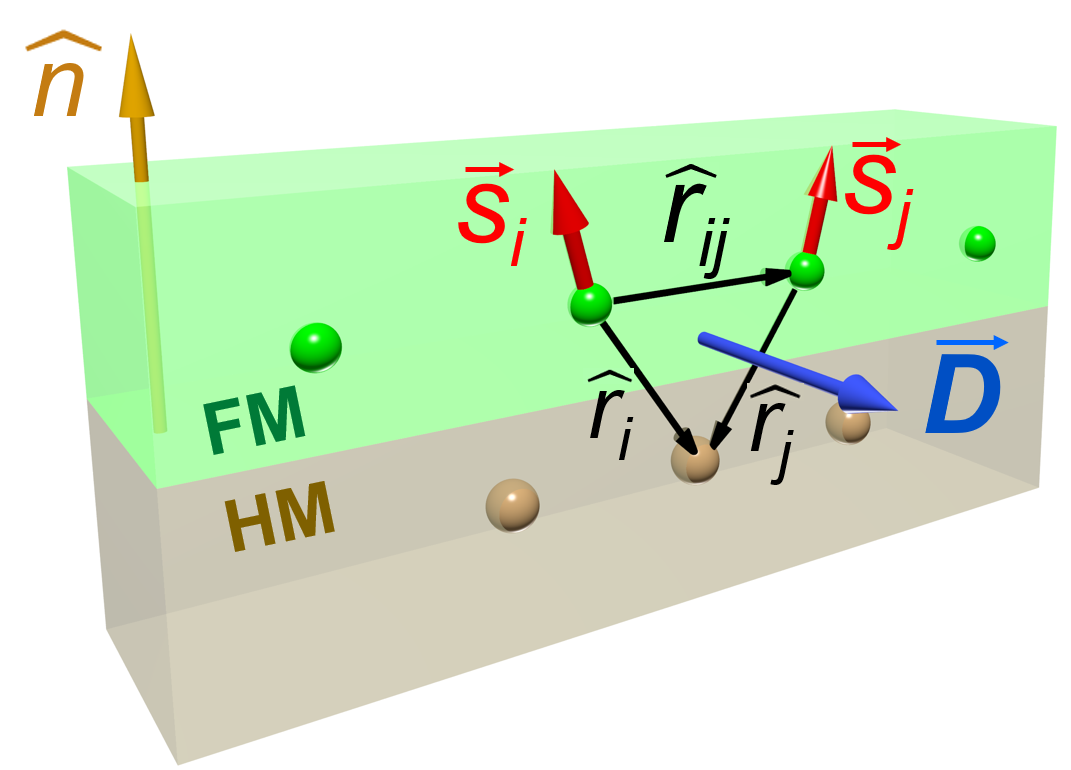}
	\caption{Diagram of a system possessing interfacial DMI. The exchange coupling between the spins $\vec{S}_i$ and $\vec{S}_j$ in a ferromagnetic (FM) layer is mediated by a heavy metal (HM) atom with large spin-orbit coupling. In this case, the local DMI vector with magnitude $D_{ij}$ is perpendicular to the plane formed by the two atoms of the ferromagnetic layer and the one of the heavy metal layer.}
	\label{fig:dmifertype}
\end{figure}

More recently, the presence of  DMI was also demonstrated in thin magnetic multilayers with perpendicular magnetic anisotropy and shown to be localised at the interface between the layers \cite{ZAK-10, JE-13, RYU-12, MOO-13}. This interfacial DMI has since received a broad attention from the magnetic community (see e.g. \textcite{HEL-17, GAL-19}). It is of particular importance in the context of systems consisting of an ultrathin ferromagnetic (FM) film with perpendicular magnetic anisotropy (PMA), in contact with a heavy metal (HM) under- or over-layer. In such a system the DMI arises due to the broken inversion symmetry at the interface between the two materials and the large SOC of the heavy metal atoms, which mediate the interaction between neighboring spins $\vec{S}_i$ and $\vec{S}_j$ of the ferromagnet. It follows that the DMI has an interfacial nature with a strength decreasing with the magnetic film thickness. If the symmetry breaking is only due to the interface and symmetry planes or axes are always normal to the film plane, according to Moriya's rules, the DMI vector $\vec{D}$, defined at the interface, has to be necessarily perpendicular to the film normal, favoring canting of the out-of-plane magnetisation (see Fig.\ref{fig:dmifertype}).

\begin{figure}
	\centering
	\includegraphics[width=0.9\linewidth]{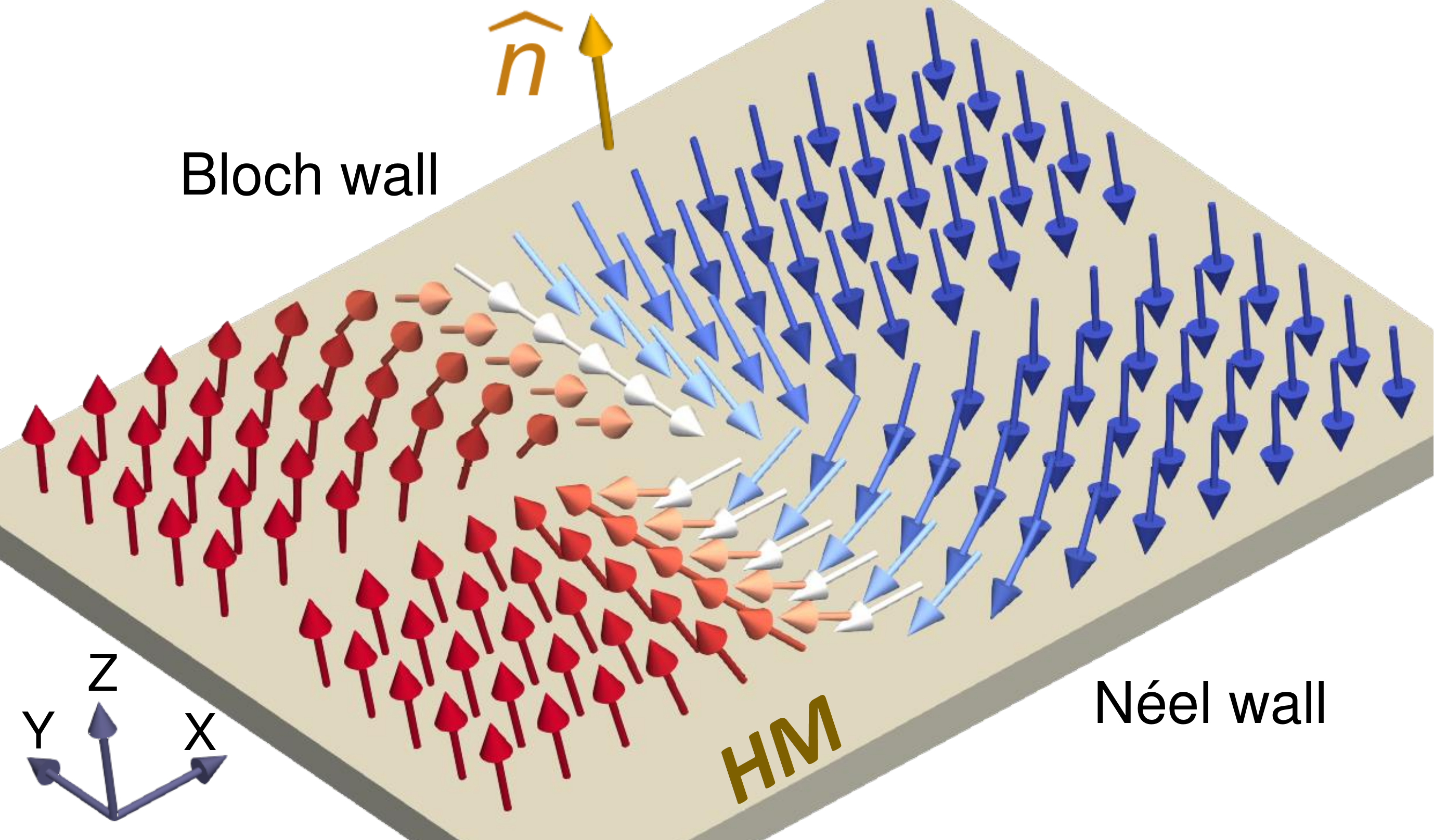}
	\caption{Bloch and N\'eel configurations for a domain wall in a perpendicularly magnetized material. The N\'eel wall has left-handed chirality ($\uparrow \leftarrow \downarrow$) considering the positive Z direction parallel and fixed to the normal, which is depicted by the arrow $\hat{n}$ and defined pointing from the substrate to the magnetic film.}
	\label{fig:neelbloch}
\end{figure}

Bilayers composed of a thin magnetic film with PMA and a heavy metal are particularly interesting from the application point of view, due to the fast domain wall motion driven by an electric current in presence of DMI. Early experiments on current-induced domain wall motion in such systems were not able to fully unravel the driving mechanisms behind such motion. The conventional spin-transfer torque (STT) theory \cite{SLO-96,BER-96} predicts motion in the direction of electron flow, opposite to many observations \cite{MOO-08,MOO-09}. On the other hand, the newly proposed spin-orbit torques (SOTs) --  caused by the inverse spin galvanic (or Rashba) effect (iSGE) \cite{MIH-10,MIR-11} and the spin Hall effect (SHE) \cite{LIU-12,HAA-13} -- did not have the correct symmetry to drive Bloch walls \cite{KHV-13}, which were believed to be present in these ultrathin PMA films from purely magnetostatic considerations. This issue was resolved when it was suggested\footnote{Here direct imaging of the domain wall chirality \cite{BOD-07} played a crucial role.} that in presence of DMI, N\'eel walls, which have a fixed chirality \cite{HEI-08,THI-12}, occur instead of the expected Bloch walls (Fig.~\ref{fig:neelbloch}). A series of experimental works later confirmed these findings \cite{EMO-13,RYU-13}, substantiating the view that DMI-stabilized N\'eel walls are mostly moved by the SHE torque. Ever since, DMI-based phenomena have created an extremely active research field, often referred to as chiral magnetism. 

Considering the ferromagnetic film itself isotropic (i.e. without special crystal symmetries, leading to an additional symmetry breaking), the DMI vector simplifies to $\vec{D} = D (\hat r_{ij} \times \hat n)$ \cite{MOR-60,CRE-98,  THI-12}, with a constant $D$ describing the strength of the DMI in the film plane ($\hat r_{ij}$ the vector between the two neighboring spins and $\hat n$ is the normal to the film interface). The interfacial DMI can then be thought of as equivalent to an effective in-plane magnetic field $H_\mathrm{DMI}$ which, acting across the domain wall, causes a reorientation of spins from the Bloch configuration into the N\'eel configuration with a fixed chirality \cite{THI-12,CHE-13}. The strength of the interfacial DMI is then measured as the DMI field \cite{THI-12}, to which it is proportional. The sign of $D$ dictates the chirality of the N\'eel domain wall, and, consequently, together with the sign of the spin Hall angle, the direction of the domain wall motion under SHE torque \footnote{The sign of $D$ depends on the convention chosen in relation with the Hamiltonian \ref{eq:hamiltonian} which can be either positive or negative. Throughout this review we use the plus sign in the Hamiltonian, so that positive $D$ corresponds to N\'eel domain walls with the right-handed (clockwise) chirality ($\uparrow \rightarrow \downarrow$ or $\downarrow \leftarrow \uparrow$), while negative $D$ corresponds to N\'eel domain walls with the left-handed (anticlockwise) chirality ($\uparrow \leftarrow \downarrow$ or $\downarrow \rightarrow \uparrow$), where up direction is defined by the normal in Fig.\ref{fig:neelbloch}}. 

Because of the crucial role that the interfacial DMI plays in stabilizing chiral spin structures as well as in attaining extremely efficient domain wall motion, it is of utmost importance for the magnetic community to be able to accurately measure and predict the magnitude and sign of $D$ in heavy metal (HM)/ferromagnet (FM) combinations in order to be able to optimize these relevant effects. However, despite a rapidly increasing number of works quantifying the DMI through different experimental techniques, a remarkable spread of data is currently present in the literature and systematic studies or reviews are still rare \cite{CHO-18}. Not only are different measurement techniques found to provide contradictory values of $D$ for nominally the same system, but controversies are also present when utilizing the same method on nominally identical stacks, or different methods on the same sample \cite{SOU-16,SHA-19}. 

The most commonly used experimental techniques that in recent years have been employed to measure $D$ in HM/FM structures can be broadly divided into three categories: 
\begin{enumerate}
\item \textit{Domain wall methods}, where $D$ is extracted by either measuring the domain wall velocity or energy as a function of an in-plane magnetic field, or by measuring domain wall spacing in stripe domain phases, or by directly measuring the domain wall internal structure;
%Measuring \textit{domain wall properties}, such as velocity or energy, and investigating their dependence on in-plane magnetic fields 
\item \textit{Spin wave methods}, where $D$ is extracted by measuring the non-reciprocity of propagating spin waves, owing to the presence of DMI, in in-plane magnetised films;
\item \textit{Spin-orbit torque methods}, where $D$ is extracted by measuring the field shift of the out-of-plane hysteresis loop under an in-plane magnetic field.
%\item Using \textit{scanning NV magnetometry}
\end{enumerate}

With this literature review we aim at providing an accurate analysis of the current state-of-the-art regarding measurements of the interfacial DMI constant $D$ based on the different techniques summarized above. Therefore, Sections II, III, and IV are devoted to analyze each of the three main classes of techniques mentioned above. In particular, in each section we start with a theoretical background, followed by an overview of the relevant experimental methodologies and then a number of detailed tables with a collection of the relevant results published in the literature. Finally, for each of the three above sections, we will provide a critical discussion of advantages and limitations of each experimental technique. The last section of this review article, Section V, is devoted to a comparison and discussion of the methods and results obtained by the three classes of methods analyzed in the previous sections, as well as to final considerations about open problems and challenges. In particular, we address crucial aspects of the current research, such as the reliability of the different techniques, the choice of material combinations to achieve large/small, negative/positive DMI, and the reasons for the spread of the results (e.g. sample/interface quality, growth conditions and reliability of the measuring techniques). 

We anticipate that, with regard to the available literature, it is out of the scope of the present work to give definitive recipes to prepare specific interfaces with a controlled value of DMI, evaluate the level of uncertainty of the different measurement methods, or classify the methods clearly according to their applicability. Therefore, one should consider our synoptic Figs.~\ref{fig:Ds_Pt_Co_X}-\ref{fig:Ds_X_CoFeB_MgO} as general guides to explore measurement methods and ranges of values of common heterostructures. We emphasize instead the need for more systematic studies and hope that this review will stimulate the community to tackle metrological aspects of the measurement of the DMI at interfaces.

\section{Domain wall methods}\label{sec:DW}

\subsection{Method overview}
In ultra-thin films with PMA, Bloch domain walls are magnetostatically favored over N\'eel DWs\footnote{However, N\'eel DWs become energetically more favorable when the film is patterned into very narrow wires \cite{KOY-11}.}. However, the interfacial DMI manifests itself as a local effective in-plane magnetic field $H_\mathrm{DMI}$ that acts on the domain wall which, when large enough to overcome the magnetostatic energy coming from the shape of the wall, converts Bloch DWs into N\'eel DWs with a chirality determined by the sign of $D$  \cite{THI-12,CHE-13}. By tuning the spin texture of the DWs towards a chiral N\'eel configuration, the interfacial DMI changes the static and dynamic properties of the DWs \cite{THI-12}. 

The direct observation of chiral magnetization states in presence of a DMI \cite{HEI-00, BOD-07} was the key for understanding the modified DW dynamics. Indeed, first estimates of the DMI strength were obtained by fitting DW energies and numerical simulations of DW imaging in FM mono-/double-layers on tungsten heavy metal substrates \cite{ZAK-10, BOD-07, HEI-08, FER-08a}. A further experimental technique used to extract a quantitative value of $D$ was the measurement of DW velocities driven by an electric current as a function of in-plane magnetic fields \cite{TOR-14}. This is not surprising, given the importance of DMI in these systems in the context of current-induced domain wall motion experiments \cite{EMO-13,RYU-13}. At the same time, it was proven that $D$ could also be inferred by measuring magnetic field driven DW velocities \cite{JE-13,HRA-14}. 
%In any case, whether the driving mechanism is current or field (for advantages and limitations - see Sec.~\ref{sec:advantagesDW}), 
$H_\mathrm{DMI}$, and thus $D$, is extracted by analyzing the dependence of the DW velocity on the in-plane magnetic fields, which - added or subtracted to the $H_\mathrm{DMI}$ - can enhance or reduce the DW speed. 

Following these early investigations \cite{RYU-13,TOR-14,JE-13, HRA-14}, several further works have used DW motion, induced by either current or field, to estimate the strength and sign of the interfacial DMI in different material systems. Nucleation of reverse domains, which is a necessary step to study the motion of DWs, was shown to be dependent on $D$ \cite{PIZ-14}, which established another path to access the magnitude and sign of $D$. Finally, also the static properties of domains were shown to be altered by the presence of DMI. This was observed from two points of view: firstly, the domain width is altered due the modification of the DW energy originated from DMI \cite{MOR-16}. Secondly, the DMI affects the stability of reversed domains and their field of annihilation \cite{HIR-14}. Both effects have been used to estimate the magnitude of $D$. 

\begin{figure}[t]
	\centering
	\includegraphics[width=0.7\linewidth]{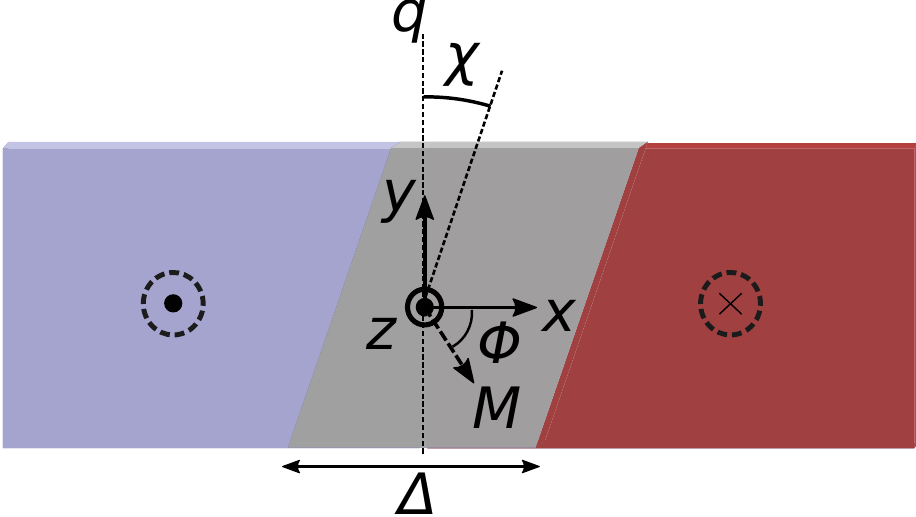}
	\caption{Sketch of the collective coordinates used to describe DW dynamics in different types of 1D models: $q$ is the domain wall position, $\Phi$ the angle of the projection of the magnetization $\mathbf{M}$ in the $x$--$y$ plane, $\Delta$ the width, and $\chi$ the tilting angle of the DW.  }
	\label{fig:1Dmodel}
\end{figure}

%\subsubsection{One-dimensional DW model}\label{sec:1Dmodel}
DW dynamics in the presence of DMI can be understood making use of analytical 1D models. One-dimensional DW models have been established as useful tools to support computationally costly micromagnetic simulations as well as to interpret experimental results. As the name suggests, the 1D models for DW dynamics are based on the approximation that the magnetization $\mathbf{M}$ varies along one direction only, namely the axis of a narrow wire, usually identified with $x$, as depicted in Fig.~\ref{fig:1Dmodel}.

The origins of the 1D models can be traced back to the original description of Bloch walls by F. Bloch \cite{BLO-32}. The simplest form of the 1D model was introduced by Walker and Slonczewski \cite{SCH-74,MAL-79} in the 1970s\footnote{ The 1D model was reported earlier by L. R. Walker, Bell Telephone Laboratories Memorandum, 1956 (unpublished).
A description of this work can be found in J.F. Dillon Jr., Domains and Domain
Walls, in: Magnetism, Vol. III, edited by G. T. Rado and H. Suhl (Academic, New York, 1963).}, to study DW dynamics in PMA materials under the influence of an applied perpendicular field. Here DW dynamics was described in terms of two time-dependent variables (see Fig.~\ref{fig:1Dmodel}): the DW position $q(t)$ along the wire axis, and the DW angle $\Phi(t)$, defined as the in-plane ($x$--$y$) angle of the internal DW magnetization $\mathbf{M}$ with respect to the positive $x$ axis\footnote{According to these conventions, $\Phi = 0$ ($\pi$) corresponds to a right (left) handed chiral N\'eel DW, while $\Phi = \pm \pi/2$ corresponds to a Bloch DW.}. The $q - \Phi$ model was later extended by  \textcite{SOB-94,SOB-95} to describe DW dynamics in PMA materials under in-plane fields. \textcite{THI-02} next adapted the 1D model to systems with in-plane magnetic anisotropy, using the domain wall width $\Delta(t)$ as an additional time-dependent variable in a revised $q - \Phi - \Delta$ model.

More recently, several other contributions have been included in the 1D model to take into account newly discovered effects, such as the STTs \cite{ZHA-04, THI-05}, the SOTs \cite{HAY-12,SEO-12,BOU-12,MAR-13a,MAR-13b} (see Eq.~\ref{eqSHE}), and the DMI \cite{THI-12,EMO-13,MAR-13a}. 
In this context, $D$ is then derived according to \citealp{EMO-13}: 
\begin{equation}
D = \mu_0 H_\mathrm{DMI} M_\mathrm{S} \Delta,
\label{eq:DMI}
\end{equation}      
where $M_\mathrm{S}$ is the saturation magnetization, and $\Delta = \sqrt{A/K_\mathrm{eff}}$ is the DW width parameter, with $A$ being the exchange stiffness and $K_\mathrm{eff} = K_\mathrm{U} - \frac{1}{2} \mu_0 M_\mathrm{S}^2$ the effective perpendicular anisotropy constant, corresponding to the intrinsic perpendicular anisotropy constant $K_\mathrm{U}$ decreased by the demagnetizing energy $\frac{1}{2}\mu_0 M_\mathrm{S}^2$.

Additions of thermal fluctuations \cite{MAR-07} and of spatially dependent pinning \cite{CON-12} helped to make the 1D model more realistic. Furthermore, considering experiments of fast current-driven DW motion \cite{RYU-12}, \textcite{BOU-13} proposed to include DW tilting as an additional time-dependent variable $\chi(t)$ (defined as the angle of the DW normal plane with respect to the positive $x$-axis), which led to the development of the $q - \Phi - \chi$ model. Very recently the 1D model was extended to implement all four collective coordinates, namely $q - \Phi - \Delta - \chi$,  with the aim of improving the agreement with experimental observations and micromagnetic simulations when large in-plane fields are applied \cite{NAS-17}. The same authors later showed that the simple two coordinate $q - \Phi$ model can grant higher accuracy when combined with an ansatz (which links collective coordinates to magnetization components) that takes into account magnetization canting within the domains under an in-plane field \cite{NAS-18}.

Here, we provide a summary for the different DW-based techniques, 
explaining the underlying models (Sec.~\ref{sec:theoryDW}), reviewing the main results achieved regarding the determination of $D$ (Sec.~\ref{sec:experimentDW}), and highlighting the respective advantages and limitations (Sec.~\ref{sec:advantagesDW}).

\subsection{Theory and models}\label{sec:theoryDW}
\subsubsection{Current-driven domain wall motion}\label{sec:theoryDWcurrent}
The driving mechanism behind the current-induced motion of DWs in heavy metal/ferromagnet bilayers with PMA is now widely believed to be due to a combination of SHE and DMI, while the iSGE torque is considered to be negligible \cite{EMO-13,RYU-13,MAR-14}, as the latter acts to stabilize Bloch walls \cite{MIR-11}, and does not have the correct symmetry to drive DWs directly \cite{EMO-13, KHV-13}. The SHE in the heavy metal converts an in-plane charge current into a transverse spin current that gives rise to spin accumulation at the interface between the two layers, with consequent spin-current diffusion into the ferromagnet. This spin-current can interact with the local magnetization by exerting a torque on it, known as the SHE spin-orbit torque (SHE-SOT)\footnote{To be precise, field-like and damping-like torques can occur as a result of iSGE or SHE effects.}. According to the 1D model, the amplitude of the effective field associated with the SHE-SOT is expressed as \cite{THI-12,KHV-13}:
\begin{equation}
H_\mathrm{SHE} = \frac{\hbar \theta_\mathrm{SH} |J_e|}{2 \mu_0 |e| M_\mathrm{S} d}\cos(\Phi), 
\label{eqSHE}
\end{equation} 
where $\theta_\mathrm{SH}$ is the spin Hall angle, $J_e$ is the electron current density, $d$ is the thickness of the ferromagnetic film, and $\Phi$ is the internal in-plane DW angle as defined in Fig.~\ref{fig:1Dmodel}. As mentioned earlier, the SHE-SOT can move DWs only if they possess a N\'eel component (i.e. $\Phi \neq \pm \pi/2$) in their spin structure, due to the interfacial DMI. The direction in which DWs move with current depends both on the sign of the spin Hall angle $\theta_\mathrm{SH}$, determined by the spin-orbit coupling constant of the heavy metal, and on the sign of $D$ (i.e. on the chirality of the DW). Micromagnetic simulations and 1D model predict that the maximum velocity of DWs driven by the SHE-SOT increases with the magnitude of $D$, and saturates for larger currents showing a clear plateau \cite{THI-12,REA-17,Lemesh2019}.

\begin{figure}[t]
	\centering
	\includegraphics[width=1\linewidth]{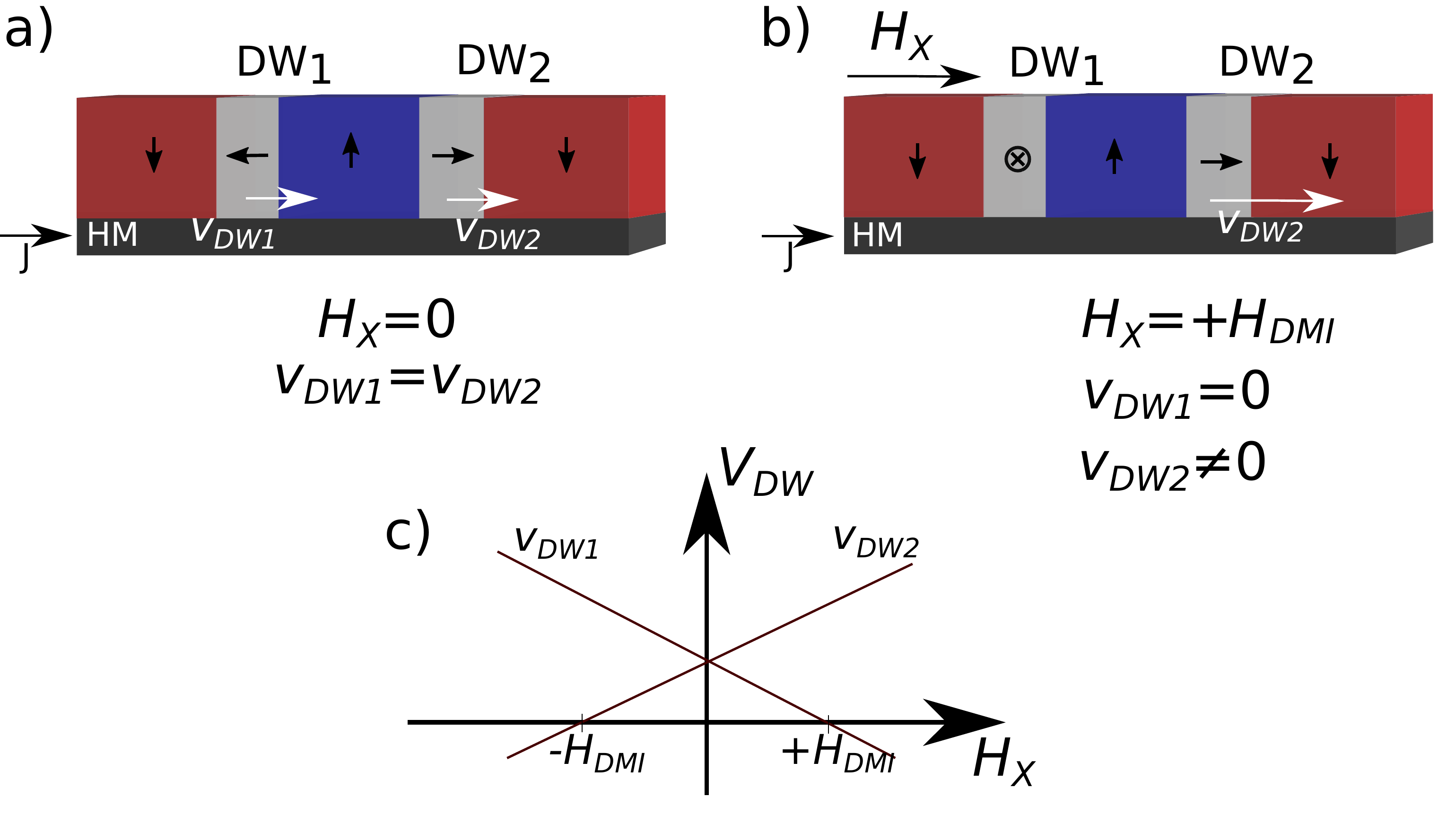}
    	\caption{Sketch of the dynamics of DWs driven by current in the absence (a) or presence (b) of an applied in-plane field $H_{x}$. The white arrows indicate DW velocity. The strength of $H_{x}$ applied in (b) matches that of the DMI field $H_\mathrm{DMI}$, thus stopping the motion of DW$_{1}$. A sketch of the velocity for the two DWs as a function of applied $H_{x}$ is illustrated in (c).}
	\label{fig:current}
\end{figure}

The current-driven DW dynamics is dramatically affected by the application of an in-plane magnetic field $H_x$ along the current direction. It is indeed due to this that DW velocity measurements as a function of $H_x$ provide a means to quantify $D$. Given a fixed $J$, it is observed that N\'eel DWs with the same chirality $\uparrow \rightarrow \downarrow$ and $\downarrow \leftarrow \uparrow$ have the same velocity when $H_{x} = 0$, while they move with different velocities under a non-vanishing $H_{x}$. In particular, for a given sign of $H_{x}$, the DW type for which $H_x$ is parallel to $H_\mathrm{DMI}$ moves faster while the other kind, where $H_x$ partially compensates $H_\mathrm{DMI}$, slows down with respect to its velocity at $H_{x} = 0$. The situation is reversed by changing the sign of the applied $H_{x}$. This remarkable behaviour has an important consequence: both DWs stop moving for a certain $|H_{x}|$, equal in strength but opposite in sign for $\uparrow \rightarrow \downarrow$ and $\downarrow \leftarrow \uparrow$ DWs respectively, as schematically shown in Fig.~\ref{fig:current}. 

The applied $H_{x}$ under which these N\'eel DWs stop moving -- typically referred to as the ``compensating'' or ``stopping'' field -- is the field that, opposing the DMI, restores a Bloch DW configuration, for which indeed no motion is expected via SHE-SOT. As such, this ``compensating" field $H_x^*$ can be considered equivalent in strength (but opposite in sign) to the $H_\mathrm{DMI}$ acting locally across the DWs. In other words, because of the change in the DW internal magnetization angle \cite{RYU-12}, N\'eel DWs move faster or slower depending on whether the effective in-plane field experienced by the DW is enhanced ($H_x + H_\mathrm{DMI}$) or decreased ($H_x - H_\mathrm{DMI}$), respectively. In the latter case, DWs stop moving for $H_x = H_x^* =  - H_\mathrm{DMI}$, when they become Bloch walls, and they reverse direction of motion for $H_x > - H_\mathrm{DMI}$, when the N\'eel configuration is re-established but with an opposite chirality. However, identifying $H_x^*$ with $H_\mathrm{DMI}$ is valid only if conventional STT, due to the spin-polarized current in the ferromagnet, \cite{SLO-96,BER-96} can be neglected. When a significant STT is present, the relationship between the $H_x^*$ and $H_{DMI}$ in the 1D model becomes \cite{EMO-13,RYU-13}:        
\begin{equation}
H_x^* = H_\mathrm{DMI} + \textrm{sgn}(\theta_\mathrm{SH})\frac{2}{\pi}\frac{\mu_\mathrm{B} P}{\gamma e M_\mathrm{S} \Delta}|J_\mathrm{e}|, \label{eqStoppingField}
\end{equation} 
where $\mu_\mathrm{B}$ is the Bohr magneton, $P$ is the spin current polarization, and $\gamma$ is the gyromagnetic ratio. Eq. \ref{eqStoppingField} implies that $H_x^*$ can depend on the amplitude of the current used to drive DWs. When only a modest dependence of $H_x^*$ on $J_e$ is observed, as for instance in \textcite{RYU-14,KAR-18a}, it is possible to conclude that the contribution from STT is small. In any case, once $H_{DMI}$ is determined, the magnitude of $D$ is derived through Eq. \ref{eq:DMI}, while its sign is inferred from the direction of the DW motion and the sign of the spin Hall angle $\theta_\mathrm{SH}$.   

The dependence of the DW velocity on an in-plane field for a fixed current density can be analytically described in the context of the 1D model, taking into account STT, SHE-SOT and DMI \cite{THI-12,EMO-13,RYU-13}. As seen in Fig.~\ref{fig:current}, the DW velocity is expected to be approximately linear with $H_x$ around the ``compensating'' field $H_x^*$, and some experimental works derive $H_\mathrm{DMI}$ by linearly fitting the data \cite{RYU-13,TOR-14}. In some cases, it has been observed that the DW velocity remains small or null in a quite large range of $H_x$ around $H_x^*$ \cite{RYU-14,LOC-15,LOC-17}, consistent with thermally activated creep regime, and strong pinning effects. To account for it, the 1D model has been extended to include an effective pinning potential both without \cite{RYU-14,LOC-17}, and with thermal fields \cite{LOC-15} to describe the influence of thermal fluctuations. In both cases, the modified 1D model has provided good agreement with the experimental data and has been used to extract $H_\mathrm{DMI}$. The range of $H_x$ values for which the DW velocity is negligible has been observed to decrease upon increasing the current density, due to a reduced influence of pinning \cite{RYU-14}. 

A few papers have shown that $D$ can also be quantified through current-driven DW dynamics by measuring the dependence of the DW depinning efficiency, rather than the DW velocity, on the in-plane field $H_x$ \cite{FRA-14,KIM-18}. The efficiency of DW depinning is defined as:
\begin{equation}
\epsilon = \mu_0 \frac{dH_\mathrm{SHE}}{dJ_\mathrm{e}},
\label{eqEff}
\end{equation} 
and is measured as the slope of the out-of-plane depinning field as a function of $J_e$. The DW depinning efficiency changes as a function of $H_x$, due to the corresponding variation of the DW internal structure. In particular, $\epsilon$ is found to vanish at a certain $H_x^*$, equal in strength but opposite in sign for $\uparrow \rightarrow \downarrow$ and $\downarrow \leftarrow \uparrow$ DWs. Similarly to what was already discussed for the DW velocity dependence on $H_x$, this in-plane field $H_x^*$ for which $\epsilon = 0$ represents the field at which a Bloch DW configuration is restored, and can thus be identified with $H_\mathrm{DMI}$.

\subsubsection{Field-driven domain wall motion}\label{sec:theoryDWfield}

\begin{figure}[t]
	\centering
	\includegraphics[width=.7\linewidth,trim=0 0 440 0,clip]{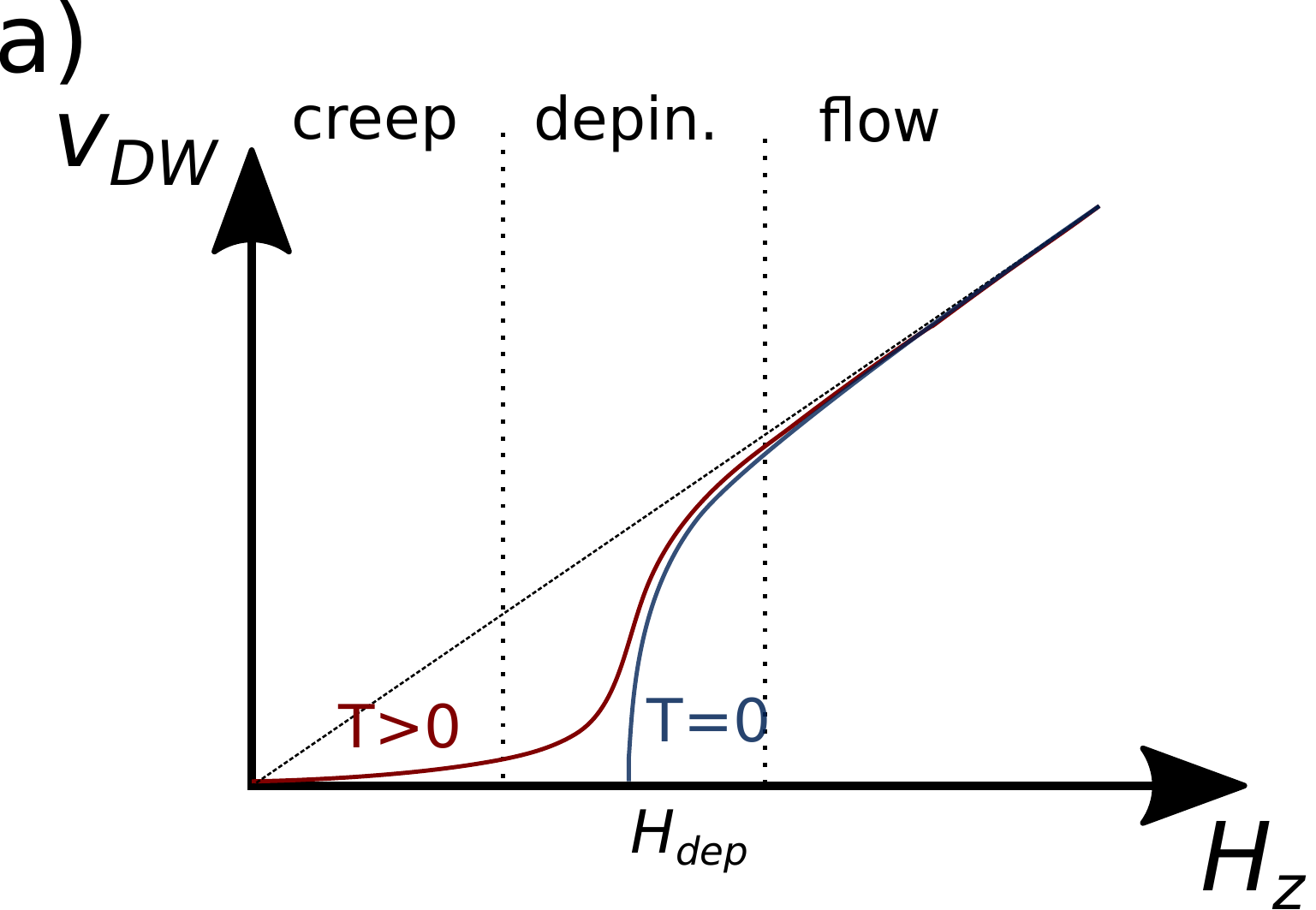}\\
	\includegraphics[width=.7\linewidth,trim=440 0 0 0,clip]{creep_dep_flowv2.pdf}\\
	\caption{Dependence of the DW velocity $v_\mathrm{DW}$ as a function of increasing applied field. In (a) the three regimes - creep, depinning and flow - are visible. The depinning transition, which is abrupt at zero temperature, shows some rounding due to thermal effects. At higher applied fields (b), two distinct regimes can be distinguished within the flow regime, namely steady and precessional flow. The drop in the velocity is called Walker breakdown.}
	\label{fig:creep-dep-flow}
\end{figure}

The simplest and most common way to move DWs in PMA materials is by applying a perpendicular field $H_{z}$. To minimize the Zeeman energy associated with $H_{z}$, domains with magnetization along the field direction expand at the expense of the others, leading to DW motion. Field-driven DW dynamics are typically classified into three distinctive regimes -- creep, depinning, and flow -- which occur in succession upon increasing $H_z$, as shown in Fig.~\ref{fig:creep-dep-flow}a.

For sufficiently low driving fields DWs move in the thermally activated creep regime where they interact strongly with disorder, and their velocity grows exponentially as $v \sim \exp (-H_z)^{-1/4}$ \cite{LEM-98,CHA-00}. Upon increasing $H_z$ above a critical value, known as the depinning field $H_\mathrm{dep}$, disorder starts to become irrelevant and the DW velocity grows as $v \sim (H_z - H_\mathrm{dep})^{\beta}$ \cite{CHA-00}, with $\beta$ being the depinning exponent at $T=0$ as in Fig.~\ref{fig:creep-dep-flow}a. Finally, for $H_z \gg H_\mathrm{dep}$ the DW enters the flow regime where the velocity increases linearly with $H_{z}$ up to the so-called Walker field $H_\mathrm{W}$, which marks a significant decrease in DW velocity, due to a change in its internal dynamics \cite{MET-07}. For $H_z \gg H_\mathrm{W}$ the DW recovers a linear flow with $H_{z}$, albeit with a reduced mobility $dv/dH$.

With the exceptions of a few works \cite{VAN-15,JUE-16,PHA-16,AJE-17,SOU-19,KRI-19}, the DMI has been mostly quantified through experiments of field-driven DW motion in the creep regime, which is addressed below. Methods to extract the DMI from DW dynamics in the flow regime will be discussed later. 

\paragraph{Creep regime}
%\subsubsubsection{Creep regime}

In the creep regime, DWs are driven by modest fields (typically down to a few percent of $H_\mathrm{dep}$) and move slowly by thermal activation, interacting strongly with disorder of various origin (pinning defects, film thickness variations, $M_\mathrm{S}$ inhomogeneities, etc). The DW creep dynamics is understood in terms of the motion of a one-dimensional elastic line in a two-dimensional disordered potential. The dependence of the DW velocity $v$ on the applied field $H_z$ is  described by the so-called creep law \cite{LEM-98, CHA-00}:
\begin{equation}
v = v_0 \exp {[-\zeta(\mu_0 H_z)^{-\mu}]},
\label{eq:Creep}
\end{equation}
where $v_0$ is the characteristic speed proportional to the attempt frequency for DW propagation, $\zeta$ is a scaling constant, and $\mu = 1/4$ is the scaling exponent for the 1D elastic line.  

\begin{figure}[t]
	\centering
	\includegraphics[width=0.9\linewidth]{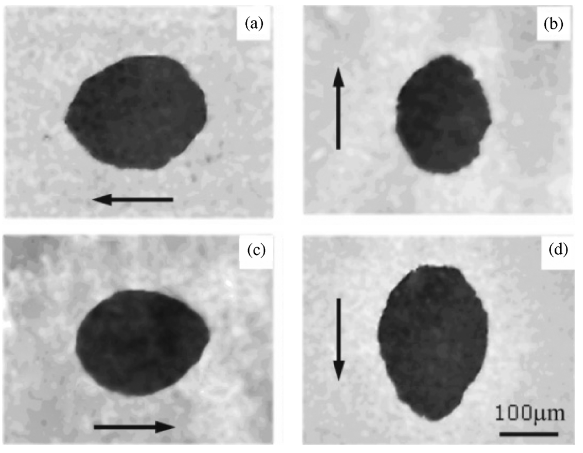}
	\caption{First observation by magneto-optical Kerr effect (MOKE) of the asymmetric expansion of a bubble under in-plane magnetic field in a continuous film of Pt/Co/Pt. Arrows show the in-plane field directions. From \citealp{KAB-10}.}
	\label{fig:KAB10}
\end{figure}

Creep motion of DWs driven by a perpendicular field $H_z$ is significantly altered by the simultaneous presence of an in-plane field $H_x$, similarly to what already discussed for the current-driven case (see Sec.~\ref{sec:theoryDWcurrent}). Indeed, it is observed experimentally that when a circular magnetic bubble expands under the application of $H_z$ only, the radial symmetry is maintained and the bubble grows isotropically -- thus retaining its original shape. However, the symmetry is broken when the bubble is expanded under the application of both $H_z$ and $H_x$, as N\'eel $\uparrow \rightarrow \downarrow$ and $\downarrow \leftarrow \uparrow$ DWs acquire different velocities along the direction of the applied $H_x$. This circumstance was first observed in \textcite{KAB-10} for continuous films of Pt/Co/Pt (see Fig.~\ref{fig:KAB10}), mentioning the interfacial DMI as a possible origin. Only later it was fully understood and modelled in the context of DW creep.

The first model proposed to explain the asymmetric motion of N\'eel DWs with the in-plane field \cite{JE-13}, suggested to modify the creep law (Eq.~\ref{eq:Creep}) by changing the scaling parameter $\zeta$ to take into account the dependence of DW energy on the in-plane field \cite{JE-13}:
\begin{equation}
\zeta(H_{x}) = \zeta_0[\sigma(H_x)/\sigma(0)]^{1/4},
\label{eq:Sigma}
\end{equation}
where $\zeta_0$ is a scaling constant and $\sigma$ is the DW energy density. The dependence of DW energy density on the in-plane field has been calculated as \cite{THI-12, JE-13}:
\begin{equation}
\sigma(H_{x}) = \sigma_0 - \frac{\pi^2\Delta\mu_0^2 M_\mathrm{S}^2}{8K_\mathrm{D}}(H_x + H_\mathrm{DMI})^2,
\label{eq:Sigma1}
\end{equation}	 
when the condition $|H_x + H_\mathrm{DMI}| < 4K_\mathrm{D}/\pi\mu_0M_\mathrm{S}$ is satisfied. In this case the effective field acting on the DW is not strong enough to fully convert it into a N\'eel DW. Otherwise, for higher fields and fully  N\'eel DWs the expression becomes:
\begin{equation}
\sigma(H_{x}) = \sigma_0 + 2K_\mathrm{D}\Delta - \pi\Delta\mu_0M_\mathrm{S}|H_x + H_\mathrm{DMI}|.
\label{eq:Sigma2}
\end{equation}	

\begin{figure}[t]
	\centering
	\includegraphics[width=0.8\linewidth]{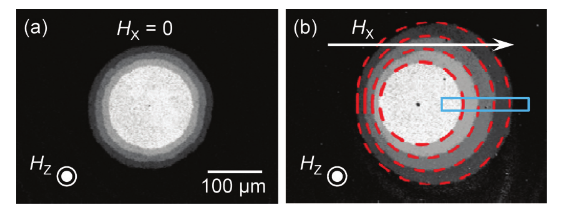}\\
	\includegraphics[width=0.8\linewidth]{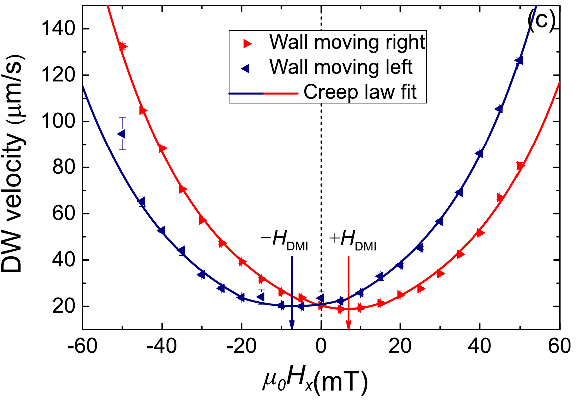}
	\caption{(a-b) Dynamics of bubble DWs driven by perpendicular field $\mu_0 H_z$=3~mT in the absence (a) or presence (b) of an in-plane field $\mu_0 H_x$=50~mT in a Ta(5~nm)/Pt(2.5~nm)/Co(0.3~nm) thin film with PMA. Each image is obtained by adding four sequential images with a fixed time step (0.4~s), which are captured using a magneto-optical Kerr effect microscope. The blue box in (b) designates the field of measurement. From \citealp{JE-13}.
    (c) A typical symmetric profile of the DW moving left and right as a function of the in-plane field $H_x$ in Ta(5~nm )/Co$_{20}$Fe$_{60}$B$_{20}$(0.8~nm)/MgO(2~nm) thin film. Here, the velocity minima occur at $|H_x| = H_\mathrm{DMI}$. From \citealp{KHA-16}.}
	\label{fig:bubble}
\end{figure}

In these equations, $\sigma_0 = 4 \sqrt{AK_\mathrm{eff}}$ is the Bloch DW energy density, $\Delta = \sqrt{A/K_\mathrm{eff}}$ is the domain wall width parameter, and $K_\mathrm{D} = N_x \mu_0 M_\mathrm{S}^2/2$ is the DW shape anisotropy energy density\footnote{$K_D$ represents the fact that Bloch DWs are magnetostatically more stable in the absence of DMI due to the high PMA of the films.}, with the demagnetising factor of the DW given by $N_x = \textrm{ln}(2) d / (\pi \Delta)$, with $d$ the magnetic film thickness \cite{TAR-98}.  
In other words, this model predicts that the effective in-plane field acting locally on the N\'eel DWs on either side of the bubble can be increased (decreased) if $H_x$ and  $H_\mathrm{DMI}$ have the same (opposite) sign, resulting in smaller (larger) DW energy $\sigma$ and thus a faster (slower) DW, just as for the current-driven case (see Fig.~\ref{fig:bubble}).

The field $H_x^* = - H_{\mathrm{DMI}}$ at which N\'eel DWs are converted into Bloch DWs is then equal in magnitude but opposite in sign for $\uparrow \downarrow$ and $\downarrow \uparrow$ DWs, respectively. It is important to notice that, differently from the current-driven case, field-driven Bloch DWs do not stop moving under the in-plane field $H_x^*$, since the perpendicular field $H_z$ keeps expanding the magnetic bubble to minimize Zeeman energy. Rather, in field-driven experiments, $|H_x^*| = H_{\mathrm{DMI}}$ corresponds to a minimum in DW velocity. In this simple model, the velocity for each type of DW should be symmetric around its own minimum, as in Fig.~\ref{fig:bubble}c, although in many experimental cases it is not. It follows that in the absence of DMI, i.e. when the bubble DW is in the Bloch configuration, both sides of the bubble have the same velocity dependence with $H_x$, show a minimum at $H_x = 0$ and are symmetric around $H_x = 0$ \cite{KIM-15}. This model can be used to fit the dependence of bubble DW velocities on in-plane field using three fitting parameters: $v_0$, $\zeta_0$, and $H_{DMI}$ itself. Alternatively, the scaling parameters $v_0$ and $\zeta_0$ can be extracted separately as the intercept and gradient of a linear fit to the curve $\textrm{ln}(v)$ vs. $ H_z^{-1/4}$ for $H_x = 0$, thus leaving $H_\mathrm{DMI}$ as the only fitting parameter of the $v$ dependence on $H_x$. Once $H_\mathrm{DMI}$ has been determined, the magnitude of $D$ is derived through Eq.~\ref{eq:DMI}, while its sign is inferred from the orientation of the bubble asymmetry with respect to the $H_x$ direction.  

\begin{figure}[t]
	%\centering
	\includegraphics[width=0.9\linewidth]{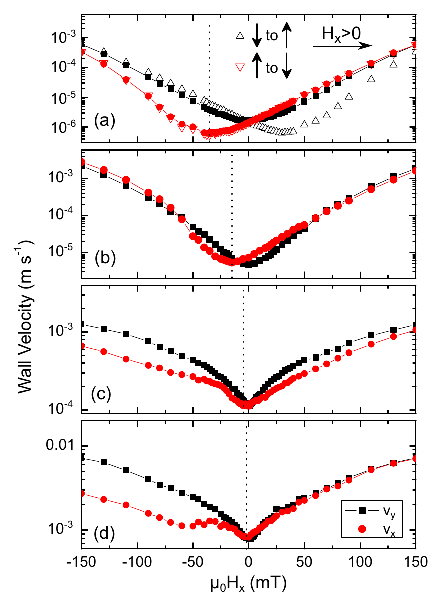}
	\caption{Example of asymmetry of the domain wall velocity profile as a function of the in-plane field (red dots) for a set of multilayers with nominal structure  Si/SiO$_2$/X($t$)/Co$_{20}$Fe$_{60}$B$_{20}$(1~nm)/MgO(2~nm)/Ta(1~nm) with four underlayers X($t$): W (2~nm, (a)), W (3~nm, (b)), TaN (1~nm, (c)), and Hf (1~nm, (d)). All cases are highly asymmetric in the velocity profile, even showing a local maximum as for the latter case (d). Note that the velocity in the direction orthogonal to the in-plane field (black squares) remains symmetric. From \citealp{SOU-16}.}
	\label{fig:asymmetric_v_profile}
\end{figure}

This modified creep model, in which the in-plane field affects DW dynamics only through a variation of domain wall energy, has been successfully applied to fit several experimental data and estimate the interfacial DMI constant $D$ \cite{JE-13,HRA-14,PET-15,YU-16,KHA-16,WEL-17,KIM-17,KUS-18, SHA-18}. However, for a growing number of experiments, as for instance in Fig.~\ref{fig:asymmetric_v_profile}, the model fails to provide an adequate description of the data, which often show an asymmetric behavior around the minimum velocity \cite{LAV-15,JUE-16a,LAU-16,SOU-16,PEL-17,CAO-18, KIM-18,SHE-18,SHA-19}, a local peak in velocity \cite{LAV-15,SOU-16,BAL-17}, or even a maximum in velocity in the flow regime \cite{VAN-15}.

To explain these non-symmetric results, several different mechanisms have been suggested. \textcite{JUE-16a} proposed to neglect any DW energy contribution, and consider a DW chirality-dependent damping, which acts as a dissipative SOT on the DW: this damping would modulate the attempt frequency for DW motion, and would thus modify the characteristic speed $v_0$ of the creep law (see Eq.~\ref{eq:Creep}). On the contrary, \textcite{LAV-15} suggested that chiral damping could not be an explanation of the asymmetry of their own results, as $v_0$ was found to be symmetric with respect to $H_{x}$. The importance to consider possible $H_{x}$ dependences on both creep parameters $v_0$ and $\zeta$ (not necessarily related to chiral damping) has been highlighted in \textcite{BAL-17,SHE-18}. While Shepley \textit{et al.} showed that the dependence on $H_{x}$ is asymmetric for both $v_0$ and $\zeta$, Balk \textit{et al.} modelled their results through a modified creep law that takes for both parameters a dependence on $H_{x}$ into account. By combining Eqs.~\ref{eq:Creep}, and~\ref{eq:Sigma}, with their expression for $v_0$, they were able to fit velocity vs $H_{x}$ curves that showed a local peak around $H_{x} = 0$, introducing a local anisotropy due to pinning $K_\mathrm{pin}$, and having $\zeta_0$ and $H_\mathrm{DMI}$ as fitting parameters. 

Several analytical and numerical studies were devoted to understanding these features. \textcite{KIM-16} attributed the asymmetry in the DW energy density $\sigma$, to the asymmetric variation of the DW width with $H_x$, later confirmed by micromagnetic simulations \cite{SAR-18}. \textcite{LAU-16} described the velocity asymmetry in terms of the Wulff construction, which yields a methodology to determine the shape of a magnetic bubble, although it does not explicitly provide a model for the velocity as a function of in-plane field. They also speculated that nucleation and annihilation of Bloch lines may be responsible for a peculiar flattening of magnetic bubbles, which could in turn cause the observed velocity asymmetry. This again was confirmed through micromagnetic simulations by \textcite{SAR-18}. \textcite{PEL-17} argued that under an applied magnetic field $H_x$, where the DW energy density $\sigma$ becomes anisotropic with respect to the DW orientation in the film plane, the correct elastic energy that should be considered to describe the creep regime does not simply identify with $\sigma$, as typically assumed in the phenomenological model of creep \cite{LEM-98}. Rather, to reproduce the asymmetry in their velocity curves, they propose that $\sigma$ should be replaced by the DW stiffness $\widetilde{\sigma} = \sigma + \sigma ^{\prime\prime}$, with $\sigma ^{\prime\prime} = \frac{\partial ^2 \sigma}{\partial \Theta^2}$, where $\Theta$ is the azimuthal angle of the DW normal. Interestingly, the $H_\mathrm{DMI}$ value extracted using this stiffness model was found to be higher than the field at which the minimum in velocity is observed. In a later work \cite{LAU-18}, the same authors proposed to upgrade the stiffness model, by taking into account a possible variation of the characteristic speed $v_0$ with $H_x$, which they also speculate could be due to a chiral damping mechanism. Through this improved model, they could fit velocity curves as a function of $H_x$ that are not only asymmetric about the minimum, but also show a crossover between DW velocities at opposite sides of the bubble. More recently, another model was advanced by \textcite{SHA-19} to explain the presence of both asymmetry and DW crossover in the velocity curves. Here, the DW depinning field $H_\mathrm{dep}$ is allowed to vary with $H_x$ in a way determined from micromagnetic simulations. Notably, this work shows that the velocity minimum underestimates $H_\mathrm{DMI}$, as was also found for the stiffness model \cite{PEL-17}. This is also confirmed by \textcite{HAR-19} and \textcite{GEH-20}: the former reconsidered the change of the DW stiffness due to deformation as an angular shape, and minimizing the energy of the system with a semi-analytical approach they could calculate the velocity profile and show that the minimum of the velocity does not occur at $H_\mathrm{DMI}$. The latter showed that $H_x$ can also modify the characteristic length scale of pinning, in strong correlation with the DW width, again implying that the minimum of the velocity could not correspond to the $H_\mathrm{DMI}$ field.

\paragraph{Flow regime}
%\subsubsubsection{Flow regime}

According to the 1D model \cite{THI-12}, the presence of DMI significantly increases the Walker field $H_\mathrm{W}$. Indeed, in samples with a strong enough DMI to convert the wall to a fully N\'eel form, $H_\mathrm{W}$ is proportional to $H_\mathrm{DMI}$ :
\begin{equation}
H_\mathrm{W} \propto \alpha H_\mathrm{DMI},
\label{Eq:H_wb} 
\end{equation}
where $\alpha$ is the Gilbert damping constant \cite{THI-12}. The DW velocity at Walker breakdown can thus be expressed as:
\begin{equation}
v_\mathrm{W} = \gamma_0 \frac{\Delta}{\alpha} H_\mathrm{W} \sim \frac{\pi}{2} \gamma \frac{D}{M_\mathrm{S}},
\label{Eq:v_wb} 
\end{equation}
where $\gamma_0$ is the gyromagnetic ratio. In contrast to the prediction of the 1D model, it is found experimentally and confirmed by 2D micromagnetic simulations that in samples with large DMI, the DW velocity does not decrease at fields larger than $H_\mathrm{W}$, but instead reaches a plateau \cite{PHA-16,YOS-16,AJE-17}. This fact originates from the complex meander-like structure that the DW adopts at velocities above $H_\mathrm{W}$, with continuously nucleation and annihilation of pairs of vertical Bloch lines \cite{YOS-16,PHA-16}. Measurements of this roughly constant velocity, which corresponds to $v_\mathrm{W}$, provide a simple way to determine $D$ for samples with large DMI. A combination of experiments and modelling has been used to show that $D$ is also associated with the value of $H_z$ at which the end of the plateau is reached \cite{KRI-19}.

In other works \cite{VAN-15,PHA-16,AJE-17,SOU-19} measurements of the minimum DW velocity as a function of $H_x$ in the flow regime have also been used to quantify $D$: in the flow regime the DW velocity depends only on DW width (not DW energy) and this assumes the minimum value when the DW is Bloch. Indeed, it has been shown that under certain conditions the 1D model provides an expression for the DW velocity that exhibits a parabolic dependence on $(H_x + H_\mathrm{DMI})$ \cite{KIM-16d}.

\subsubsection{Equilibrium stripe domain pattern}\label{sec:theorydwwidth}
\begin{figure}[t]
\centering
\includegraphics[width=.9\linewidth]{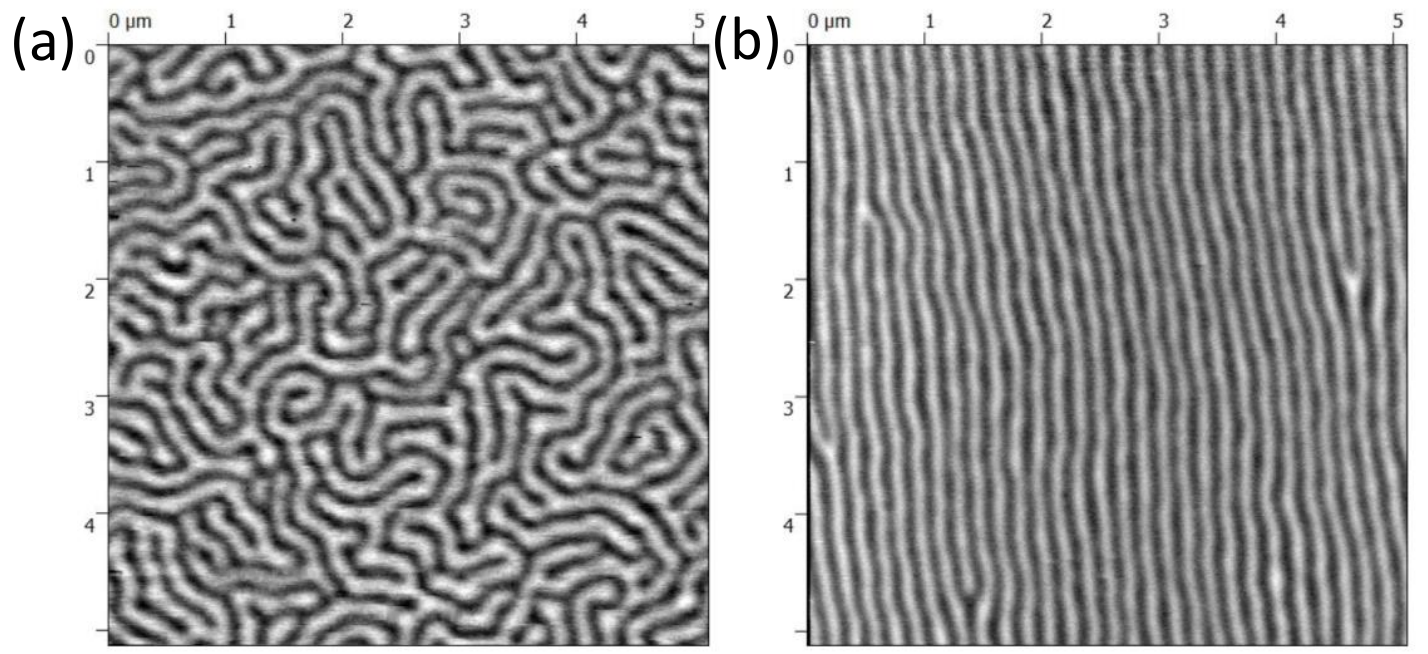}
\caption{MFM domain patterns for an (a) out-of-plane and (b) in-plane demagnetized multilayer composed of  [Pt(1~nm)/Co(0.8~nm)/Al$_2$O$_3$(1~nm)]$_{20}$. From Supplementary Materials of \citealp{LEG-18}.}
\label{fig:domainlab}
\end{figure}

The demagnetized state of PMA materials consists of a complex domain pattern, usually in the form of labyrinth structures of domains pointing either up or down. If the thickness of the sample is very small, the equilibrium domain width can exceed the sample dimensions and such structures are not expected. Examples of the typical patterns are shown in Fig.~\ref{fig:domainlab} for a multilayer of Pt/Co/Al$_2$O$_3$, taken from \textcite{LEG-18}. The exact demagnetized pattern depends on the direction and the history of the applied field: when an out of plane field is applied,  a maze domain structure is created (Fig.~\ref{fig:domainlab}a), while for an in plane field,  domain walls remain almost parallel and typically form a stripe structure (Fig.~\ref{fig:domainlab}b). In the latter, the width and the density of domains can be used to estimate the domain wall energy. As suggested by the theory developed for infinite parallel stripe domains \cite{MAL-58,KOO-60}, the domain width is a function of the domain wall energy, which includes magnetostatic, anisotropy, Zeeman, and exchange terms. N\'eel walls, favored by DMI, have a reduced energy and, consequently the DMI markedly affects the equilibrium domain width. 

For domain widths much larger than the domain wall width, as is generally the case for PMA materials, the magnetostatic contribution comes only from the surfaces charges at the top and bottom surfaces, so that DW energy is calculated as:
\begin{equation}
\label{eq:energy_domain_wall}
\sigma_{DW} = \frac{2A}{\Delta}+2K_\mathrm{eff}\Delta-\pi \cos(\Phi)|D| 
\end{equation}
where $\Phi$ is the same angle of the internal DW magnetization defined in Fig.~\ref{fig:1Dmodel}. The minimization of the energy with respect to the domain width $\Delta$ and core angle $\Phi$ gives the value of the equilibrium domain width: a critical value $D_\mathrm{c}$ is obtained. For $D > D_\mathrm{c}$ the preferred configuration is a N\'eel wall and for $D \leq D_\mathrm{c}$ it gradually transforms into a Bloch wall as $D$ approaches zero. 

Initially, this method was limited to N\'eel walls only, with $D>D_\mathrm{c}$, and hence $\cos(\Phi)=1$, so that the DW energy density simplifies to $\sigma_\mathrm{DW}=\sigma_0-\pi|D|$. Later, the inclusion of the angle of the wall $\Phi$ in Eq.\ref{eq:energy_domain_wall} removed this limitation \cite{MEI-17,LEM-17}. A further improvement took into account the dipolar terms coming from the internal structure of the DW or/and from the DW interaction \cite{LEM-17}. This is important in thicker samples, as the dipolar energy causes the internal magnetization to vary along the thickness, even in the presence of DMI. To account for all these situations, the model was extended allowing a different angle $\Phi$ for each layer of the sample \cite{LEM-18a}. 

The analytical expression for the DW energy $\sigma_\mathrm{DW}$ as a function of the periodicity $w$ of the domains, which is twice the domain width, is
\begin{equation}
\label{eq:energy_domain_width}
\frac{\sigma_{DW}}{\mu_0 M_\mathrm{S}^2 d}= \frac{w^2}{d^2\pi^3}\displaystyle\sum_{\mathrm{odd}\; n=1}^{\infty} \frac{1-\left(1+\frac{2 \pi  n d}{w}\right)e^{-\frac{2 \pi n d}{w}}}{n^3}
\end{equation} 
where $d$ is the thickness. From the experimentally obtained domain width one can then calculate the domain wall energy, and estimate the DMI constant using one of the previously derived approximations. 

This general method has been applied in three different ways: comparisons to micromagnetic simulations \cite{MOR-16,LEG-18}, analytical estimations \cite{WOO-16,YU-17,WOO-17,WON-18}, and scaling of the energy of an experimental image \cite{BAC-19}.

\subsubsection{Magnetic stripe annihilation}\label{sec:theoryannihilation}
\begin{figure}[t]
\centering
\includegraphics[width=.9\linewidth]{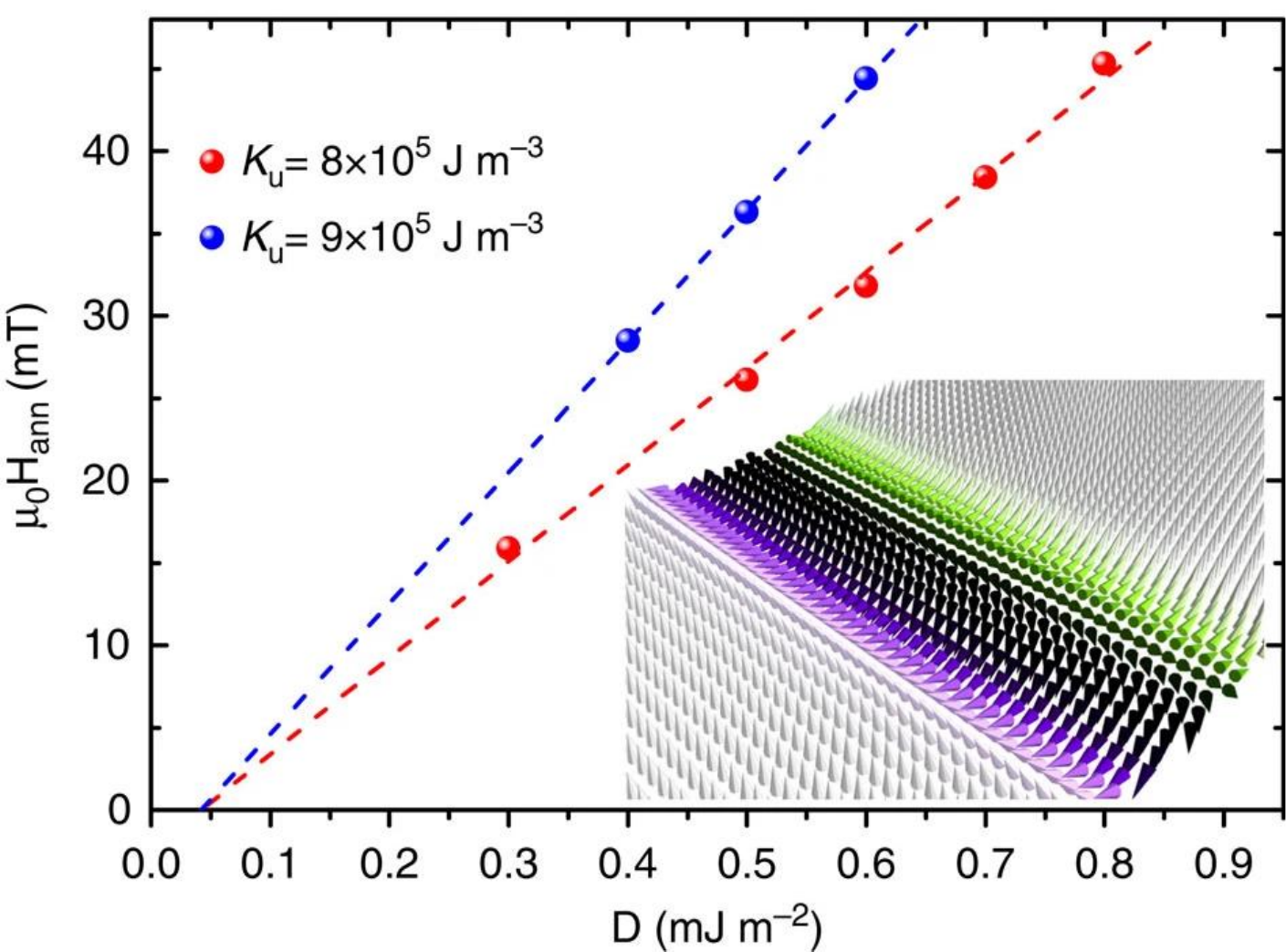}
\caption{Annihilation field of two homochiral walls as a function of the DMI value calculated from micromagnetic simulations for two different anisotropy values. Inset: Magnetic configuration of the two N\'{e}el walls squeezed together by the perpendicular field.  From \citealp{BEN-15}.}
\label{fig:annil}
\end{figure}

Two parallel domain walls in a thin film with strong enough DMI are homochiral N\'{e}el walls.  Since the chirality of the domain walls is the same, either left-handed or right-handed, the core magnetization of those parallel domain walls point in opposite directions. Such a pair of domains constitutes a topological structure. This situation is different from the case of absence of DMI, where the core magnetization of the Bloch walls point in the same direction but not in a fixed orientation. The annihilation of these two parallel N\'eel walls depends on the strength of the DMI due to the topological configuration. This fact was confirmed by simulations \cite{MAR-14a,HIR-14}. To annihilate the walls an out of plane field is applied that reduces the size of the domain disfavored by the field. This domain achieves a minimum size until the walls meet as a so-called ``winding pair'' to form a $360^\circ$ wall, which annihilates when a given value of the out-of-plane field is exceeded. From the determination of the field of annihilation and the minimum domain width the DMI can be extracted when compared with the corresponding simulations \cite{HIR-14}. An example from \cite{BEN-15} of the annihilation field dependence on DMI calculated by means of micromagnetic simulations can be seen in Fig.~\ref{fig:annil}. More recently, a formula for the minimum width of stripe domains was derived \cite{LEM-17}, using the analytical formulation discussed in the previous section. This allowed the extraction of the DMI value without the performance of systematic micromagnetic simulations.

The analytical formula for the domain wall energy $\sigma_\mathrm{DW}$ obtained from the minimum domain width $w_\mathrm{min}$ is given by:  
\begin{equation}
\label{eq:energy_domain_width_min}
\frac{\sigma_\mathrm{DW}}{\mu_0 M_\mathrm{S}^2 d}=\frac{1}{2\pi}(\ln[1 + (w_\mathrm{min}/d)^2] + (w_\mathrm{min}/d)^2 \ln[1 + (w_\mathrm{min}/d)^{-2}]).
\end{equation}
As in the previous method this allows extracting the value of the domain wall energy density which is also a function of the DMI constant. Using the same expressions for the energy density as given in the previous section, one can estimate its value.

\begin{figure*}[t]
\centering
\includegraphics[width=.7\textwidth]{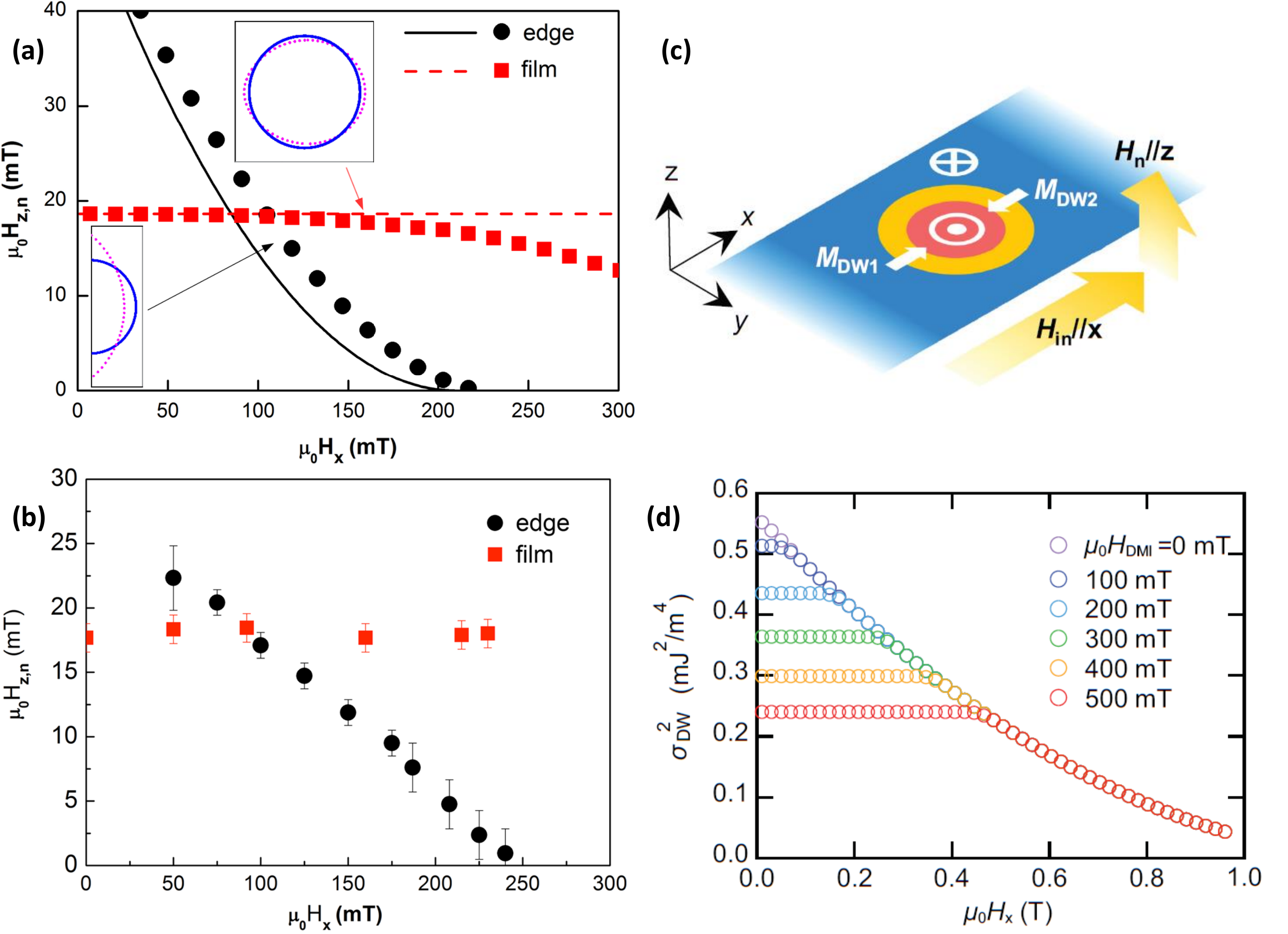}
\caption{(a) Theoretical nucleation field as a function of the magnetic field applied perpendicular to the edge of the sample for edge and bubble (film) nucleation as a function of the field. (b) Experimental nucleation field as a function of the magnetic field applied perpendicular to the edge for edge and bubble (film) nucleation. (c) Schematics of the nucleation of bubble domain (red color) showing the orientation of the wall core parallel and anti-parallel to the applied field for in-plane fields below the DMI field. (d) Calculated domain wall energy as a function of the in-plane field for different DMI fields.
(a) and (b) are from~\citealp{PIZ-14}, (c) and (d) from~\citealp{KIM-17b}.}
\label{fig:nucleation}
\end{figure*}

\subsubsection{Nucleation field}\label{sec:theorynucleation}
As stated above, the energy of a domain wall is reduced by the presence of DMI. This modification also affects the energy barrier relevant for nucleation of reversed domains. \textcite{PIZ-14} showed that the nucleation process is affected by the presence of DMI. They distinguished between nucleation at the edge of the patterned sample and in the center of the magnetic film. They showed that the nucleation of reversed domains at the edge depends on the value of the in-plane applied field and it is asymmetric with respect to the combination of the DMI sign and the in-plane field direction. The nucleation is favored at the edge of the sample having a nucleated wall with the core magnetization in the direction of the in-plane field. 
The half-droplet model \cite{VOG-06}, applied in the paper, describes the nucleation of a magnetic domain at the side edge under the application of an in-plane magnetic field. The out-of-plane nucleation field for a reversed domain is:
\begin{equation}
\label{eq:nucleationfield}
H_\mathrm{n}=\frac{\pi\sigma^2d}{2\mu_0 M_\mathrm{s}pk_\mathrm{B}T}
\end{equation}
where $\sigma$ is the domain wall energy associated with the bubble, $T$ is the temperature and $p$ is the factor related to the waiting time according to $\tau=\tau_0 \exp(p)$ where $\tau_0$ is the attempt frequency. The DW energy is a function of the in-plane field and the DMI constant. From the best adjustment of their numerical model to the experimental results, they were able to estimate the DMI constant in Pt/Co/AlOx.

In the case of a bubble domain away from the edges, \textcite{PIZ-14} concluded that the Zeeman energy gained within the half droplet having a DW magnetization component parallel to the in-plane field is compensated by the loss of energy within the half droplet with opposite magnetization (see the orientation of the magnetization in Fig.~\ref{fig:nucleation}(b)), while this is not true for an incomplete bubble nucleated at the edge. Therefore, the nucleation field for a complete bubble is independent of the in-plane field value. This fact agrees well with their model shown in Fig.~\ref{fig:nucleation}(a) and their measurements shown in Fig.~\ref{fig:nucleation}(b).

More recently, \textcite{KIM-17b} realized that such an argument does not hold for bubbles in films above a critical in-plane field value, corresponding to the DMI field, where the energy of the bubble is altered due to the fact that the magnetization at its boundary aligns with the in-plane field. Hence, the picture in Fig.~\ref{fig:nucleation}(c) is no longer valid.  This fact leads to a reduced value of the bubble magnetic energy and an associated reduction of the nucleation field when the in-plane field value is increased beyond the DMI field. This was in agreement with their numerical calculations as shown in Fig.~\ref{fig:nucleation}(d). This fact was not observed in \textcite{PIZ-14} due to the large DMI field in the samples studied in their work.
 
Therefore, the critical field above which a change from a constant nucleation field to a field dependent nucleation is observed is a measure of the DMI constant, since it coincides with the DMI field $H_{DMI}$. Experimentally, this can be determined from the dependence of the out-of-plane nucleation field on the in-plane field.

\begin{figure}[t]
\centering
\includegraphics[width=.8\linewidth]{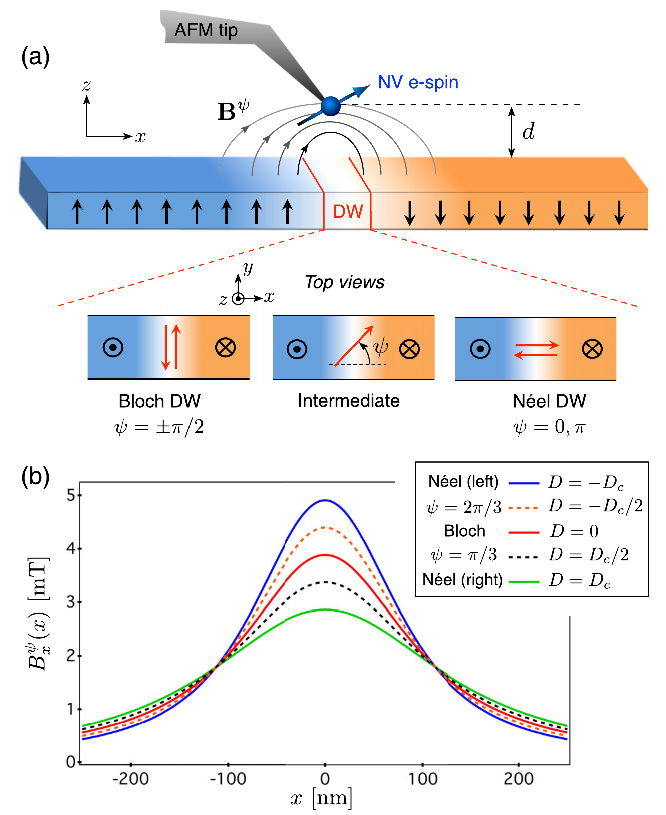}
\caption{Schematic of the stray field measurement using a NV magnetometer. The internal DW magnetization angle $\psi$ ($\Phi$, in our notation) depends on the DMI constant, resulting in different profiles of the stray field component $B_x^\psi(x)$. From \citealp{GRO-16}.}
\label{fig:NV_center}
\end{figure}

\subsubsection{Domain wall stray fields}

Another method taking into account the static DW structure is the direct measurement of the stray field in the DW. The stray field has a typical profile along the axis perpendicular to the DW including information on the angle $\Phi$ and the strength and sign of $D$ (see Fig.~\ref{fig:NV_center}). This method is limited to low values of $D \leq D_\mathrm{c}$, i.e. when a significant angle is present. $D_\mathrm{c}= 2\mu_0 M_\mathrm{S}^2 d \ln 2/\pi^2$ is the critical $D$ value above which formation of fully oriented N\'eel walls occurs ($d$ is the film thickness). Therefore only samples with relatively small $D$ can be measured since $D_\mathrm{c}$ is typically on the scale of 0.2~mJ/m$^2$ (see Tab.~\ref{table:Tablenvcenter}). For samples with larger DMI, where fully oriented N\'eel walls are present, only a lower limit of $D$ can be given together with its sign \cite{TET-15}. Since DW widths are of the order of 5-10~nm, the stray field profile determination requires an excellent spatial resolution, which was experimentally tackled by nitrogen vacancy (NV) magnetometry \cite{GRO-16}. 

\subsubsection{Domain wall internal structure imaging}

With the progress in the development of advanced microscopy techniques with nanometric resolution and magnetic contrast, it has become possible to access not only the size and distribution of magnetic domains, but also to study the detailed structure of domain walls. As discussed in the previous sections, the domain wall structure is influenced by the presence of DMI, so that from a detailed identification of the wall structure one can in principle quantify the DMI strength. Basically four methods for spin sensitive imaging were used in literature: pioneering works were performed by spin-polarized scanning tunneling microscopy/spectroscopy (SP-STM/STS), the most quantitative works were done by spin-polarized low-energy electron microscopy (SPLEEM) and in a few more recent works Scanning electron microscopy with polarization analysis (SEMPA) and Lorentz-transmission electron microscopy (L-TEM) were employed.

\begin{figure}[t]
\centering
\includegraphics[width=.7\linewidth]{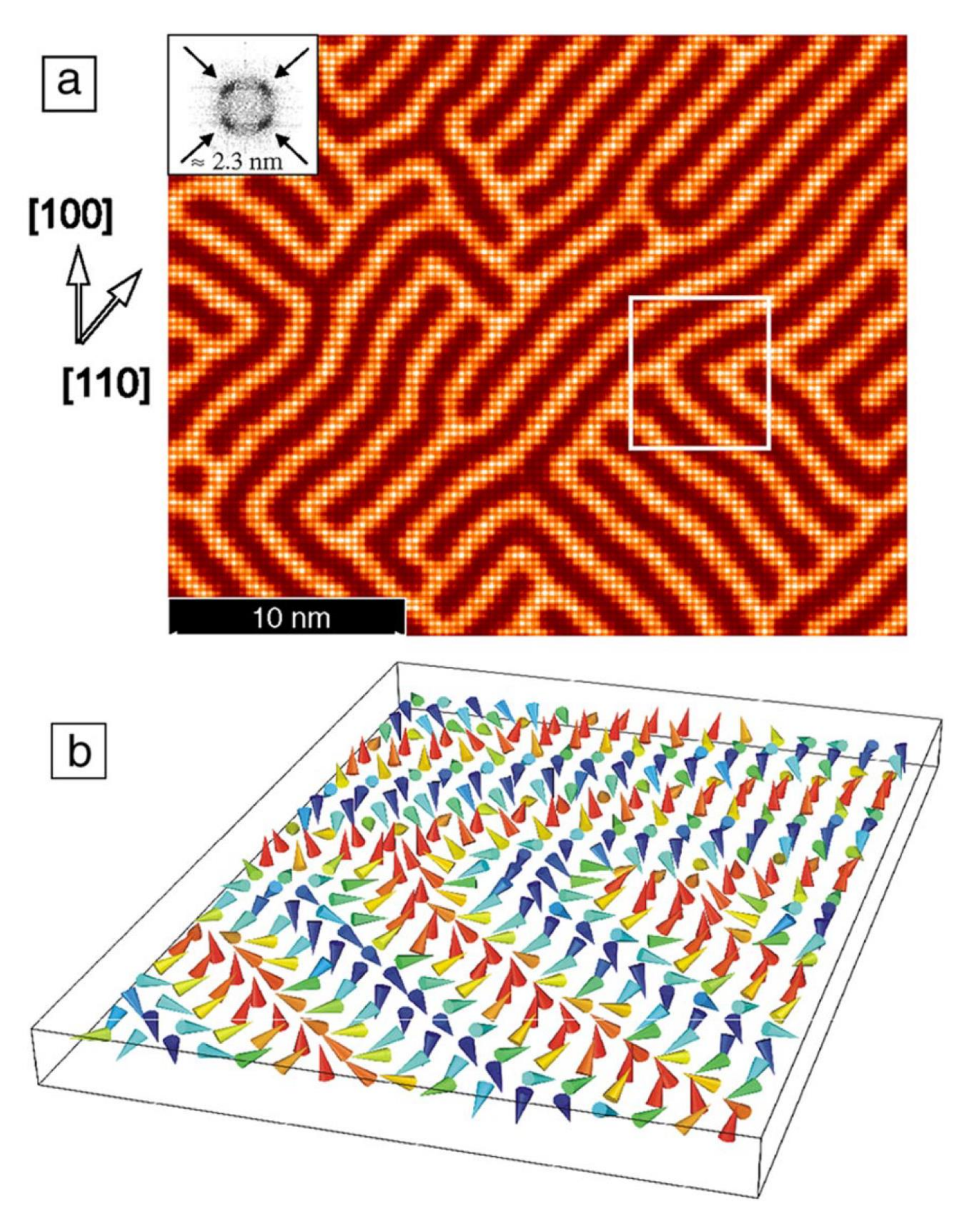}
\caption{a) Simulations of SP-STM images based on Monte Carlo calculations \cite{HEI-06} of a monolayer of Mn on W(001) at 13~K (the inset shows the Fourier transform). b) Detailed spin structure for the white square indicated in a) where the helical spinstructure can be directly seen in the illustration. From \citealp{FER-08a}.}
\label{fig:simSTM}
\end{figure}

Some of the earliest evidences that the DMI inﬂuences domain wall energies and alters DW conﬁgurations was obtained from SP-STM of ultrathin magnetic films deposited on heavy metal substrates such as single atomic layers of Mn or Fe on W(110) substrates \cite{BOD-07, HEI-08, HEI-11}. These methods provide information on the magnetic order at the atomic scale. In order to extract the parameters entering the total energy of the observed non-collinear spin structures, such as the DMI constant, simulations of the images (see Fig.\ref{fig:simSTM}) based on models describing the tunneling process in the tip \cite{TER-98} and density functional theory were performed \cite{WOR-01}. Since the computational cost of ab initio models is large, simpliﬁed approaches were devised, combining them with Monte Carlo or micromagnetic calculations \cite{HEI-06, HEI-08}. Successively, using SP-STM, \textcite{PIE-11, MEC-09, MEC-09a} concentrated on spin spiral domains in these bilayers and showed that the magnetic structure of the Fe double layer grown on W(110) is an inhomogeneous right-rotating cycloidal spin spiral. \textcite{MEC-09} extracted the magnitude of the Dzyaloshinskii-Moriya vector from the experimental data using analytical micromagnetic calculations. The result was then conﬁrmed by comparison of the measured saturation ﬁeld along the easy axis to the respective value as obtained from Monte Carlo simulations.

%They found that the Dzyaloshinskii-Moriya interaction is too weak to destabilize the single domain state, but it can deﬁne the sense of rotation and the type of the cycloidal spiral once the single domain state is destabilized by dipolar interaction. 

More recently, spin-polarized low-energy electron microscopy (SPLEEM) has been successfully exploited to achieve a quantification of the DMI strength in different systems \cite{CHE-13, CHE-13a, CHE-20, CHE-17, YAN-18}. While first works give only an estimate of the DMI constant, a remarkable development towards quantitative values is observed in literature. Here, the HM thickness dependent transition from chiral Néel walls to achiral Bloch walls is investigated in wedge samples. From the critical thickness a quantitative estimate of the DMI strength is obtained.  
In addition to SP-STM and SPLEEM, also L-TEM \cite{GAR-19} and SEMPA \cite{COR-17, KLO-19, MEI-20a} have been used to quantify the DMI strength by comparing the measured and the calculated profiles of the domain walls. In most cases micromagnetic analytical calculations or simulations of the DW energy or strayfield are employed.

\subsection{Experimental results}\label{sec:experimentDW}

\subsubsection{Current-driven domain wall motion}\label{sec:experimentDWcurrent}

\begin{table*}
\caption{Overview of DMI measurements for \textbf{Co/Ni/Co} thin films via \textbf{current-induced domain wall motion} experiments. FM and NM stand for ferromagnetic and non-magnetic layer, respectively. $H_\mathrm{DMI}$ is the DMI field, $D$ is the interfacial DMI constant and $D_\mathrm{s} = D \cdot d$, with $d$ being the thickness of the ferromagnetic film. Numbers in roman were quoted in the reviewed papers, while numbers in italics were either extracted from figures or calculated using the parameters provided.} 
\label{table:Current1}
\begin{tabular}{c|c|c|c|c|c|c}
\hline 
\hline
\textbf{FM} & \textbf{Bottom NM} & \textbf{Top NM}  & $\mid$\textbf{$\mu_0$H$_\mathbf{{DMI}}$}$\mid$ & $\mathbf{D}$ & $\mathbf{D_s}$ & \textbf{Ref}\\ 
(nm) & (nm) & (nm) & (mT) & (mJ/m$^2$) & (pJ/m) & \\
\hline 
\hline
\multirow{4}{*}{\centering Co(0.3)/Ni(0.7)/Co(0.15)} & Pt(1) & \multirow{4}{*}{\centering TaN(5)} & \textit{60} & n.a. & n.a. & \multirow{4}{*}{\hfil \centering \citealp{RYU-13}}\\
 & Pt(1.5) & & \textit{190} & n.a. & n.a. & \\
 & Pt(3) & & \textit{240} & n.a. & n.a. & \\
 & Pt(5) & & \textit{220} & n.a. & n.a. & \\
\hline
\multirow{3}{*}{\centering Co(0.3)/Ni(0.7)/Co(0.15)} & Pd(5) & \multirow{3}{*}{\centering TaN(5)} & 12 & n.a. & n.a. & \\
 & Ir(3) & & 18 & n.a. & n.a. & {\centering \citealp{RYU-14}}\\
 & Pt(5) & & 140 & n.a. & n.a. & \\
\hline 
\hline

\end{tabular}
\end{table*}

\begin{table*}
\caption{Overview of DMI measurements for \textbf{CoFeB} thin films via \textbf{current-induced domain wall motion} experiments. FM and NM stand for ferromagnetic and non-magnetic layer, respectively. $H_{DMI}$ is the DMI field, $D$ is the interfacial DMI constant and $D_\mathrm{s} = D \cdot d$, with $d$ being the thickness of the ferromagnetic film. Numbers in roman were quoted in the reviewed papers, while numbers in italics were either extracted from figures or calculated using the parameters provided.} 
\label{table:Current2}

\begin{tabular}{c|c|c|c|c|c|c|c}
\hline 
\hline
\textbf{FM} & \textbf{Bottom NM} & \textbf{Top NM}  & $\mid$\textbf{$\mu_0$H$_\mathbf{{DMI}}$}$\mid$ & $\mathbf{D}$ & $\mathbf{D_s}$ & \textbf{Sign} & \textbf{Ref}\\ 
(nm) & (nm) & (nm) & (mT) & (mJ/m$^2$) & (pJ/m) && \\
\hline 
\hline
\multirow{4}{*}{\centering Co$_{20}$Fe$_{60}$B$_{20}$(1)} & Hf(2.6--6) & \multirow{4}{*}{\centering MgO(2)} & n.a. & \textit{0.38 -- 0.05}\footnotemark[1] & \textit{0.38 -- 0.05}\footnotemark[1] &-& \multirow{4}{*}{\hfil \centering \citealp{TOR-14}}\\ 
 & Ta(0.5--1.3) &  & n.a. & \textit{0.08 -- 0.07}\footnotemark[1] & \textit{0.08 -- 0.07}\footnotemark[1] &-& \\
 & TaN(0.4--6.6) &  & n.a. & \textit{0.04 -- 0.20}\footnotemark[1] & \textit{0.04 -- 0.20}\footnotemark[1] &+& \\
 & W(2.1--3.6) &  & n.a. & \textit{0.24 -- 0.37}\footnotemark[1] & \textit{0.24 -- 0.37}\footnotemark[1] &+& \\
\hline 
Co$_{20}$Fe$_{60}$B$_{20}$(1) & Ta(5) & MgO(2) & 7.8 & 0.06 & \textit{0.06} &+& \citealp{LOC-15} \\ 
\hline
Co$_{20}$Fe$_{60}$B$_{20}$(0.8) & Ta(5) & MgO(2) & n.a. & 0.03 & \textit{0.02} &-& \citealp{KAR-18a} \\ 
\hline 
\hline
\end{tabular}
\footnotetext[1]
{The highest and lowest DMI values reported do not necessarily correspond to the extremes of the bottom layer thickness range (for instance the highest DMI value could occur at mid thickness range).}
\end{table*}
Measurements of current-driven DW motion are performed in wires with typical widths ranging between 1~$\mu$m and 5~$\mu$m and lengths of a few tens of $\mu$m. Usually, the magnetization in the wire is initially saturated by applying an out-of-plane field $H_z$. $H_z$ is then removed and a current pulse (with duration of a few tens of ns) is applied either directly to the wire or to an Oersted line patterned on top of the wire. In either case the current pulse nucleates a reversed magnetic domain and a DW is thus injected into the wire. Alternatively, one can use a magnetic field pulse to generate domains in injection pads, and inject them in the wires.

For measurements of DW velocities, current pulses are applied to the wire to move a DW, while the DW position along the wire is imaged by magneto-optical Kerr effect (MOKE) microscopy in polar configuration. The number and duration of the applied current pulses is chosen to obtain a significant displacement of the DW, at least over a few $\mu$m. Typical values of current densities fall in the range between 10$^{10}$~A/m$^2$ and 10$^{12}$~A/m$^2$. The velocity of the DW is calculated as the ratio between the current-induced DW displacement and the total duration of the pulses. Finally, velocities are measured for both $\uparrow \downarrow$ and $\downarrow \uparrow$ DWs under different strength of $H_x$, keeping the current density fixed, in order to determine the ``compensating'' field $H_x^*$ of Fig.~\ref{fig:current}.  

\begin{table*}
\caption{Overview of DMI measurements for \textbf{Co} thin films via \textbf{current-induced domain wall motion} experiments. FM and NM stand for ferromagnetic and non-magnetic layer, respectively. $H_\mathrm{DMI}$ is the DMI field, $D$ is the interfacial DMI constant and $D_\mathrm{s} = D \cdot d$, with $d$ being the thickness of the ferromagnetic film. Numbers in roman were quoted in the reviewed papers, while numbers in italics were either extracted from figures or calculated using the parameters provided.} \label{table:current3}
\begin{tabular}{c|c|c|c|c|c|c|c}
\hline 
\hline
\textbf{FM} & \textbf{Bottom NM} & \textbf{Top NM}  & $\mid$\textbf{$\mu_0$H$_\mathbf{{DMI}}$}$\mid$ & $\mathbf{D}$ & $\mathbf{D_s}$ & \textbf{Sign} & \textbf{Ref}\\ 
(nm) & (nm) & (nm) & (mT) & (mJ/m$^2$) & (pJ/m) && \\
\hline 
\hline
Co(0.36) & Pt(4) & Pt(1) & 37 &\textit{0.24} & \textit{0.09}&-& \multirow{5}{*}{\hfil \centering \citealp{FRA-14}}\\
Co(0.36) & Pt(4) & Pt(2) & 12.5 & \textit{0.10}& \textit{0.04}&-& \\
Co(0.5) & Pt(4) & Pt(2) & 11 & \textit{0.09}& \textit{0.04}&-& \\
Co(0.5) & Pt(2) & Pt(4) & 3 &\textit{0.03} & \textit{0.02}&-& \\
Co(0.8) & Pt(4) & AlO$_x$(1.9) & $\gg$40 & n.a. & n.a. && \\
\hline
Co(1) & Pt(5) & Gd(2) & 280 & \textit{1.00} & 1.00 &-& \citealp{VAN-15} \\ 
\hline 
Co(0.93) & \multirow{3}{*}{\centering Pt(4)} & \multirow{3}{*}{\centering AlO$_x$(2)} & 99 & 0.54 & \textit{0.50} &-& \\ 
Co(1.31) &  &  & \textit{54} & \textit{0.48} & \textit{0.63} &-& {\centering \citealp{LOC-17}}\\ 
Co(1.37) &  &  & \textit{48} & \textit{0.47} & \textit{0.64} &-&  \\ 
\hline 
Co(0.9) & \multirow{4}{*}{\centering Pt(2.5)} & Al(2.5) & 107 & 0.87 & \textit{0.78} &-& \multirow{4}{*}{\hfil \citealp{KIM-18}}\\ 
Co(0.9) &  & Ti(2.5) & \textit{197} & 1.42 & \textit{1.28} &-&\\ 
Co(0.9) &  & W(2.5) & \textit{183} & 1.35 & \textit{1.22} &-& \\ 
Co(0.5) &  & Pt(1.5)& 0 & 0 & 0 &&\\
\hline
\hline
\end{tabular}
\end{table*}

For measurements of DW depinning efficiencies, a constant current density $J_e$ is applied continuously to the wire for a given in-plane field $H_{x}$, while the out-of-plane field $H_z$ is ramped up until DW motion is detected by MOKE. The DW depinning efficiency $\epsilon$ for a given $H_{x}$ is then determined from the slope of the depinning field $H_{z\mathrm{,dep}}$ as a function of $J_e$. Modest current values are used, up to about 10$^{10}$~A/m$^2$, in order to exclude Joule heating effects. Finally, $\epsilon$ is measured for both $\uparrow \downarrow$ and $\downarrow \uparrow$ DWs under different strength of $H_x$, and $H_\mathrm{DMI}$ is determined as the ``compensating'' field $H_x^*$ for which $\epsilon = 0$.

Measurements of current-driven DW motion as a function of the in-plane field $H_x$ have been used to extract $H_\mathrm{DMI}$ in several material stacks, with Co/Ni/Co, Co$_{20}$Fe$_{60}$B$_{20}$ or Co as ferromagnetic films, different heavy metals as underlayers and either a heavy metal or an oxide as overlayers. Tables \ref{table:Current1}, \ref{table:Current2}, and \ref{table:current3} present a summary of the DMI values measured for Co/Ni/Co, CoFeB, and Co, respectively. 

For Co/Ni/Co films (see Table \ref{table:Current1}), the DMI seems to increase upon increasing the thickness of the Pt underlayer \cite{RYU-13}, while it is very low when using Pd or Ir as underlayers \cite{RYU-14}. 

For CoFeB films with Ta and MgO as underlayer and overlayer (see Table \ref{table:Current2}), the DMI values, generally small, are spread not only in magnitude, but also in sign \cite{TOR-14, LOC-15, KAR-18a}. This can be explained by the fact that the two interfaces contribute with small and opposite DMI values, and that the DMI at the CoFeB/oxide interface strongly depends on its oxidation state. Furthermore, \textcite{KAR-18a} find different signs of DMI for Ta/CoFeB/MgO depending on whether they measure DW motion driven by current or field (see table \ref{tab:TableField2} for the field-driven case). Among the underlayers used with CoFeB, W provides the highest (positive) DMI \cite{TOR-14}. 

Finally, for Co films, Pt is the only underlayer investigated, while overlayers are either oxides or metals (see Table \ref{table:current3}). In the symmetric structure Pt/Co/Pt, where no DMI is expected if the two interfaces are identical as experienced with epitaxial layers, either negative \cite{FRA-14} or vanishing \cite{KIM-18} DMI values are measured. This lack of perfect compensation depends on growth conditions, oxidation of the interfaces, and many other experimental conditions. For all the other stacks the DMI is always negative and largest in magnitude when using Ti as overlayer \cite{KIM-18}.  

\subsubsection{Field-driven domain wall motion}\label{sec:experimentDWfield}

\begin{figure*}[t]
\centering
\includegraphics[width=.9\textwidth]{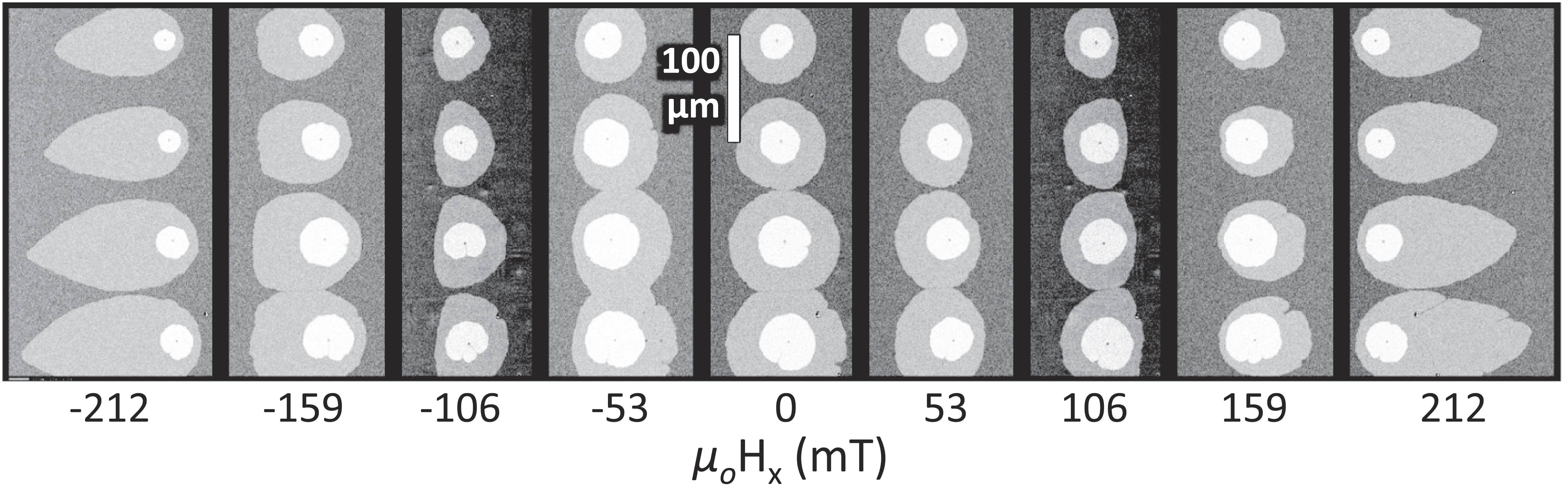}
\caption{Evolution of bubble domain shapes in an applied in-plane magnetic field in Co/Ni multilayers measured by perpendicular Kerr effect. From \citealp{LAU-16}.}
\label{fig:LAU-16}
\end{figure*}

Experiments of field-driven DW dynamics are mostly performed in continuous films where magnetization reversal proceeds by nucleation and growth of magnetic bubble domains. Only few works report studies of planar DW dynamics (i.e. straight walls) in continuous films \cite{PEL-17,KIM-17} or in $\upmu$m-wide wires \cite{JUE-16, KIM-18}. %In any case, the measurement technique is identical. 
The magnetization in the continuous film/wire is initially saturated by applying a perpendicular field $H_z$. A bubble domain (or a domain separated by a planar DW) is then nucleated by applying a short $H_z$ pulse in the opposite direction, through either a coil or an electromagnet. The bubble domain (or the planar domain) is expanded under simultaneous application of continuous $H_x$ (from an electromagnet) and pulsed or continuous $H_z$ (from a coil or an electromagnet). The initial and final positions of the DW are imaged through MOKE microscopy in polar configuration. Typical bubble growths are in the range of at least a few tens of $\upmu$m, to avoid coalescence with other bubbles. The velocity of the DW is measured along the direction of the applied $H_x$ 
%(typically along the left--right direction in a magnetic bubble) 
and is calculated as the ratio between the DW displacement and the total time during which $H_z$ is applied -- whether in pulses or continuously. Finally, velocities are measured for both $\uparrow \downarrow$ and $\downarrow \uparrow$ DWs (i.e. DWs on opposite sides of the bubble) under different strength of $H_x$, upon keeping $H_z$ constant. In this way velocity vs $H_x$ curves are constructed for both DWs and $H_\mathrm{DMI}$ is determined either through fitting with one of the modified creep formula previously discussed (\ref{sec:theoryDWfield}), or simply by finding the field for which the velocity is minimum. 
  
Regarding the strength of the applied fields, values differ greatly depending on the material and DW motion regime investigated. For measurements in the creep regime only modest $\mu_0 H_z$ are needed -- usually of a few mT depending on the depinning field -- while $\mu_0 H_z$ of hundreds of mT are used to drive DWs in the flow regime. The maximum values of applied $\mu_0 H_x$ depend instead on the strength of the DMI. In samples with large DMI in-plane fields up to 350~mT have been used \cite{HRA-14,LAV-15,CAO-18}. However, since the use of in-plane fields can be accompanied by artefacts (e.g. the nucleation of bubbles can be influenced, misalignments of the fields and crosstalk between perpendicular and in-plane electromagnets might be present), measuring large DMI values can be problematic. A simple scheme has been proposed to overcome the need for high $H_x$ values \cite{KIM-17}, whereby the DW velocity is measured at an angle $\theta > 0$ with respect to the in-plane field direction. In this way, the minimum of the DW velocity scales by a factor $\textrm{cos} \theta$, and becomes measurable also for samples with large DMI by applying moderate in-plane fields.

Field-driven DW dynamics in the creep regimes have been investigated to extract $H_\mathrm{DMI}$ in material systems with Co/Ni multilayers, CoFeB or Co as ferromagnetic layers, and different combinations of heavy metals and oxides as underlayers and overlayers. Regarding the flow regime, instead, $H_\mathrm{DMI}$ has been measured only for Co films as shown in Table \ref{tab:TableField4}. 
Tables \ref{tab:TableField1}, \ref{tab:TableField2} and \ref{tab:TableField3} present a summary of DMI measurements in the creep regime for Co/Ni, CoFeB and Co, respectively. In the following we highlight some of the most interesting findings and observations of these experimental studies. 

For Co/Ni multilayers (see Tab.~\ref{tab:TableField1}), \textcite{YU-16} studied the effect of different capping layers with a Pt underlayer and found values for $D$ ranging from $D=$~0.05~mJ/m$^2$ ($D_s$=~0.05~pJ/m) for MgO to $D=$~0.39~mJ/m$^2$ ($D_s$=~0.35~pJ/m) to Ta. Separately, three works by the same group \cite{LAU-16, PEL-17, LAU-18} have reported on the asymmetry of DW velocity curves about their minima and on the crossover between DW velocities for $\uparrow \downarrow$ and $\downarrow \uparrow$ sides of magnetic bubbles, corresponding to a morphological change from flattened to teardrop bubble shapes (Fig.~\ref{fig:LAU-16}). As previously mentioned, these velocity curves have been fitted through a model that takes into account both DW stiffness and a field-dependent prefactor $v_0$, showing that measurements of the minimum in velocity can be misleading to quantify the DMI \cite{LAU-18}. The maximum value corresponds to an underlayer of Pt and a top layer of Ta/TaN with $D=$~0.52~mJ/m$^2$ ($D_s$=~0.94~pJ/m). The range of values are larger than in \textcite{YU-16} but further analysis is needed to identify its origin.

Regarding CoFeB films (see Tab. \ref{tab:TableField2}), both the respective Co and Fe compositions \cite{JAI-17} and the post-growth annealing temperature \cite{KHA-16} have been shown to play a crucial role in the magnitude of DMI. Furthermore, for low DMI values (namely, $D \lesssim$~0.2~mJ/m$^2$), the DW motion technique disagrees with the values obtained by Brillouin Light Scattering (BLS) \cite{SOU-16}. Also worth mentioning is a recent study by \textcite{DIE-19a} which showed that the DMI can be tuned in a Ta/CoFeB/MgO system through light He$^+$ irradiation, due to an increasing interface intermixing mostly between Ta and CoFeB layers.     

Co is the most widely studied material, specifically in combination with Pt as underlayer, which can provide high DMI magnitudes depending on the overlayer choice. DMI measurements performed in the creep regime (see Tab.~\ref{tab:TableField3}) indicate that the nominally symmetric stack Pt/Co/Pt can have positive values of $D_\mathrm{s}$ as high as 0.58~pJ/m \cite{HRA-14}, although the same work shows that upon ensuring epitaxial growth, $D_\mathrm{s}$ reduces to almost 0. An even larger value of 0.71~pJ/m was calculated from the experimental data by \textcite{HAR-19} \footnote{In this paper the minimum velocity is found at $\mu_0 H_x \sim$~50~mT, a value compatible with other publications which report much smaller $D/D_s$ values as e.g. in \textcite{KIM-15}. The difference is the model used to evaluate $D$ from the velocity minimum. \textcite{HAR-19} calculate $\mu_0 H_{DMI} \approx$~170~mT, resulting in the high value of $D_s =$~0.71~pJ/m.}. Two more studies report a vanishing DMI \cite{PHA-16,AJE-17} for Pt/Co/Pt layers measured in the flow regime (see Tab.~\ref{tab:TableField4}), while a small negative $D_\mathrm{s}$ is found in \textcite{SHA-18}. 

Another system widely investigated in the literature is the Pt/Co/Ir stack, which also gives rise to some controversial results. First principle calculations predict opposite DMI signs for the Pt/Co and Ir/Co interfaces, which would result in an additive effect for the Pt/Co/Ir stack \cite{YAN-15,YAM-16}. However, \textcite{HRA-14} show that $D$ decreases when introducing a thin Ir top layer in a Pt/Co/Pt stack, and even changes sign upon increasing the Ir thickness. Similarly, \textcite{SHA-19} find that the magnitude of $D$ is smaller in Pt/Co/Ir/Ta than in Pt/Co/Ta, suggesting that Pt/Co and Co/Ir interfaces contribute to the net DMI with opposite signs. On the other hand, a huge increase of DMI has been observed in Pt/Co/MgO layers, upon insertion of a very thin Mg film between Co and MgO \cite{CAO-18}, providing among the highest values of $D _s=~2.32~pJ/m$ ($D =$~2.32~mJ/m$^2$) reported in the literature. 

All these results show how important the quality of the interfaces is for the determination the DMI. Perfect compensation is rarely obtained for nominally symmetric interfaces, and large variations are possible. The ``controlled damage'' of the interfaces by  Ar$^{+}$ ion irradiation can even be an effective way to tune the sign of DMI in Pt/Co/Pt films \cite{BAL-17}. The key aspect is the influence of the growth conditions on DMI, as extensively investigated in \textcite{LAV-15,WEL-17}. Both studies report a dramatic variation of the DW velocity dependence with in-plane field, observed upon changing sputter-deposition conditions, namely Ar gas pressure, substrate temperature or chamber base pressure. Indeed, it is speculated that such different growth conditions may lead to different degrees of interfacial intermixing and/or quality, resulting in a huge range of measured DMI values, which can even change sign \cite{WEL-17}.

\begin{table*}
\caption{Overview of DMI measurements for \textbf{Co/Ni multilayers} via \textbf{field-induced domain wall motion} experiments in the  \textbf{creep} regime. FM and NM stand for ferromagnetic and non-magnetic layer, respectively. $H_\mathrm{DMI}$ is the DMI field, $D$ is the interfacial DMI constant and $D_\mathrm{s} = D \cdot d$, with $d$ being the thickness of the ferromagnetic film. Numbers in roman were quoted in the reviewed papers, while numbers in italics were either extracted from figures or calculated using the parameters provided.} 
\label{tab:TableField1}

\begin{tabular}{c|c|c|c|c|c|c|c}
\hline 
\hline
\textbf{FM} & \textbf{Bottom NM} & \textbf{Top NM}  & $\mid$\textbf{$\mu_0$H$_\mathbf{{DMI}}$}$\mid$ & $\mathbf{D}$ & $\mathbf{D_s}$ & \textbf{Sign} & \textbf{Ref}\\ 
(nm) & (nm) & (nm) & (mT) & (mJ/m$^2$) & (pJ/m) && \\
\hline 
\hline
\multirow{4}{*}{\centering Co(0.1)/[Ni(0.1)/Co(0.1)]$_4$} & \multirow{4}{*}{\centering Pt(4)} & MgO(2) & 15.59 & 0.05 & \textit{0.05} &-& \multirow{4}{*}{\hfil\centering \citealp{YU-16}}\\
 & & Cu(2) & 19.23 & 0.12 & \textit{0.11} &-& \\
 & & Pt(2) & 34.88 & 0.20 & \textit{0.18} &+& \\
 & & Ta(2) & 103.86 & 0.39 & \textit{0.35} &-& \\
\hline
 [Co(0.2)/Ni(0.6)]$_2$/Co(0.2) & Pt(2.5) & Ta(0.5)/TaN(3) & 60 & \textit{0.21}  & \textit{0.38}  &-& \citealp{LAU-16}\\
\hline
 [Co(0.2)/Ni(0.6)]$_2$/Co(0.2) & Pt(2.5) & Ta(0.5)/TaN(6) & 106 & \textit{0.37} & \textit{0.67} &-& \citealp{PEL-17}\\
\hline
\multirow{4}{*}{\centering [Co(0.2)/Ni(0.6)]$_2$/Co(0.2)} & Pt(1.2) & Ta(0.8)/TaN(6) &n.a. & \textit{0.52} & \textit{0.94} &-& \multirow{4}{*}{\hfil\centering \citealp{LAU-18}}\\ 
 & Ir(1.2) & Ta(0.8)/TaN(6) & n.a. & \textit{0.07} & \textit{0.13}&-& \\ 
 & Pt(2.5) & Ir(2.5) &n.a. & 0.31 & \textit{0.56} &-& \\ 
 & Ir(2.5) & Pt(2.5) &n.a. & 0.21 & \textit{0.38} &+& \\ 	
\hline
\hline
\end{tabular}
\end{table*}

\begin{table*}
\caption{Overview of DMI measurements for \textbf{CoFeB} thin films via \textbf{field-induced domain wall motion} experiments in the \textbf{creep} regime. FM and NM stand for ferromagnetic and non-magnetic layer, respectively. $H_\mathrm{DMI}$ is the DMI field, $D$ is the interfacial DMI constant and $D_\mathrm{s} = D \cdot d$, with $d$ being the thickness of the ferromagnetic film. Numbers in roman were quoted in the reviewed papers, while numbers in italics were either extracted from figures or calculated using the parameters provided.} 
\label{tab:TableField2}

\begin{tabular}{c|c|c|c|c|c|c|c}
\hline 
\hline
\textbf{FM} & \textbf{Bottom NM} & \textbf{Top NM}  & $\mid$\textbf{$\mu_0$H$_\mathbf{{DMI}}$}$\mid$ & $\mathbf{D}$ & $\mathbf{D_s}$ & \textbf{Sign} & \textbf{Ref}\\ 
(nm) & (nm) & (nm) & (mT) & (mJ/m$^2$) & (pJ/m) && \\
\hline 
\hline
Co$_{20}$Fe$_{60}$B$_{20}$(0.8) & Ta(5) & MgO(2) & \textit{6.2 -- 16.3}\footnotemark[1] & \textit{0.02 -- 0.06}\footnotemark[1] &  \textit{0.02 -- 0.05}\footnotemark[1] &+& \citealp{KHA-16}\\ 
\hline
\multirow{4}{*}{\centering CoFeB(1)} & W(2) & \multirow{4}{*}{\hfil\centering MgO(2)} & 35 & 0.23 & \textit{0.23} &+& \multirow{4}{*}{\hfil\centering \citealp{SOU-16}}\\
& W(3) & & 15  & 0.12 & \textit{0.12} &+& \\
& TaN(1) & & 5 & 0.05 & \textit{0.05} &+& \\
& Hf(1) & & 2 & 0.01 & \textit{0.01} &+& \\ 
\hline
Co$_{20}$Fe$_{60}$B$_{20}$(0.6) & \multirow{2}{*}{\centering W(5)} & MgO(2) & 93 & 0.68 & \textit{0.41} &+& \multirow{2}{*}{\hfil\centering \citealp{JAI-17}}\\
Co$_{40}$Fe$_{40}$B$_{20}$(0.6) & &MgO(2)/Ta(5) & 4 & 0.03 & \textit{0.02} &+& \\
\hline
Co$_{20}$Fe$_{60}$B$_{20}$(0.8) & Ta(5) & MgO(2) & 8.8 & 0.03 & \textit{0.02} &+& \citealp{KAR-18a}\\ 
\hline
Co$_{20}$Fe$_{60}$B$_{20}$(1) & Ta(5) & MgO(2) & 2.6 -- 16\footnotemark[2] & \textit{0.02 -- 0.08}\footnotemark[2] & \textit{0.02 -- 0.08}\footnotemark[2],&+& \citealp{DIE-19a}\\
\hline
\hline
\end{tabular}
\footnotetext[1]{Different DMI values correspond to different annealing temperatures.}
\footnotetext[2]{Different DMI values correspond to different doses of He$^+$ ion irradiation.}
\end{table*}

%tablefield3
\begin{table*}
\caption{Overview of DMI measurements for \textbf{Co} thin films via \textbf{field-induced domain wall motion} experiments in the \textbf{creep} regime. FM and NM stand for ferromagnetic and non-magnetic layer, respectively. $H_\mathrm{DMI}$ is the DMI field, $D$ is the interfacial DMI constant and $D_\mathrm{s} = D \cdot d$, with $d$ being the thickness of the ferromagnetic film. Numbers in roman were quoted in the reviewed papers, while numbers in italics were either extracted from figures or calculated using the parameters provided.} 
\label{tab:TableField3}

\begin{tabular}{c|c|c|c|c|c|c|c}
\hline 
\hline
\textbf{FM} & \textbf{Bottom NM} & \textbf{Top NM} & $\mid$\textbf{$\mu_0$H$_\mathbf{{DMI}}$}$\mid$ & $\mathbf{D}$ & $\mathbf{D_s}$ & \textbf{Sign} & \textbf{Ref}\\ 
(nm) & (nm) & (nm) & (mT) & (mJ/m$^2$) & (pJ/m) && \\
\hline 
\hline
Co(0.3) & Pt(2.5) & Pt(1.5) & 26.5 & \textit{0.11} & \textit{0.03} &+& \citealp{JE-13}\\
\hline
\multirow{3}{*}{\centering Co(0.7)} & \multirow{3}{*}{\centering Pt(5)} & Pt(3) & \textit{104} & \textit{0.83} & \textit{0.58} &+& \multirow{3}{*}{\hfil\centering \citealp{HRA-14}} \\
 & & Ir(0.23)/Pt(3) & \textit{10} & \textit{0.08} & \textit{0.06} &+& \\
 & & Ir(0.69)/Pt(3) & \textit{155} & \textit{1.23} & \textit{0.86} &-& \\
\hline
Co(0.8) & Pt(10) & Pt(3) & 50 & n.a. & n.a. & n.a.& \citealp{PET-15}\\
\hline
 \multirow{2}{*}{\centering Co(0.4)} & Pt(2) & Pt(2) & 83 & \textit{0.33} & \textit{0.13}  &+& \multirow{2}{*}{\hfil\centering \citealp{KIM-15}}\\
& Pt(2) & Pd(2) & $>$200 & n.a.  & n.a.  &n.a.&  \\
\hline
\multirow{2}{*}{\centering Co(0.6)} & \multirow{2}{*}{\centering Pt(3)} & AlO$_x$(1.6) & 138 & n.a. & n.a. &n.a.& \multirow{2}{*}{\hfil\centering \citealp{KIM-17}}\\
&  & MgO(2) & 483 & n.a. & n.a. &n.a.&  \\
\hline
Co(0.54) & Au(4) & NiO(10) & n.a. & \textit{2.04} & 1.11 &+& \citealp{KUS-18}\\
\hline
Co(0.56) & \multirow{2}{*}{\centering Pt(4)} & \multirow{2}{*}{\centering Ir(5)} & \textit{53} & \textit{0.31} & \textit{0.17} &-& \multirow{2}{*}{\hfil\centering \citealp{SHE-18}}\\
Co(1.05) & &  & \textit{8.5} & \textit{0.12} & \textit{0.13} &-& \\
\hline
\multirow{2}{*}{\centering Co(1)} & \multirow{2}{*}{\centering Pt(3)} & MgO(0.65)/Pt(5) & \textit{65.8} & 0.77 & \textit{0.77} &n.a.& \multirow{2}{*}{\hfil\centering \citealp{CAO-18}}\\
 & & Mg(0.2)/MgO(1.5-2)/Pt(5) & \textit{311} & 2.32 & \textit{2.32} &n.a.& \\
\hline
Co(1.8) & Pt(5) & W(1)/Pt(1) & 25 & 0.19 & \textit{0.34} &n.a.& \citealp{LIN-18}\\
\hline
\multirow{3}{*}{\centering Co(0.6)} & \multirow{3}{*}{\centering Pt(3)} & Pt(3) & \textit{7.8} &  0.07 & \textit{0.04}  &-& \multirow{3}{*}{\hfil\centering\citealp{SHA-18}}\\
& & Pt$_{50}$Au$_{50}$(3) & \textit{48} & 0.35 & \textit{0.21} &-& \\
& & Au(3) & \textit{175} & 1 & \textit{0.6} &-& \\
\hline
\multirow{2}{*}{\centering Co(0.8)} & \multirow{2}{*}{\centering Pt(2.2)} & Ta(4) & \textit{140}\footnotemark[1] & \textit{1.12}\footnotemark[1] & \textit{0.9}\footnotemark[1] &-& \multirow{2}{*}{\hfil\centering \citealp{SHA-19}}\\
 & & Ir(0.2 -- 2)/Ta(4) & \textit{64.2 -- 104.2}\footnotemark[2] & \textit{0.49 -- 0.93}\footnotemark[2] & \textit{0.39 -- 0.74}\footnotemark[2] &-&  \\
 \hline
 Co(0.6) & {\centering \multirow{3}{*}{Ta(4)/Pt(4)}} & Pt(4) & 170 & \textit{1.20} & \textit{0.71} &-\footnotemark[3]& \multirow{3}{*}{\hfil\centering \citealp{HAR-19}}\\
 Co(0.8--1.2) &  & Gd(3)/Pt(2) & 217--77 &  \textit{0.43--0.23} & \textit{0.34} &-\footnotemark[3]& \\
 Co(0.8--1) &  & Ir(4) & 156--92 &  \textit{0.39--0.23} & \textit{0.23} &-\footnotemark[3]& \\
 \hline
 Co(1) & {\centering \multirow{3}{*}{Ta(4)/Pt(4)}} & Gd(3)/Pt(4) & 280 & 0.37 & \textit{0.37} &n.a.& \multirow{3}{*}{\hfil\centering \citealp{CAO-20a}}\\
 Co(1) &  & Gd(3)/Ta(4) & 255 &  0.24 & \textit{0.24} &n.a.& \\
 Co(1)/Gd(3)/Co(1) &  & Ta(4) & 138 &  0.90 & \textit{0.45} &n.a.& \\
\hline
\multirow{3}{*}{\centering Co(0.9)} & Pt(5) & Pt(5) & 0 &0 & 0&n.a& \multirow{3}{*}{\hfil \centering \citealp{GEH-20}}\\
& Pt(5) & Au(5) & 105 & 0.87& \textit{0.78}&-& \\
& Au(5) & Pt(5) & 78 & 0.59& \textit{0.53}&+& \\
\hline
\hline
\end{tabular}
 \footnotetext[1]{For the Pt/Co/Ta system, the value provided in the table is obtained considering $H_\mathrm{DMI}$ to be the minimum in the velocity curves. However the work estimates the DMI also using a model that assumes $H_\mathrm{dep}$ = $H_\mathrm{dep}(H_x)$. In this case $D =$~-2~mJ/m$^2$ and $D_\mathrm{s} =$~-1.6~pJ/m.}
 \footnotetext[2]{The highest and lowest DMI values reported do not necessarily correspond to the extremes of the Ir thickness range.}
\footnotetext[1]{Sign convention used here is opposite to the used in the article.}
\end{table*}

%table flow regime
\begin{table*}
\caption{Overview of DMI measurements for \textbf{Co} thin films via \textbf{field-induced domain wall motion} experiments in the \textbf{flow} regime. FM and NM stand for ferromagnetic and non-magnetic layer, respectively. $H_\mathrm{DMI}$ is the DMI field, $D$ is the interfacial DMI constant and $D_\mathrm{s} = D \cdot d$, with $d$ being the thickness of the ferromagnetic film. The subscripts 1 and 2 indicate values of $D$ and $D_s$ extracted from $H_{DMI}$ and $v_{sat}$, respectively. Numbers in roman were quoted in the reviewed papers, while numbers in italics were either extracted from figures or calculated using the parameters provided.} 
\label{tab:TableField4}

\begin{tabular}{c|c|c|c|c|c|c|c|c|c}
\hline 
\hline
\textbf{FM} & \textbf{Bottom NM} & \textbf{Top NM} & $\mid$\textbf{$\mu_0$H$_\mathbf{{DMI}}$}$\mid$ & $\mathbf{D_1}$ & $\mathbf{D_2}$ & $\mathbf{D_{s1}}$ & $\mathbf{D_{s2}}$ & \textbf{Sign} & \textbf{Ref}\\ 
(nm) & (nm) & (nm) & (mT) & (mJ/m$^2$) & (mJ/m$^2$) & (pJ/m) & (pJ/m) &&\\
\hline 
\hline
Co(1) & Pt(5) & Gd(5) & 180 & 1.48 & n.a. & 1.48 & n.a. &-& \citealp{VAN-15}\\
\hline
Co(0.8) & \multirow{4}{*}{\centering Pt(4)} & AlO$_x$(3) & 220 & 1.63 / 1.91\footnotemark[1] & n.a. & \textit{1.3 / 1.53}\footnotemark[1] & n.a. &-& \multirow{4}{*}{\hfil\centering \citealp{PHA-16}}\\
Co(1) & & GdOx(4) & 200 & 1.48 / 1.73\footnotemark[1] & n.a. & \textit{1.48 / 1.73}\footnotemark[1] & n.a. &-& \\
Co(1) & & Gd(3) & 300 & 1.52 / 1.78\footnotemark[1] & n.a. & \textit{1.52 / 1.78}\footnotemark[1] & n.a. &-& \\
Co(1) & & Pt(4) & 0 & 0 & n.a. & 0  & n.a. && \\
\hline
\multirow{5}{*}{\centering Co(0.6)} & Pt(2) & Al(2) & 200 / 212 & \textit{1.52 / 2.02}\footnotemark[2] & \textit{1.38 / 2.2}\footnotemark[2] & 0.91 / 1.21\footnotemark[2] & 0.83 / 1.32\footnotemark[2]&-& \multirow{5}{*}{\hfil\centering \citealp{AJE-17}} \\
 & Pt(2) & Ir(2) & 96 / 125 & \textit{0.5 / 0.92 }\footnotemark[2] & \textit{0.37 / 0.67}\footnotemark[2] & 0.3 / 0.55\footnotemark[2] & 0.22 / 0.4\footnotemark[2] &-& \\
 & Pt(2) & Cu(2) & 200 & \textit{0.93} & \textit{1.03} & 0.56 & 0.62 &-& \\
 & Pt(2) & Pt(2) & 0 & 0 & 0 & 0 & 0 && \\
 & Ir(2) & Pt(2) & 106 & \textit{1.08} & \textit{1.08} & 0.65 & 0.65 &+& \\
\hline
Co(0.8) & Pt(30) & AlO$_x$(1-3) & n.a. & n.a. & \textit{1 -- 1.79}\footnotemark[3] & n.a. & 0.8 -- 1.43\footnotemark[3] &-& \multirow{2}{*}{\hfil\centering \citealp{SOU-19}}\\
Co(1) & Pt(4) & GdOx(1-3) & 150 -- $>$260\footnotemark[3] & \textit{0.6 -- 1.42}\footnotemark[3] & \textit{0.62 -- 1.34}\footnotemark[3] & 0.6 -- 1.42\footnotemark[3] & 0.62 -- 1.34 \footnotemark[3] &-& \\
\hline
Co(1) & \multirow{4}{*}{\centering Pt(4)}  & Gd(4) & n.a. & n.a. & 1.45 & n.a. & \textit{1.45}& - & \multirow{4}{*}{\hfil\centering \citealp{KRI-19}} \\
 Co(1) &  & GdOx(4) & n.a. & n.a. & 1.5 & n.a. & \textit{1.5}& - & \\
 GdCo(4) &  & Ta & n.a. & n.a. & 0.2 & n.a. & \textit{0.8}& - & \\
 GdCo(4.8) &  & Ta & n.a. & n.a. & 0.2 & n.a. & \textit{0.96}& - & \\
 \hline
 \hline
\end{tabular}
\footnotetext[1]{The two values of DMI reported derive from two different values for the exchange constant $A$.}
\footnotetext[2]{The two values of DMI reported are due to different growth temperatures for the Co layer, either room temperature or $100 ^{\circ}$C.}
\footnotetext[3]{The highest and lowest DMI values reported do not necessarily correspond to the extremes of the top NM thickness range.}

\end{table*}

\subsubsection{Equilibrium stripe domain pattern}\label{sec:domainwidthexperiments}
In the case of the equilibrium stripe domain pattern method, the DMI constant is estimated by measuring magnetic domain widths from nanometric magnetic microscopy images. The values and systems analyzed using this method are presented in
Table~\ref{table:Tabledomaincofeb} for CoFeB and Ni/Fe and in
Table~\ref{table:Tabledomainco} for Co and FeCo thin films. As the uncertainty of the measurement depends heavily on the precision of the imaging techniques, submicrometer range resolution is indispensable. The most commonly used experimental techniques are: Scanning Transmission X-ray Microscopy (STXM) \cite{MOR-16,LEM-18}, Magnetic Transmission X-ray Microscopy (MTXM) \cite{WOO-16,WOO-17}, Magnetic Force Microscopy (MFM) \cite{BUT-17,SOU-17,SCH-18,LEG-18,BAC-19,CAS-19,DAV-19,DUG-19,AGR-19,KOZ-20}, and in a few cases MOKE \cite{YU-17,WON-18,SCH-21}. Most of the materials analyzed with this method have large DMI values \cite{MOR-16,WOO-16,SOU-17}. For example, \textcite{SOU-17} reported that the replacement of the Co layer by a layer of FeCo increases substantially the value of $D$, up to a maximum of $D_s=$~2.18~pJ/m for Fe(0.6)/Co(0.6). The maximum of $D$ corresponds to $D=$~2.1~mJ/m$^2$ for a thinner magnetic layer Fe(0.4)/Co(0.4). 

Many of the analysed films are multilayers because the additional dipolar contribution between layers stabilizes skyrmions and increases the sizes of stable skyrmions \cite{MOR-16}. \textcite{LEG-18} studied the variation of the DW internal structure along the multilayer thickness, due to the competition of dipolar field and DMI in Co multilayers. The domain walls in such structures have hybrid character between Bloch and N\'eel types with different N\'eel chiralities at bottom and top surfaces. Albeit being hybrid domain walls, the DMI can be extracted comparing the measured domain width with the one obtained using micromagnetic simulations.
With the same technique symmetric Pd/Co/Pd samples were also analyzed \cite{DAV-19,DUG-19, KOZ-20}. The origin of the DMI in such symmetric trilayers is a different residual stress in the top and bottom of the Co/Pd interfaces due to different lattice matching \cite{DAV-19}. \textcite{DUG-19} were able to optimize the stack with the insertion of W between Co and Pd, resulting in an increase of $D$, with an optimal size for a W thickness of 0.2~nm and $D_s=$~0.65~pJ/m ($D=$~1.3~mJ/m$^2$). \textcite{KOZ-20} found a strong variation of the estimated $D$ value, when values of the exchange constant in the range $A=23$-30~pJ/m were used in the numerical model. The range of DMI for different exchange constants increases with the increase of the Co thickness. 

Wedges of Ni on top a Fe layer with a Cu(001) substrate were analyzed with Threshold Photoemission Magnetic Circular Dichroism with PhotoEmission Electron Microscopy (TP-MCD-PEEM) by \textcite{MEI-17}. In these samples, the DMI originates at the Ni/Fe interface due to the lack of inversion symmetry, but the obtained $D$ values are small because of the lack of a source of strong SOC. $D$ increases when those bilayers are capped with a Pt layer. Using this method, the dependence of DMI on temperature was also measured for Pt/Co/Cu multilayers up to a temperature of 500~K \cite{SCH-18}. It was found that DMI has a stronger dependence on temperature than other magnetic properties, like magnetocrystalline anisotropy. Polar MOKE was used to measure the variation of $D$ induced by electric fields in Pt(3~nm)/Co(0.49~nm)/AlO$_x$(6~nm) thin films \cite{SCH-21}. The electric field also produces variations of the magnetization and anisotropy constant, which need to be quantified to obtain the DMI energy. $D$ was measured only under electric field and the variation of $D$ measured for an electric field of 133~MV/m was 0.14~mJ/m$^2$ and 0.26~mJ/m$^2$ assuming an exchange constant of 7.5~pJ/m and 16~pJ/m, respectively.

%Table domain pattern CoFeB NI/Fe
\begin{table*}
\caption{Overview of DMI measurements for \textbf{CoFeB and Ni/Fe} thin films via \textbf{equilibrium stripe domain pattern}. FM and NM stand for ferromagnetic and non-magnetic layer, respectively. $D$ is the interfacial DMI constant and $D_\mathrm{s} = D \cdot d$, with $d$ being the thickness of the ferromagnetic film. Numbers in roman were quoted in the reviewed papers, while numbers in italics were either extracted from figures or calculated using the parameters provided.} 
\label{table:Tabledomaincofeb}

\begin{tabular}{c|c|c|c|c|c}
\hline 
\hline
\textbf{FM} & \textbf{Bottom NM} & \textbf{Top NM} & $\mid$$\mathbf{D}$$\mid$ & $\mid$$\mathbf{D_s}$$\mid$ & \textbf{Ref}\\ 
(nm) & (nm) & (nm) & (mJ/m$^2$) & (pJ/m) & \\
\hline 
\hline
[Pt(3)/Co$_{40}$Fe$_{40}$B$_{20}$(0.8)/MgO(1.5)]$_{\times 20}$ & Ta(3) & Ta(2) & 1.66 & \textit{1.33}&\citealp{WOO-17}\\
\hline
[Pt(2.7)/Co$_{60}$Fe$_{20}$B$_{20}$(0.8)/MgO(1.5)]$_{\times 15}$& Ta(2.3)/Pt(3.7) & & 1.5 & \textit{1.2}&\citealp{BUT-17}\\
\hline
Co$_{40}$Fe$_{40}$B$_{20}$(1.2) & Ta(5) & TaO$_x$(5) & 0.17 & \textit{0.2} & \citealp{YU-17}  \\
\hline
Ni(6--12 ML)/Fe(1--3 ML)& \multirow{4}{*}{\centering (001)Cu} &  & n.a.  & 0.28 $\pm$ 0.14 &\multirow{4}{*}{\hfil\centering  \citealp{MEI-17}}\\
Ni(9 ML)/Fe(1-–3 ML)/Ni(4-–14 ML)&  & & n.a. & 0 & \\
Fe(1 ML)/Ni(6-–12 ML) &  &   & n.a. & 0.38 $\pm$ 0.14 & \\
Fe(1 ML)/Ni(6-–12 ML) &  & Pt(0.4)  & n.a. & 0.6 $\pm$ 0.2 & \\
\hline
 Co$_{20}$Fe$_{60}$B$_{20}$(1.2) &  MgO(1) &  Ta(5) & 0.65 $\pm$ 0.08 & \textit{0.78}& \citealp{WON-18}\\
 \hline
[Pt(2.7)/CoFeB(0.86)MgO(1.5)]$_{x15}$& Ta(3.6)/Pt(1) & Pt(2.7)  & 1.76 & \textit{1.51}&\citealp{LEM-18}\\
\hline
$[$MgO(2)/Co$_{20}$Fe$_{60}$B$_{20}$(1)/Ta(5)$]_{\times 15}$& & Ta(5) & 0.08 $\pm$ 0.03 & 0.08$\pm$0.03&\multirow{3}{*}{\hfil\centering  \citealp{CAS-19}}\\
$[$MgO(2)/Co$_{20}$Fe$_{60}$B$_{20}$(0.6)/W(5)$]_{\times 15}$ & & Ta(5) &0.61$\pm$0.03 &0.37$\pm$0.02& \\
$[$MgO(1.4)/Co$_{60}$Fe$_{20}$B$_{20}$(0.8)/Pt(3.4)$]_{\times 15}$ & Ta(5.7) & Ta(5) & 1.0$\pm$0.1 & 0.80$\pm$0.08&\\
\hline
[Pt(2.5-–7.5)/Co$_{60}$Fe$_{20}$B$_{20}$(0.8)/MgO(1.5)]$_{\times 13}$ & Ta(3) & Ta(2) & 1.6 $\pm$ 0.2  & \textit{1.28}  & \citealp{AGR-19}  \\
\hline
\hline
\end{tabular}
\end{table*}

\subsubsection{Magnetic stripe annihilation}\label{sec:annihilationexperiments}
With this method, the first step is to nucleate two domain walls in a perpendicularly magnetized sample. The two walls are then manipulated with an out-of-plane field to minimize their distance until the two domain walls collapse. This annihilation field and the minimum width of the domain separating the walls before its collapse depend on $D$. The values of $D$ extracted using this method are shown in Table~\ref{table:Tableanihil}. As for the equilibrium stripe domain pattern technique (Sec.~\ref{sec:domainwidthexperiments}), this method is based on magnetic imaging. For this purpose, MTXM \cite{JAI-17,LIT-17}, STXM \cite{WOO-16,WOO-17} and MOKE \cite{YU-16a} have been used.
All the samples are magnetic thin films or wide patterned tracks, to allow high number of domains and the manipulation of the walls. In the original study \cite{BEN-15}, the nucleated domain walls were parallel and analysed using a combination of Lorentz transmission electron microscopy (L-TEM) and polar Kerr. In later studies \cite{JAI-17,WOO-16,YU-16a,WOO-17,LIT-17}, the annihilation field was studied independently of the shape and boundary of domains. \textcite{BEN-15} measured a value for $D$ of 0.33~mJ/m$^2$ ($D_s$=~0.27~pJ/m) in a film Pt/Co/AlO$_x$, but they considered it a lower limit because of a possible underestimation of the annihilation field due to thermally activated processes. A DMI value of $D_s$=~0.44~pJ/m ($D$=0.73~mJ/m$^2$) was measured in W/Co$_{20}$Fe$_{60}$B$_{20}$/MgO \cite{JAI-17}, while a smaller value of $D_s$=~0.25~pJ/m ($D$=0.25~mJ/m$^2$) was obtained when the same composition was sandwiched between Ta layers \cite{YU-16a}.

%table domain co
\begin{table*}
\caption{Overview of DMI measurements for \textbf{Co and FeCo} thin films via \textbf{equilibrium stripe domain pattern}. FM and NM stand for ferromagnetic and non-magnetic layer, respectively. $D$ is the interfacial DMI constant and $D_\mathrm{s} = D \cdot d$, with $d$ being the thickness of the ferromagnetic film. Numbers in roman were quoted in the reviewed papers, while numbers in italics were either extracted from figures or calculated using the parameters provided. The method does not provide the sign. All the $D$ values are absolute values unless noted.} 
\label{table:Tabledomainco}

\begin{tabular}{c|c|c|c|c|c|c}
\hline 
\hline
\textbf{FM} & \textbf{Bottom NM} & \textbf{Top NM}  & $\mathbf{D}$ & $\mathbf{D_s}$ & \textbf{Sign} & \textbf{Ref}\\ 
(nm) & (nm) & (nm) & (mJ/m$^2$) & (pJ/m) && \\
\hline 
\hline
 Co(0.6)/Pt(1)/$[$Ir(1)/Co(0.6)/Pt(1)$]_{\times 10}$ & Pt(10) & Pt(3) & 1.6$\pm$0.2 & \textit{0.96}& n.a. &\multirow{2}{*}{\centering \citealp{MOR-16} }\\
Co(0.6)/Pt(1)/$[$Co(0.6)/Pt(1)$]_{\times 10}$ & Pt(10) & Pt(3) & 0.2$\pm$0.2 & \textit{0.12} & n.a. &\\	
\hline
$[$Pt(3)/Co(0.9)/Ta(4)$]_{\times 15}$ & Ta(3) & & 1.5$\pm$0.2 & \textit{1.35}& n.a. &\citealp{WOO-16}\\
\hline
$[$Ir(1)/Co(0.6)/Pt(1)$]_{\times 20}$  &\multirow{8}{*}{\hfil\centering Ta(3)/Pt(10) } & \multirow{8}{*}{\hfil\centering Pt(2) } & \textit{1.67}  & \textit{1}  & n.a. & \multirow{8}{*}{\hfil\centering \citealp{SOU-17} }\\
$[$Ir(1)/Fe(0.2)/Co(0.6)/Pt(1)$]_{\times 20}$ &   &  & \textit{1.8}  & \textit{1.44}  & n.a. & \\
$[$Ir(1)/Fe(0.3)/Co(0.6)/Pt(1)$]_{\times 20}$  &  &  & \textit{1.88}  & \textit{1.69}  & n.a. & \\
$[$Ir(1)/Fe(0.2)/Co(0.5)/Pt(1)$]_{\times 20}$  &  &  & \textit{1.98}  & \textit{1.39}  & n.a. & \\
$[$Ir(1)/Fe(0.4)/Co(0.4)/Pt(1)$]_{\times 20}$  &   & & \textit{2.1}  & \textit{1.68}  & n.a. & \\
$[$Ir(1)/Fe(0.4)/Co(0.6)/Pt(1)$]_{\times 20}$  &   & & \textit{1.99}  & \textit{1.99}  & n.a. & \\
$[$Ir(1)/Fe(0.5)/Co(0.5)/Pt(1)$]_{\times 20}$   &   & & \textit{1.96}  & \textit{1.96}  & n.a. & \\
$[$Ir(1)/Fe(0.6)/Co(0.6))/Pt(1)$]_{\times 20}$  &   & & \textit{1.82}  & \textit{2.18}  & n.a. & \\
\hline
$[$Ir(1)/Co(0.6)Pt(1)$]_{\times 5}$ & Pt(10) & Pt(3) & 2.30 & \textit{1.38}  &+\footnotemark[1]& \multirow{11}{*}{\hfil\centering  \citealp{LEG-18}} \\
$[$Ir(1)/Co(0.8)/Pt(1)$]_{\times 5}$ & Pt(10) & Pt(3) & 2.00 & \textit{1.6} &+\footnotemark[1]&  \\
$[$Co(0.8)/Ir(1)/Pt(1)$]_{\times 5}$ & Pt(11) & Pt(3) & 1.37  & \textit{1.1} &-\footnotemark[1]& \\
$[$Co(0.8)/Ir(1)/Pt(1)$]_{\times 5}$ & Ta(5)/Pt(10) &  Pt(3) & 1.63 & \textit{1.3} &-\footnotemark[1]&  \\
$[$Co(0.8)/Ir(1)/Pt(1)$]_{\times 10}$ & Pt(11) & Pt(3) & 1.52 & \textit{1.22} &-\footnotemark[1]&  \\
$[$Co(0.8)/Ir(1)/Pt(1)$]_{\times 10}$ & Ta(5)/Pt(10) & Pt(3) & 2.06 & \textit{1.65} &-\footnotemark[1]&  \\
Co(0.8)/$[$Pt(1)/Ir(1)/Co(0.8)$]_{\times 10}$ & Ta(15) & Pt(3) & 1.14 & \textit{0.91} &+\footnotemark[1]&  \\
$[$Pt(1)/Co(0.6)/Al$_2$O$_3$(1)$]_{\times 20}$ & Ta(10)/Pt(7) & Pt(3) & 1.29 & \textit{0.77} &-\footnotemark[1]&  \\
$[$Pt(1)/Co(0.8)/Al$_2$O$_3$(1)$]_{\times 20}$ & Ta(10)/Pt(7) & Pt(3) & 1.01 & \textit{0.81} &-\footnotemark[1]&  \\
$[$Al$_2$O$_3$(1)/Co(0.6)/Pt(1)$]_{\times 20}$ & Ta(10) & Pt(7) & 1.94 & \textit{1.16} &+\footnotemark[1]&  \\
$[$Al$_2$O$_3$(1)/Co(0.8)/Pt(1)$]_{\times 20}$ & Ta(10) & Pt(7) & 1.69 & \textit{1.35} &+\footnotemark[1]&  \\
\hline
\multirow{2}{*}{\centering $[$Pt(2)/Co(1.1)/Cu(1)$]_{x15}$ } & \multirow{2}{*}{\centering Ta(3) } & \multirow{2}{*}{\centering Pt(2) } &  \textit{1.55 (298 K)}  & \textit{1.7}  & n.a. &  \multirow{2}{*}{\hfil\centering \citealp{SCH-18}}\\
 & & &  \textit{0.47 (423 K)}  & \textit{0.52}  & n.a. & \\
\hline
Co(0.6)/Pt(1)/$[$Ir(1)/Co(0.6)/Pt(1)$]_{\times 5}$ & Pt(10) & Pt(3) & 1.97$\pm$0.02 & \textit{1.18} & n.a. &\citealp{BAC-19}\\
\hline
$[$Pt(2.5-–7.5)/Co(0.8)/Pt(1.5)$]_{\times 13}$ & Ta(3) & Ta(2) & 0 $\pm$ 0.1  & 0 & n.a. & \citealp{AGR-19} \\
\hline
$[$Co(0.8)/Pd(2)$]_{\times 5}$ &  \multirow{3}{*}{\hfil\centering Cu(2)/Pd(3) }  & \multirow{3}{*}{\hfil\centering Pd(3) }  & 1.6 $\pm$ 0.35\footnotemark[2] & \textit{1.28}\footnotemark[2] &-\footnotemark[3]&\multirow{3}{*}{\hfil\centering  \citealp{DAV-19}}\\
$[$Co(0.8)/Pd(2)$]_{\times 10}$ & & & 1.85 $\pm$ 0.45\footnotemark[2] & \textit{1.48}\footnotemark[2] &-\footnotemark[3]&\\
$[$Co(0.8)/Pd(2)$]_{\times 20}$ & & & 2.3 $\pm$ 0.5\footnotemark[2] & \textit{1.84}\footnotemark[2]  &-\footnotemark[3]&\\
\hline
\multicolumn{3}{c|}{$[$Pd(1)/Co(0.5)/Pd(1)$]_{\times 15}$} & \textit{0.3} $\pm$ \textit{0.1} & \textit{0.15}&  n.a.  & \multirow{5}{*}{\hfil\centering  \citealp{DUG-19}} \\
\multicolumn{3}{c|}{$[$Pd(1)/Co(0.5)/W(0.1)/Pd(1)$]_{\times 15}$} & \textit{1.1} $\pm$ \textit{0.2} & \textit{0.55}&  n.a.  &\\
\multicolumn{3}{c|}{$[$Pd(1)/Co(0.5)/W(0.2)/Pd(1)$]_{\times 15}$} & \textit{1.3} $\pm$ \textit{0.2} & \textit{0.65}&  n.a.  &\\
\multicolumn{3}{c|}{$[$Pd(1)/Co(0.5)/W(0.3)/Pd(1)$]_{\times 15}$} & \textit{0.4} $\pm$ \textit{0.1} & \textit{0.2}&  n.a.  &\\
\multicolumn{3}{c|}{$[$Pd(1)/Co(0.5)/W(1)/Pd(1)$]_{\times 15}$} & \textit{0.4} $\pm$ \textit{0.1} & \textit{0.2}&  n.a.  &\\
\hline
\hline
\end{tabular}
\footnotetext[1]{The sign was measured using CD-XRMS (circular dichroism in x-ray
resonant magnetic scattering). Sign convention used here is opposite to the used in the article.}
\footnotetext[2]{Values assuming $A=20$~pJ/m. Other values are also assumed in the article.} 
\footnotetext[3]{The negative sign of DMI is inferred in combination with other experiments.}
\end{table*}

%table anhil
\begin{table*}
\caption{Overview of DMI measurements of \textbf{Co and CoFeB} thin films via \textbf{magnetic stripe annihilation}. FM and NM stand for ferromagnetic and non-magnetic layer, respectively. $D$ is the interfacial DMI constant and $D_s = D \cdot d$, with $d$ being the thickness of the ferromagnetic film. Numbers in roman were quoted in the reviewed papers, while numbers in italics were either extracted from figures or calculated using the parameters provided.} 
\label{table:Tableanihil}

\begin{tabular}{c|c|c|c|c|c}
\hline 
\hline
\textbf{FM} & \textbf{Bottom NM} & \textbf{Top NM}  & $\mid$$\mathbf{D}$$\mid$ & $\mid$$\mathbf{D_s}$$\mid$ & \textbf{Ref}\\ 
(nm) & (nm) & (nm) & (mJ/m$^2$) & (pJ/m) & \\
\hline 
\hline
 Co(0.8) & Pt(3) & AlO$_x$(3) & 0.33 $\pm$ 0.05 \footnotemark[1]& \textit{0.27} \footnotemark[1]& \citealp{BEN-15}\\
\hline
[Pt(3)/Co(0.9)/Ta(4)]$_{\times 15}$ & Ta(3) &  & 1.1 $\pm$ 0.2 &\textit{1} &\citealp{WOO-16}\\
\hline
Co$_{20}$Fe$_{60}$B$_{20}$(1)& Ta(5)& Ta(0.74-0.9)& 0.25 &\textit{0.25} &\citealp{YU-16a}\\
\hline
 Co$_{20}$Fe$_{60}$B$_{20}$(0.6) & W(5) & MgO(2) & 0.73 $\pm$ 0.5& \textit{0.44} & \citealp{JAI-17}\\
\hline
[Pt(3)/Co$_{40}$Fe$_{40}$B$_{20}$(0.8)/MgO(1.5)]$_{\times 20}$ & Ta(3) & Ta(2) & 1.35 &\textit{1.08} &\citealp{WOO-17}\\
\hline
[Pt(3.2)/CoFeB(0.7)/MgO(1.4)]$_{x15}$ & Ta(3) &  & 1.35 $\pm$ 0.05 &\textit{0.95} &\citealp{LIT-17}\\
%\hline
%CoFeB(1)& W(0.2-1) & Mg0(2) & 1 & 0.01--0.25 &\textit{0.01--0.25} &\citealp{REA-17}\\
\hline
\hline
\end{tabular}
\footnotetext[1]{Lower limit}
\end{table*}

\subsubsection{Nucleation field}\label{sec:nucleationexperiments}
A reversed domain is nucleated in a perpendicular magnetized material and the out-of-plane nucleation field as a function of the in-plane applied field is analyzed. The latter can be particularly large, up to 1~T as in \textcite{KIM-17b}. This type of measurement can be divided in two groups: edge nucleation in patterned wires \cite{PIZ-14} or asymmetric microstructures (e.g. triangles)  \cite{HAN-16} and bubble nucleation in extended films \cite{KIM-17b,KIM-18c}. In all the experiments, the magnetic images were obtained using MOKE. In \textcite{HAN-16}, the hysteresis loop of asymmetric microstructures measured by wide-field polar Kerr shows an asymmetry due to DMI similar to exchange biasing. This asymmetry is attributed to the asymmetric nucleation when an in-plane field is present and is independent of the structure size. The values obtained for $D$ are presented in Table \ref{table:Tablenucleation}, mostly single layers of Co \cite{PIZ-14, HAN-16,KIM-18c} or trilayers of Co/Ni/Co \cite{KIM-17b}. \textcite{PIZ-14} obtained a value of $D_s$=-1.32~pJ/m ($D$=-2.2~mJ/m$^2$) for Pt/Co/AlO$_x$. \textcite{HAN-16} measured the opposite sign for the reversed structure  AlO$_x$/Co/Pt with a value of $D_s$=1.62~pJ/m ($D=$1.43~mJ/m$^2$). When the structure was reversed and the  AlO$_x$ was replaced by Ir a value of $D_s$=-2.03~pJ/m ($D=$-1.69~mJ/m$^2$) was measured, indicating opposite contributions for Pt and Ir. This method was used to obtain the value of $D$ as a function of temperature in a Co layer \cite{KIM-18c}, obtaining a decrease from 1.18~mJ/m$^2$ at 100 K to 0.48~mJ/m$^2$ at 300~K.

%Table nucleation
\begin{table*}
\caption{Overview of DMI measurements of \textbf{Co and Co/Ni} thin films via \textbf{nucleation field} dependence. FM and NM stand for ferromagnetic and non-magnetic layer, respectively. $H_\mathrm{DMI}$ is the DMI field, $D$ is the interfacial DMI constant and $D_\mathrm{s} = D \cdot d$, with $d$ being the thickness of the ferromagnetic film. Numbers in roman were quoted in the reviewed papers, while numbers in italics were either extracted from figures or calculated using the parameters provided. Edge nucleation provides $D$
sign, while bubble nucleation provides only the magnitude. Values with $^*$ are according to the convention used in this paper and opposite to that in the original manuscript.} 
\label{table:Tablenucleation}

\begin{tabular}{c|c|c|c|c|c|c|c|c}
\hline 
\hline
\textbf{FM} & \textbf{Bottom NM} & \textbf{Top NM}  & $\mid$\textbf{$\mu_0$H$_\mathbf{{DMI}}$}$\mid$ & $\mathbf{D}$ & $\mathbf{D_s}$ & \textbf{Sign} & \textbf{Nucleation type}& \textbf{Ref}\\ 
(nm) & (nm) & (nm) & (mT) & (mJ/m$^2$) & (pJ/m) && & \\
\hline 
\hline
 Co(0.6) & Pt(3) & AlO$_x$(2) &  & 2.2$^*$ & \textit{1.32}$^*$  &-& Edge & \citealp{PIZ-14}\\
\hline
 Co(1.15) & AlO$_x$(2.5) &  Pt(4) &  & 1.43$^*$ $\pm$ 0.06 & 1.62$^*\pm$0.07  &+& \multirow{2}{*}{\centering Edge} & \multirow{2}{*}{\centering \citealp{HAN-16}}\\
 Co(1.2) & Pt(4) &  Ir(4) &  & 1.69$^*$ $\pm$ 0.03 & 2.03$^*$ $\pm$ 0.04  &-&  & \\
\hline
 Co(0.3)/Ni(0.6)/Co(0.3) & Pt(2) & MgO(1) & 228$\pm$60 & 0.45 $\pm$ 0.15 & \textit{0.54}  &n.a.& Bubble & \citealp{KIM-17b}\\
\hline
 \multirow{4}{*}{\centering Co(0.5)} & \multirow{4}{*}{\centering Pt(2)} & \multirow{4}{*}{\centering MgO(2)} & 372$\pm$30 & \textit{1.18} & \textit{0.59} &n.a.& Bubble (100K)& \multirow{4}{*}{\centering \citealp{KIM-18c}} \\
  &&& 324$\pm$15 & \textit{1.03} & \textit{0.52} &n.a.& Bubble (150K) & \\
  &&& 245$\pm$45 & \textit{0.75} & \textit{0.37} &n.a.& Bubble (200K) & \\
  &&& 166$\pm$50 & \textit{0.48} & \textit{0.24} &n.a.& Bubble (300K) &\\
\hline
\hline
\end{tabular}

\end{table*}

\subsubsection{Domain wall stray fields} \label{sec:NVmagnet}
In NV magnetometry, the value of $D$ is obtained by measuring the magnetic stray field generated by a 180$^\circ$ Bloch wall in PMA materials \cite{TET-15,GRO-16}. The NV magnetometer measures the Zeeman shift in the electronic spin sublevels of a NV defect in a diamond crystal in presence of a small magnetic field. The diamond nanocrystal is placed on the tip of an AFM (atomic force microscope) and scanned across the DW at a distance of about 100~nm from the surface (see Fig.~\ref{fig:NV_center}). The Zeeman shift is proportional to the projection of the external magnetic stray field (arising from the domains adjacent to the wall, in the case of a Bloch wall, and of the domains adjacent to the wall and the wall itself, in the case of a Ne\'el wall) on the quantization axis of the NV center ($B_\mathrm{NV}$). Therefore, Bloch and Ne\'el walls can be easily distinguished by the Zeeman shift profile perpendicular to the wall. Since the stray field depends on several parameters such as the distance from the surface, $M_\mathrm{S}$ and the DW width, any error in these values reflects in an uncertainty of $D$. The main source for uncertainty is usually considered to arise from $A$, the exchange stiffness, but also from inhomogeneities in $M_\mathrm{S}$ or thickness variations. 

%Table nv center
\begin{table*}
\caption{Overview of DMI measurements of \textbf{CoFeB} thin films by \textbf{stray field using NV magnetometry}. FM and NM stand for ferromagnetic and non-magnetic layer, respectively, $D$ is the interfacial DMI constant and $D_\mathrm{s} = D \cdot d$, with $d$ being the thickness of the ferromagnetic film. Numbers in roman were quoted in the reviewed papers, while numbers in italics were either extracted from figures or calculated using the parameters provided.} 
\label{table:Tablenvcenter}

\begin{tabular}{c|c|c|c|c|c|c}
\hline 
\hline
\textbf{FM} & \textbf{Bottom NM} & \textbf{Top NM}   & $\mathbf{D}$ & $\mathbf{D_s}$ & \textbf{Sign} & \textbf{Ref}\\ 
(nm) & (nm) & (nm)  & (mJ/m$^2$) & (pJ/m) &&  \\
\hline 
\hline
Co$_{40}$Fe$_{40}$B$_{20}$(1) & Ta(5) & \multirow{3}{*}{\centering MgO(2)}  & 0$\pm$0.01 & \emph{0$\pm$0.01}  & & \multirow{3}{*}{\centering \citealp{GRO-16}}\\
 \multirow{2}{*}{\hfil \centering Co$_{20}$Fe$_{60}$B$_{20}$(1)} & TaN$_{0.7\%}$(4) &   & 0.03$\pm$0.01 & \emph{0.03$\pm$0.01}  &+& \\
  & TaN$_{0.7\%}$(1) &   & 0.06$\pm$0.02 & \emph{0.06$\pm$0.02}  &+& \\
\hline
\hline
\end{tabular}
\end{table*}

\subsubsection{Domain wall internal structure imaging}

\begin{figure}[t]
\centering
\includegraphics[width=.8\linewidth]{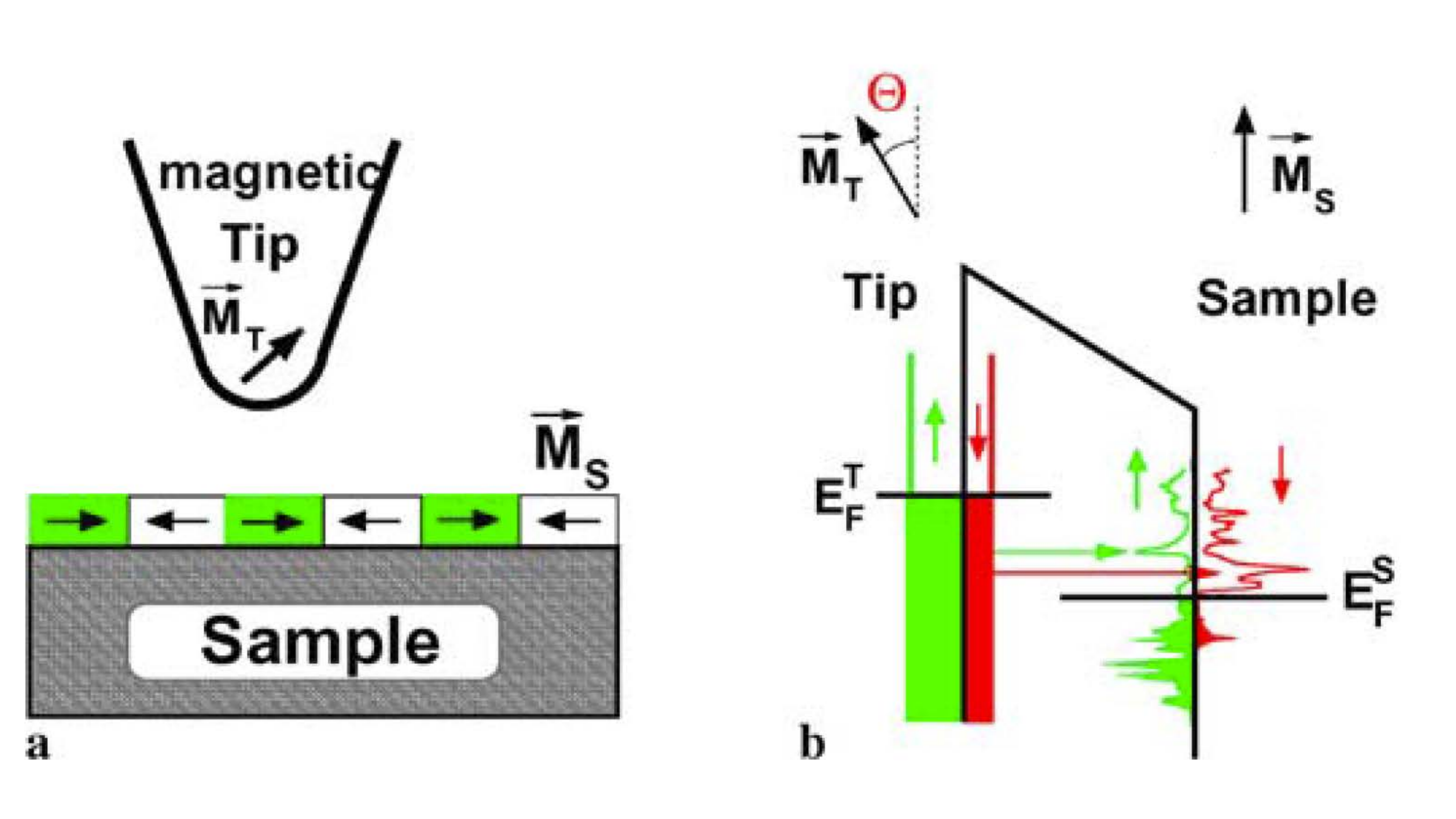}
\caption{Schematics of spin-polarized scanning tunneling microscope (SP-STM) using a magnetically coated scanning tunneling microscope tip: (a) geometry of the experimental set-up including the magnetization axes of tip $\mathbf{M}_\mathrm{T}$ and sample $\mathbf{M}_\mathrm{S}$. (b) electronic structure of tip and sample. In the case of a perfect parallel alignment of $\mathbf{M}_\mathrm{T}$ and $\mathbf{M}_\mathrm{S}$, only majority electrons can tunnel into unoccupied majority states and only minority electrons into unoccupied
minority states. From \citealp{HEI-06}.}
\label{fig:SPSTM}
\end{figure}

In \textcite{BOD-07} the spin-polarized scanning tunneling microscopy (SP-STM) technique was used to investigate the role of the DMI in systems with chiral spin structure. The specific chirality of the moments in single atomic layers of Mn on a W(110) substrate was observed directly and the different energy contributions of the cycloid structure were calculated by employing density functional theory with a generalized gradient approximation and full-potential linearized augmented plane waves. The long-range homogeneous spiral structure was accounted for by the generalized Bloch theorem adding spin-orbit coupling as a perturbation. The SP-STM experiments (see scheme in Fig.\ref{fig:SPSTM}) were performed in ultrahigh vacuum at $T= 13\pm1$~K in constant current mode. For spin resolved measurements Cr- and Fe-covered W tips were used. The tunneling current $I$ is sensitive to the relative orientation of the magnetization of the tip ($\mathbf{M}_\mathrm{T}$) and the sample ($\mathbf{M}_\mathrm{S}$), according to  $I= I_0 + I_\mathrm{sp} \mathbf{M}_\mathrm{T} \cdot \mathbf{M}_\mathrm{S}$, where $I_\mathrm{sp}$ is the spin polarized part of $I$. This allows for an atomic scale imaging of magnetic nanostructures \cite{HEI-00}. This work was continued in \textcite{FER-08a}, yielding a nearest-neighbor DMI of 4.6~meV, and $D_\mathrm{s}= 2.33$~pJ/m with an approximate lattice constant of that of W (3.165~$\AA$). In \textcite{HEI-11} Fe monolayers on Ir(111) were studied, showing similar values. Double layers of Mn were studied in \textcite{YOS-12} and Fe islands on Cu(111) in \textcite{FIS-17}. Similar studies exploiting SP-STM were also presented by \textcite{MEC-09} where they found that the Dzyaloshinskii-Moriya interaction is too weak to destabilize the single domain state, but it can deﬁne the sense of rotation and the type of the cycloidal spiral once the single domain state is destabilized by dipolar interaction. More recent SP-STM studies are found in \textcite{HSU-18} where the effect of loading double layers of Fe is investigated, in \textcite{PER-18} epitaxial films of Co/Ir(111) and Pt/Co/Ir(111) are studied making use of density functional theory (DFT) and shows that in the latter DMI comes almost only from the interface with Pt, in \textcite{ROM-18} Rh/Fe atomic bilayers on Ir(111) were studied showing that higher-order exchange interactions may compete with interfacial DMI.   

\begin{figure}[t]
\centering
\includegraphics[width=\linewidth]{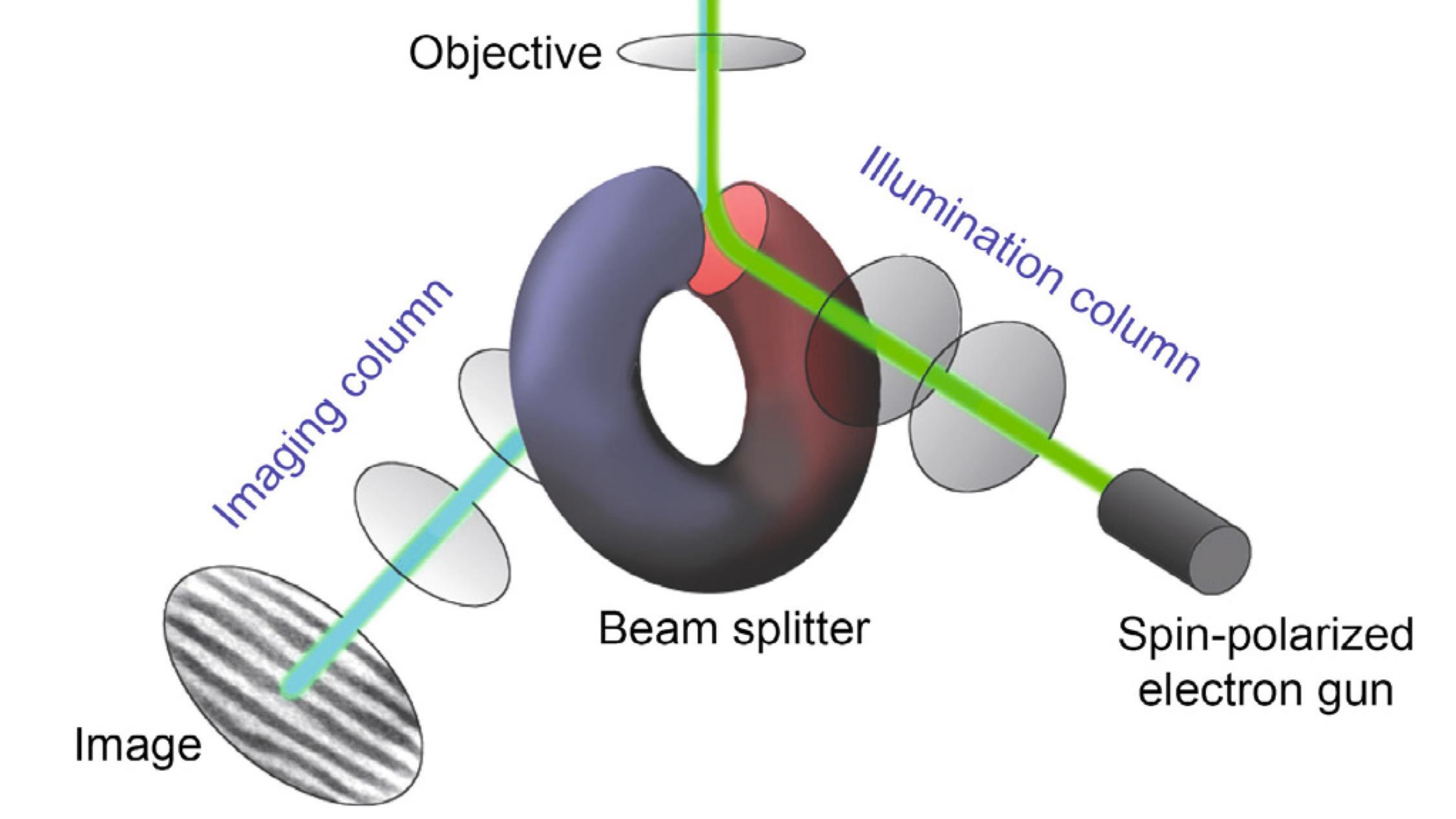}
\caption{Schematics of a SPLEEM microscope.
Spin-polarized electrons, photoemitted from a GaAs photocathode,
are injected into a spin manipulator where azimuthal
and polar orientation of the polarization is adjusted. Then, the electron beam passes through an illumination column, before being decelerated in the objective
lens. Electrons finally hit the surface with normal incidence.
Electrons that are backscattered elastically are collected in an
imaging column and focused on a phosphorous screen, where
a magnified image of the surface is obtained. The incoming
and reflected electron beams are separated in a magnetic beam
splitter using the Lorentz force. From \textcite{ROU-10}.}
\label{fig:SPLEEM}
\end{figure}

In addition to the above results obtained by SP-STM, also the SPLEEM technique has yielded a number of quantitative studies of DMI in layered systems. In this experimental approach, spin-polarized, low-energy electrons are projected towards the sample surface through an illumination column and reach the sample at normal incidence \cite{ROU-10}. A magnetic beam splitter is exploited to separate the incoming electrons from the backscattered ones and a magniﬁed image of the (magnetic) surface is obtained by passing the backscattered beam through an imaging column, similar to that found in electron microscopes (a schematic is shown in Fig.\ref{fig:SPLEEM}). This technique is extremely surface-sensitive, since it employs electrons of a few eV energy that do not penetrate more than a few atomic layers, while the lateral resolution is of the order of several nanometers.  Therefore, SPLEEM requires in-situ analysis of samples prepared in ultra-high vacuum and it is usually not suitable for samples prepared elsewhere. By a careful comparison of SPLEEM images on films of different thicknesses, the film thickness dependent transition from chiral Néel walls (dominated by the DMI) and non-chiral Bloch walls (dominated by the stray field) can be directly analyzed, so that one can quantify the DMI relying on the fact that the DMI energy and the stray field energy are comparable near the transition thickness. This is because the stray field energy will increase with increasing film thickness whereas the DMI energy is usually independent of the film thickness. Thus, the DMI strength can be quantified by calculating the stray field energy for a wall structure with a given film thickness. This approach has been followed in a number of quantitative studies that are collected in Table\ref{table:imaging}.

As for Lorentz-TEM microscopy, it was exploited to derive the value of the DMI \cite{GAR-19} looking at the mixed Bloch-N\'{e}el chiral spin textures in Co/Pd multilayers. An analysis of the observed intensities under varied imaging conditions coupled to corroborative micromagnetic simulations permitted to quantify different magnetic parameters, including the domain wall width, the exchange stiffness and the DMI.
Finally, also SEMPA was used to quantify the DMI in thin film systems. For instance, in \textcite{COR-17}, using the experimentally determined wall angle and width parameter, a span for the DMI strength of 0.8~meV/Co$ <d<$4.3~meV/Co has been derived, which was in agreement with previous ab initio calculations for this system.  Also, in a recent SEMPA study of Fe$_3$GeTe$_2$ films \cite{MEI-20a} a lower bound of $D >$0.09-0.2~mJ/m$^2$ for the DMI term, films was achieved.  A more refined and quantitative result was achieved by \cite{KLO-19}, studying the domain wall profiles of Co/Ir(111) films as a function of Co thickness via SEMPA. They found that below a cobalt thickness of 8.8~MLs, the magnetic domain walls are purely N\'{e}el oriented and show a clockwise sense of rotation. For larger thicknesses the plane of rotation changes and the domain walls show a signiﬁcant Bloch-like contribution, allowing one to calculate the strength of the Dzyaloshinskii-Moriya interaction (DMI) from energy minimization. In particular, from the angle between the plane of rotation and the domain-wall normal an interfacial DMI parameter $D_s = -$1.07 $\pm$ 0.05~pJ/m was determined.

The pioneering approach of these methods was among the first allowing an experimental determination of the DMI and observing directly its influence on the magnetic structure, and therefore the area remained very active and substantial progress towards a quantitative determination of the DMI was achieved (see Table\ref{table:imaging}).

\begin{table*}
\caption{Overview of DMI measurements by \textbf{Domain wall internal structure imaging}. The measurements in \textcite{CHE-13, CHE-13a, CHE-17, CHE-20, YAN-18} were performed by \textbf{SPLEEM}, \textcite{MEC-09} by \textbf{SP-STM}, \textcite{GAR-19} by \textbf{L-TEM} and \textcite{KLO-19} by \textbf{SEMPA}. FM and NM stand for ferromagnetic and non-magnetic layer, respectively, $D^*$ is the DMI value given in meV/atom, $D$ is the interfacial DMI constant and $D_\mathrm{s} = D \cdot d$, with $d$ being the thickness of the ferromagnetic film. Thicknesses are given either in nm or atomic monolayers (ML). Numbers in roman were quoted in the reviewed papers, while numbers in italics were either extracted from figures or calculated using the parameters provided. Signs with $^*$ are according to the convention used in this paper and opposite to that in the original manuscript. \emph{n.a.} means that the value was not indicated in the original paper nor were parameters available to calculate it.} 
\label{table:imaging}
\begin{tabular}{c|c|c|c|c|c|c|c}
\hline 
\hline
\textbf{FM} & \textbf{Bottom NM} & \textbf{Top NM}  & $\mathbf{D^*}$ & $\mathbf{D}$ & $\mathbf{D_s}$ & \textbf{Sign} & \textbf{Ref}\\ 
 &  &  & (meV/atom) & (mJ/m$^2$) & (pJ/m) && \\
\hline 
\hline
Fe(1.7ML) & (110)W & & & & 1.2 & +$^*$ & \citealp{MEC-09}\\
\hline
Ni(2ML)/Fe(2.5ML) & (001)Cu & & 0.12-0.17 & 0.08-0.12& & - &  \citealp{CHE-13a}\footnote[1]{The values for $D$ given in mJ/m$^2$ were kindly provided by the authors of the original paper. In some cases the sign or thickness values were corrected with respect to the original work.}\\
\hline
\multirow{3}{*}{\centering Ni(2ML)/[Co(1ML)/Ni(2ML)]$_2$} & (111)Pt & \multirow{3}{*}{\centering  } & 1.05 & 0.44 &  &-& \multirow{3}{*}{\hfil \centering \citealp{CHE-13}\footnotemark[1]}\\ 
 & (111)Ir &  & 0.12 & 0.14 &  &+& \\
 & (111)Pt/Ir(0.6-3ML) &   & 0.47-0.12  &0.28-0.14 & &from - to +& \\
\hline 
Ni(15ML)/Fe(3.3ML) & (110)W &  & 0.53 & 0.32 &  &+& \citealp{CHE-17}\footnotemark[1] \\ 
\hline
Co(2.4ML) & \multirow{2}{*}{\hfil \centering (0001)Ru} & - & 0.05 & 0.18 &  &+& \multirow{2}{*}{\hfil \centering \citealp{YAN-18}\footnotemark[1]}\\ 
 Co(3.9ML-8.4ML)&  & Graphene & 0.11 & 0.25 &  &-& \\
\hline
 Co(0.7nm)[Pd(0.5nm)/Co(0.7nm)]$_{10}$ & MgO(2nm)/Pt(4nm) & Pt(4nm) & & 1.05 & & n.a.& \citealp{GAR-19}\\
\hline
 Co(3.5ML-9.7ML) & (111)Ir & & & & 1.07 & +$^*$ & \citealp{KLO-19}\\
\hline 
Co(3ML)/Ni(2ML)/[Co(1ML)/Ni(2ML)]$_8$ & (111)Pd & - & 1.44 & 0.54 &  &-& \multirow{2}{*}{\hfil \centering \citealp{CHE-20}\footnotemark[1]}\\ 
Co(3ML)/Ni(1ML) & (110)W/Pd(2.6ML) & O(0.2ML) & 0.63 & 0.4 &  &+&  \\ \hline 
Co(3ML)/Ni(1ML) & (110)W/Pd(2.1ML) & H(0.6ML) & 0.01 & 0.01 &  &+& \citealp{CHE-21}\footnotemark[1] \\ 
\hline
\hline
\end{tabular}
\end{table*} 

%Instead Fe/Ni bilayers on Cu(100) were investigated by SPLEEM \cite{CHE-13a}, performing an analysis by Monte Carlo simulations. The complex analysis together with the fact that these techniques are limited to few atomic layers are drawbacks, however, an unprecedented microscopic insight into the relation between non-collinear spin configurations and the magnetic energy contributions involved was reached.

\subsection{Advantages and limitations}\label{sec:advantagesDW}
In general, methods based on domain walls require an accurate estimation of several magnetic parameters in order to evaluate the interfacial DMI (see. Eq.~\ref{eq:DMI}). Among these, the exchange stiffness $A$, which enters both in the expression of the DMI field and in the DW energy, is notoriously difficult to measure for ultrathin films leading to an uncertainty in the quantification of $D$. \textcite{HAN-16} pointed out that in the nucleation method a variation of $A$ from 5 to 15~pJ/m for Co results in a 25\% difference for $D$ obtained using the intermediate value 10~pJ/m. This is not the case for methods based on spin waves (see Sec.~\ref{sec:spin waves}), where $A$ is not needed to extract the interfacial DMI. Furthermore, it is worth mentioning that methods that rely on domain walls can be applied to measure $D$ only in perpendicularly magnetized materials, while methods based on spin waves (see Sec.~\ref{sec:spin waves}) are able to quantify $D$ also for systems with in-plane magnetization.  Below we review the main strengths and weaknesses of all methods discussed in this Section.

%\subsubsection{DW motion}
Both, measurements of current-driven and field-driven DW dynamics in presence of an in-plane magnetic field allow quantifying not only the magnitude of $D$ but also its sign, which gives information on the DW chirality. 
On the other hand, in the creep regime, the dependence of the DW velocity on the in-plane field does not always follow the simple linear dependence when driven by current (see  Fig.~\ref{fig:current}), or the symmetric shape with a well defined minimum in the field-driven case (see Fig.~\ref{fig:bubble}). Deviations from these simple theoretical predictions are in fact quite common (see Fig.~\ref{fig:asymmetric_v_profile} as an example) and there is no consensus on a general model able to predict and fit all the observed cases. Several models predict that the minimum velocity does not correspond to the DMI field but agreement on its calculation is still missing, and can lead to quite different determination of the $D$ value. 
The main reason is that a full and general microscopic description of the domain wall dynamics is still under investigation. While some systems seem to follow the quenched Edward-Wilkinson equation \cite{ALB-21,GRA-18}), there are also experimental evidences that other system are better described by a quenched-KPZ equation \cite{DUR-21}, having different exponents and critical dynamics. The consequences on the DMI estimation of these two universality classes is still unknown. 

Current-driven DW motion is overall a more complex technique to determine the interfacial DMI,  both for experimental realization and for the interpretation of the results. A standard lithography process is required to pattern films into wires of appropriate geometry and high in-plane fields, especially for samples with large DMI, need to be applied and properly aligned.
% In case of very narrow wires, a misalignment of these in-plane fields with respect to the wire axis is possible, giving rise to unwanted field components that could alter the motion of DWs and consequently the measurement of their velocity. 
Furthermore, the torque generated by large applied currents can also result in a tilt of the DW \cite{EMO-14}. Concerning the interpretation of the results, a clear definition of $H_{DMI}$ might be hard due to the interplay present in complex systems between different current-induced spin torques (STT, SHE-SOT, iSGE torque), whose collective effect on DW dynamics is still not fully understood. 

Field-driven DW motion has the advantage of being a relatively simple method to implement experimentally, as it allows the measurement of interfacial DMI in continuous films, without the need for any lithographic patterning. However, the application of high in-plane fields, particularly for samples with significant DMI, requires a perfect alignment of the in-plane field, with no out-of-plane components. Indeed, in the creep regime any unwanted component of the field perpendicular to the sample plane would influence the DW velocity in an exponential manner (see Eq. \ref{eq:Creep}). Several precautions should be taken into account to achieve an optimal alignment of the in-plane field \cite{JE-13,LAV-15,SOU-16,CAO-18}. A perpendicular stray field component may still arise in MOKE set-ups, due to a cross-talk between in-plane and perpendicular electromagnets or due to not completely non-magnetic objectives, which would alter the field distribution in a way difficult to take into account.

Regarding measurements of DW velocities in the flow regime, here the main difficulty lies in the generation of large and short magnetic field pulses and the measurement of very large velocities. The determination of the DMI constant from the Walker DW velocity (or the saturation velocity after the Walker field) does not need the knowledge of the exchange constant (see Eq.~\ref{Eq:v_wb}), but on the other hand, may not be accessible when hidden in the creep regime. 
% In any case, aside from all these experimental difficulties, the creep/flow field-driven technique suffers important limitations due to inherent problems in the interpretation of the results. 
%While avoiding the complications due to different torque effects, which arise when DWs are driven by current, a unifying model able to understand the wide variety of bubble morphology observed and related DW velocity trends is still under investigation.

%\subsubsection{DW energy}
Among the methods that are based on the DW energy to evaluate the interfacial DMI, imaging the equilibrium stripe domain pattern is particularly straightforward due to its experimental simplicity. Indeed, this method is compatible with any magnetic imaging technique like Kerr microscopy or MFM -- which has to be chosen according to the expected domain width --  and does not require the application of perpendicular nor in-plane fields. On the other hand, imaging the equilibrium domain configuration does not provide information on the sign of $D$. For multilayer samples with large DMI, small $A$ or any situation that yields small values of the left hand side of Eq.~\ref{eq:energy_domain_width}, high resolution imaging is needed in order to resolve precisely the small domain widths. Experimentally, several problems can emerge that may render it unusable (see Supplementary material of \citealp{LUC-17}): for instance, the demagnetization procedure is unavoidable, since the as-deposited samples may not be in the ground state. In patterned systems, the domain wall width, the domain size and configuration may depend on the geometry, due to confinement and dipolar effects, adding discrepancies with respect to analytical formulations. Other disadvantages include the fact that the simple analytical model to extract $D$ may not work for too thick samples which yield neither N\'eel nor Bloch wall types (hybrid wall with N\'eel caps and Bloch core), although the DMI is expected to be low in these systems. 

From the experimental point of view, the domain widths measured may be different for samples demagnetized with in-plane or out-of-plane fields, up to 20\% \cite{LEG-18,DAV-19,KOZ-20}. For this reason, the in-plane demagnetized configuration is preferred, as it is closer to the parallel stripes described by the analytical theory. In principle, these problems may be solved by using the proper analytical model. The analytical estimations are derived for perfect parallel stripe domains, while the real domain configuration consists of rounded meandering structures. Moreover, refined expressions for the DW energy may yield slightly different values of DMI \cite{LEM-17}. This disagreement in the modeling complicates the extraction of DMI and the interpretation of the experimental results.

Similar considerations hold for the method based on magnetic stripe annihilation. Here too, it is only possible to extract information on the absolute value of $D$, and not its sign. From an experimental point of view, the original application of this technique requires the presence of parallel DWs, which may not apply to the case of materials with irregular wall profiles. Those irregular walls can appear for materials that can host skyrmion phases, due to the intrinsic faceting of the DWs in such films. This fact can lead to a deviation from the analytical theory, which was derived for parallel stripe domains. Another drawback is that as the domain walls are closer together the corresponding energy barrier for thermally induced domain wall annihilation is reduced and is also affected by the presence of defects, which can yield an error in the determination of the DMI value. 

Since it relies on the nucleation of a reversed domain, the process of bubble nucleation to extract the interfacial DMI is intrinsically a statistical process and several repetitions of the same experiment need to be carried out to achieve a reliable measurement of the nucleation field. In contrast, for instance, imaging the equilibrium domain configuration in a film can automatically provide averaged information of the domain width, from which the DMI is quantified, if a large enough area is imaged. Bubble nucleation experiments can be performed both at the edge of patterned structures, such as nucleation pads or wires, as well as in continuous films. In the former case, the method can be used to determine both the absolute value of $D$ as well as its sign. Specifically, the sign of $D$ is related to the nucleated bubble position with respect to the direction of the applied in-plane field. On the other hand, bubble nucleation measurements in continuous films provide only the absolute value of $D$.

Concerning the methods based on domain and domain wall imaging, it has been shown on different magnetic systems (see Table\ref{table:imaging}) that SP-STM and SPLEEM (and, to a lesser extent, SEMPA and L-TEM) can be used to achieve a quantitative estimation of the DMI strength, looking at the domain structure or at the transition from Néel to Bloch wall as a function of thickness. One should consider that SP-STM and SPLEEM require a complex apparatus with ultra-high vacuum and in-situ operation, since they are sensitive to the first atomic layers, so that they cannot be employed for samples grown elsewhere and the sample has to be perfectly clean. Use of DFT or Monte Carlo simulations may help in refining the extraction of DMI, but requires a remarkable computational effort. However, the main advantage of these methods is the direct determination of the structure and chirality of the DW.

\section{Spin wave methods}\label{sec:spin waves}

\subsection{Method overview}

\begin{figure*}[ht!]
\centering
\includegraphics[width=1\textwidth, trim=0 0 0 0,clip]{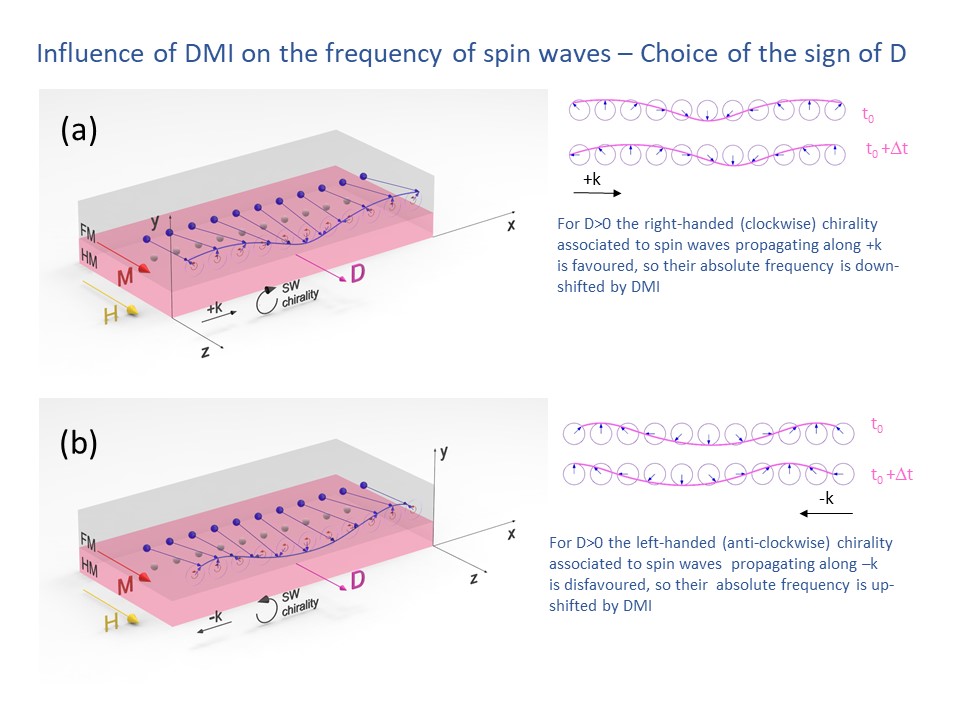}
\caption{ (a) Sketch of a spin wave, in the Damon-Eshbach configuration, propagating towards the positive $x$ direction, with a wave vector +$k$, in the FM film. All the individual magnetic moments (blue arrows) are precessing anticlockwise around the $z$-axis, i.e. around the direction of the static magnetization and of the external field $H$. Due to the phase delay from one spin to the next, moving from left to right, the chirality associated with the SW is clockwise, that is favoured by a positive DMI (magenta $D$ vector). In thin bilayer films of FM/HM, the latter couples two neighbouring spins via a three-site exchange mechanism with the underlying atom of the HM (white atoms). As a consequence, the absolute frequency of spin waves with +$k$ is down-shifted in frequency. (b) SW propagating along the negative $x$-direction. i.e. with wave vector -$k$, are characterized by an anti-clockwise chirality, that is disfavoured by a positive $D$, resulting in an up-shift of their absolute frequency.}
\label{fig:SW1}
\end{figure*}

Since their formation and propagation depend on the magnetic energies present in a material or a thin film, spin waves (SWs) are used as a sensitive tool to extract magnetic properties such as susceptibility, anisotropy, Gilbert damping, exchange stiffness, etc. These properties, together with the geometry of the sample and the excitation, determine the spin wave frequency, linewidth, relative amplitude and attenuation length. The DMI does not only modify the equilibrium magnetic ground state \cite{BAK-80} but also the spin wave dynamics \cite{MEL-73, KAT-87, ZHE-99}. SWs propagating in opposite directions have opposite chirality, so that the DMI contribution to the SW energy may either decrease or increase the SW frequency and results in an intrinsic non-reciprocity of the SW propagation.\footnote{This does not occur for arbitrary directions of DMI vector and spin wave propagation as we will explain further below, and special experimental geometries have to be realized.} This is illustrated in Fig.~\ref{fig:SW1}. Note that, different from the case of the domain walls analysed in the previous section, where the static magnetization is usually normal to the film plane,  here the static component of the magnetization is forced to be in-plane by the external magnetic field. Therefore, in the BLS community it is popular to assume a different  reference frame, as we do in this section, where both the applied field $H$ and the static magnetization lie along the $z$ direction as shown in Fig.~\ref{fig:SW1} while the dynamic component of the magnetization (small blue arrows in Fig.~\ref{fig:SW1}) exhibit a chirality that can be either the same or opposite to the DMI-favoured chirality, depending on the propagation direction of SWs. For magnetic films this DMI-induced non reciprocity was investigated the first time by \textcite{UDV-09} and later a series of theoretical investigations followed \cite{COS-10,MOO-13,COR-13,KOS-14a}. Experimentally, the relatively strong asymmetry in spin wave dispersion is of interest, since the frequency of spin wave spectra can be measured with high accuracy. A pioneering investigation of DMI-induced non-reciprocity in spin wave propagation was performed analyzing SWs with large wavenumber \cite{ZAK-10} by spin polarized electron loss spectroscopy (SPEELS). Although a high wavenumber $k$ is advantageous\footnote{As we will show further below, the DMI -induced frequency shift is linear with wave vector k of the involved spin waves.}, leading to large DMI-induced frequency changes, the most versatile and exploited technique to determine DMI-related effects on SW propagation is Brillouin light scattering (BLS), which relies on the inelastic scattering of photons by spin waves with micrometric or sub-micrometric wavelength. BLS is becoming more and more popular, thanks to the good compromise between sensitivity to SWs in a wide range of $k$-vectors and relatively simple experimental apparatus. Two techniques that have also been applied to the study of SW propagation are the Time Resolved Magneto-Optical Kerr Effect (TR-MOKE) and Propagating Spin Wave Spectroscopy (PSWS). In the following we first recall some theoretical background about the influence of DMI on SW characteristics and we will review the main results achieved by the four above mentioned experimental techniques, comparing their advantages and limitations.

\subsection{Theory and models}
In the literature we find basically two theoretical approaches to describe spin wave spectra in thin film samples with DMI. One is based on the classical theory of magnetostatic spin waves (see e.g. \textcite{GAL-19}), and this is the approach mainly utilized for analyzing experimental results obtained by BLS, PSWS or TR-MOKE. The other is based on the quantum spin wave theory, which is more suitable for high $k$ measurements in ultrathin films as done in SPEELS, where exchange interaction dominates the spin wave dispersion \cite{UDV-09,DOS-18,DOS-20}. In most experimental cases, the two formalisms are equivalent and slight differences derive only from the choice of energy terms. Some authors use even a mixed approach calculating the spin wave spectra by a quantization of the linearized Landau-Lifshitz equation \cite{UDV-03}. The models used to analyze the experimental results \cite{MOO-13} derive typically the dispersion relation and do not consider the specific sensing technique, so that they are applicable to all four techniques presented in the following section. 

\subsubsection{Quantum spin wave theory}

The usual quantum formalism for spin waves uses the Holstein Primakoff method~\cite{HOL-40} and starts directly from the Schr\"odinger equation with a Hamiltonian in a crystal lattice of spins $\vec{S}_j$ that considers the Heisenberg exchange energy at a spin site $j$ with the nearest neighbors at position $j+\delta$  and the Zeeman energy in a magnetic field $B_0 \hat{z}$
\begin{equation}
\mathcal{H}= -2\frac{J}{\hbar^2} \sum_{j,\delta} \vec{S}_j\cdot \vec{S}_{j+\delta} - \frac{g \mu_B B_0}{\hbar} \sum_{j} S_{j_z}.
\end{equation}
where $J$ is the exchange constant, $g$ the Land\'e factor, $\mu_B$ the Bohr magneton. As described in textbooks \cite{STA-91}, using the spin raising and lowering operators one can obtain the eigenstates and energies analytically in certain simplified cases (e.g. low temperature approximation) by employing the Holstein-Primakoff transformation and using the Fourier transform of the harmonic oscillator raising and lowering operators as creation and annihilation operators for the quantized spin waves (magnons). For $k a \ll 1$ in a cubic lattice of spins $s$ with lattice constant $a$, and considering only nearest neighbour interaction, one obtains the known quadratic dispersion relation
\begin{equation}
\hbar \omega_k=g \mu_B B_0 + 4 J s a^2 k^2.
\label{eq:SW1}
\end{equation}

As mentioned already in the introduction, in a microscopic approach the DMI is considered a generalized exchange interaction leading to a non-collinear spin configuration described by $\vec{D} \cdot (\vec{S_1} \times \vec{S_2})$, where the direction of $\vec{D}$ depends on the crystal symmetry. This leads to an additional term in the Hamiltonian \cite{ZHE-99,MOO-13}
\begin{equation}
\mathcal{H}_{DMI}=2\frac{D}{\hbar^2} \sum_j \hat{z} \cdot (\vec{S}_j \times \vec{S}_{j+1})
\end{equation}
where $\hat{z}$ is perpendicular to the axis of symmetry breaking.
The related term which has to be added in the spin wave dispersion (Eq.~\ref{eq:SW1}) is linear in $k$:
\begin{equation}
\hbar \omega_{DMI}=4 D s a k.
\end{equation}

A first study of DMI in bulk Ba$_2$CuGe$_2$O$_7$, an antiferromagnet with a weak helimagnetic distortion \cite{ZHE-99}, was performed already in 1999 using the quantum approach, showing low energy spin wave spectra. This approach does not consider the sample geometry but only crystal symmetry, so it is suitable for bulk materials. In the case of ultrathin films, it was shown theoretically about ten years later that the chiral degeneracy of the magnons can be lifted due to the presence of the DMI \cite{UDV-09}. The employed method is based on relativistic first principle calculations of the magnetic ground state similar to \textcite{BOD-07,FER-08a}, able to identify domain wall chirality. Therefore it was proposed to exploit this spin wave asymmetry for measuring DMI in ultrathin films. Although it was known already for a decade that the DMI stabilizes chiral spin structures in bulk materials with a certain crystal symmetry \cite{CRE-98,BOG-01}, the work of \textcite{UDV-09} was stimulated by the discovery of homochirality of domain walls in two monolayers of Fe on W(110) \cite{HEI-08,KUB-03}.  

A quantum approach was also used in \textcite{COS-10} where, going back to the microscopic origin of the DMI, the  effects of spin-orbit coupling (SOC) on spin wave spectra are studied. It is shown that the DMI leads to a linear term in $k$ in the dispersion relation and that the linewidth of spin wave modes is increased by the spin-orbit coupling. The method used here goes beyond the adiabatic approximation and directly operates in the wave vector space, avoiding calculations in real space, taking into account large numbers of neighbor shells. The starting point is the multiband Hubbard model with Hamiltonian \cite{COS-03} where a spin-orbit interaction is added. The spin wave dispersion is then obtained from the dynamic susceptibility. In essence, the quantum approach shows clearly the microscopic origin based on SOC and derives an additional $k$-linear term in the spin wave dispersion relation for certain relative orientations of spin wave propagation and direction of the DMI vector.

\subsubsection{Classical spin wave theory}

Since in most experimental investigations (as in BLS) the wavelength of the detected SWs are in the range between a few microns and a few hundreds of nanometers, one can ignore the discrete nature of the spins. Therefore, a classical formulation, assuming a continuum medium, is more suitable. Traditionally, spin waves in magnetic thin films are treated as magnetostatic spin waves, neglecting exchange interaction \cite{PRA-09}. Considering different geometries and their related boundary conditions one obtains either Forward Volume for perpendicular-to-plane magnetization or Backward Volume and Surface or Damon-Eshbach (DE) Waves for in-plane magnetization. The latter, where the wave vector $\vec{k}$ is perpendicular to the applied static field $\vec{H}$, is the preferred configuration when exciting spin waves via antennas due to their more efficient transduction.
However, as pointed out in several works, for the typical samples with DMI the simple magnetostatic solution is not sufficient and often anisotropy energy or exchange energy have to be considered, as in \textcite{GUR-96, KAL-81, KAL-86}. Here the complete problem is solved in the thin film limit for the magnetization $\vec{M}$ including the magnetostatic and dynamic regime using mixed boundary conditions, and going beyond the plane-wave approximation. The exchange term is often neglected when comparing theory to experiments, arguing that the wavenumber $k$ obtained in the experimental conditions is low. However, in order to be precise, the spin wave modes obtained in nanostructured films require the complete solution, being in the transition range from magnetostatic to exchange spin waves (dipole-exchange spin waves). One of the critical energy terms is the dipolar field term, which can be obtained from this approach by adding the exchange boundary conditions to the electrodynamics boundary conditions. As mentioned in \textcite{MOO-13}, the dipolar field term contains local and non-local contributions, and can be divided into a stray field term, related to the dipolar interactions between the spins in the SW, and a dipole or magnetostatic field term, related to the demagnetizing field. In most cases, as in \textcite{MOO-13}, the dipolar term for unpinned exchange at the film surface (i.e. $\frac{\partial}{\partial y} \vec{m}=0$ at the FM film surface) is applied \cite{KAL-94}. However, \textcite{KOS-14a} points out that the DMI pins the circular components of the magnetization at the surface (interface) of the magnetic film and mixed boundary conditions have to be used, which require a numerical solution.

\begin{figure}[ht!]
\centering
\includegraphics[width=\linewidth]{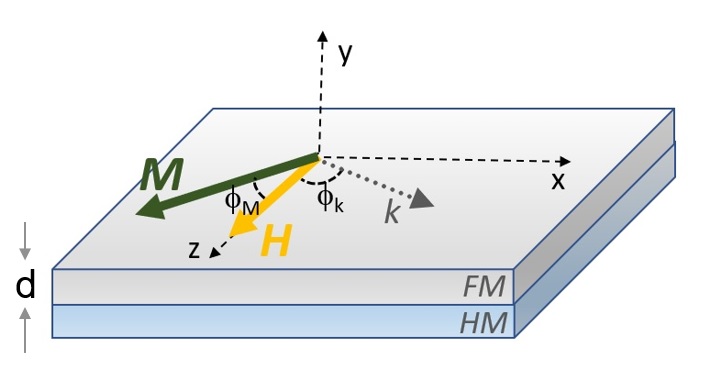}
\caption{Schematic diagram of the used geometry and notation. $k$ is the wave vector and indicates the propagation direction of the spin wave, $H$ is the applied in-plane field and $M$ the equilibrium magnetization, pointing slightly out-of-plane. The DMI frequency depends on the angles $\phi_k$ and $\phi_M$ as given in Eq.~\ref{eq:omegadmi}.}
\label{fig:SW2}
\end{figure}

Considering a sample geometry as shown in Fig.~\ref{fig:SW1}, (static component of the magnetization aligned along the z direction, i.e. parallel to the applied field $H$) the dispersion relation for small amplitude spin waves is derived from the linearized Landau-Lifshitz (LL) equation
\begin{equation}
    \frac{\partial \vec{M}}{\partial t}= - \gamma (\vec{M} \times \vec{H}_{eff})
\end{equation}
The vector $\vec{M}$ has to be decomposed in one large static component, of modulus $M_s$ directed along the $z$-axis and two small dynamic components that describe the precession around the equilibrium direction:
$\vec{M} = M_s\hat{m}(\vec{x},t)=M_s (m_x\hat{x}, m_y\hat{y},\hat{z})
$ with $|m_x|,|m_y|\ll 1$ and  $\hat{x}$,$\hat{y}$,$\hat{z}$ are the unitary vectors of the reference frame, chosen such that the $y$-axis is perpendicular to the sample plane, according to Fig.~\ref{fig:SW1}.
The effective field to be considered in the LL equation consists of different contributions, reflecting the different energy terms: 
\begin{equation}\label{eq:Heff}
\vec{H}_{eff}=-\frac{1}{\mu_0}\frac{\delta E_{tot}}{\delta \vec{M}}=\pm H \hat{z} +J\vec{\nabla}^2 \vec{m}+\vec{H}_{dip}+\vec{H}_{ani}+\vec{H}_{DMI} 
\end{equation}
where  $J=2A/\mu_0 M_s$ is the exchange constant. In case of unpinned exchange boundary conditions the dipolar field for a thin film can be written as \cite{ARI-99,KAL-94}: 
\begin{eqnarray}
\vec{H}_{dip}=-M_s m_x P(kd) \hat{x} - M_s m_y (1-P(kd)) \hat{y} \text{,} \\
\text{with} \hspace{1cm}  P(kd)=1-\frac{1-e^{-|kd|}}{|kd|} \end{eqnarray}
that, in the case of ultrathin films with thickness $d$, where $kd\ll1$, reduces to  $P(kd)=|kd|/2$ when performing a series expansion of the exponential function.
The anisotropy field, written for the case of a uniaxial perpendicular anisotropy constant $K_u$ out-of-plane, is
   $\vec{H}_{ani}=\frac{2K_u}{\mu_0 M_s} m_y\hat{y}$

\begin{figure}[ht!]
\centering
\includegraphics[width=\linewidth]{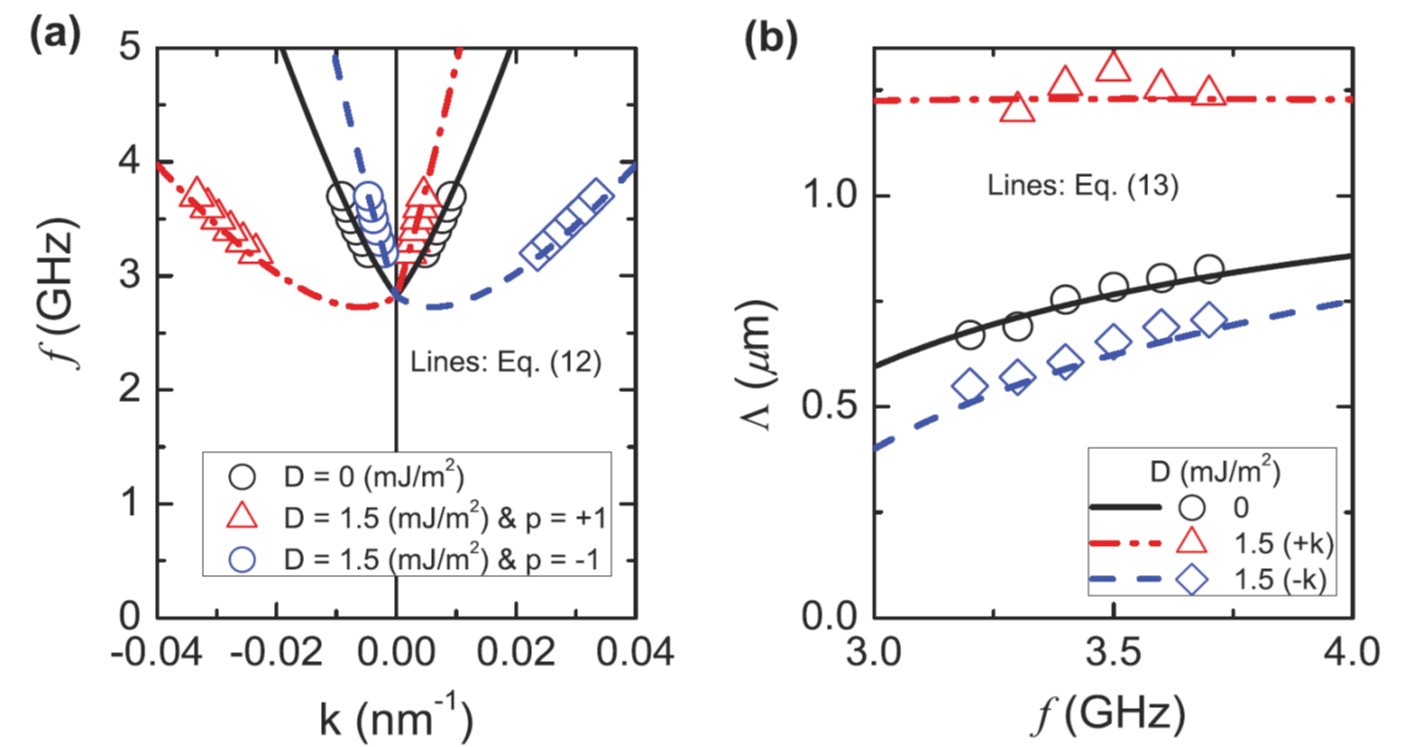}
\caption{DMI-induced asymmetric spin-wave propagation in the small-$k$ limit. (a) Dispersion relation. (b) Attenuation length as a function of the frequency $f$. From \textcite{MOO-13}.}
\label{fig:moon}
\end{figure}

The correct expression for the DMI field $H_{DMI}$ in the continuum theory is not that straightforward to obtain as it depends on the crystal symmetry considered. For bulk, the first approach from a spin to a continuum model was performed by Dzyaloshinskii himself \cite{DZY-57}, calculating a thermodynamic potential for certain crystal symmetry classes. This approach was used later by \textcite{BAK-80} for MnSi and FeGe, and by \textcite{BOG-94, BOG-01} using the Lifshitz invariants $\vec{\mathcal{L}}(\vec{M},\vec{\nabla} \times \vec{M)}$, where the $k$-th component of $\vec{\mathcal{L}}$ is $\mathcal{L}_{ij}^{(k)}=M_i\frac{\partial{M_j}}{\partial{x_k}}-M_j\frac{\partial{M_i}}{\partial{x_k}}$. The DMI energy can be described as a combination of Lifshitz invariants, depending on the crystal symmetry \cite{CRE-98, COR-13}. For certain symmetry classes, as the rotational tetrahedral $\textbf{T}$\footnote{Schoenflies notation} (e.g. MnSi), the energy density due to DMI is a combination of Lifshitz invariants with a single coefficient $D$. Only in these cases the DMI strength is sufficiently described by a scalar. Another example is the cyclic class $\textbf{C}_{nv}$\footnotemark[12]. Its Lifshitz invariants are a suitable choice for planar systems (bilayers) with perpendicular anisotropy, where symmetry breaking occurs only along the perpendicular axis and only gradients in plane contribute to the DMI induced chirality \cite{BOG-89, BOG-01}. For a planar geometry in the $xz$ plane with perpendicular anisotropy along $y$, the DMI energy can than be described by
\begin{equation}
E_{DMI}=D (\mathcal{{L}}_{xy}^x+\mathcal{{L}}_{zy}^z)   
\end{equation}
With $\vec{M} = M_s\hat{m}(\vec{x},t)=M_s (m_x\hat{x}, m_y\hat{y},\hat{z})$ the DMI field can then be obtained as 
\begin{equation}
\label{eq:hdmi}
\vec{H}_{DMI}=-\frac{1}{\mu_0}\frac{\delta E_{DMI}}{\delta \vec{M}}=\frac{2D}{\mu_0 M_s}\left( \frac{\partial m_y}{\partial x},-\frac{\partial m_x}{\partial x},0 \right)
\end{equation}
The dispersion relation is then given by
\begin{widetext}
\begin{equation} 
\label{eq:dispKost}
\omega(k)=\gamma \mu_0 \sqrt{(H_0+J k^2+M_s P(kd) sin^2\phi_k)(H_0+J k^2 - H_{ani} + M_s (1-P(kd)))} + \omega_{DMI} =  \omega_0 + \omega_{DMI} 
\end{equation}
\end{widetext}
that is valid also for values of $\phi_k$ different from $\pi/2$, as shown in Fig.~\ref{fig:SW2}.
Moreover, it can be shown that the general expression of $\omega_{DMI}$ for an interfacial DMI is given by \cite{GAL-19}.:
\begin{equation}
\omega_{DMI} = \frac{ 2\gamma\ D} {M_s}\: k\: sin \phi_k\: cos \phi_M
\label{eq:omegadmi}
\end{equation}
where the angles are defined in Fig.~\ref{fig:SW2}.
This means that, similar to the results of the quantum approach reported in the previous paragraph, the DMI leads to a shift of the dispersion relation, linear in $k$, and depending on the sign of the wave vector, i.e. a frequency non-reciprocity due to DMI (see Fig.~\ref{fig:moon}). This non reciprocity depends on the directions of both the sample magnetization and the spin wave propagation direction and is absent for $k=0$, i.e. the condition for ferromagnetic resonance (FMR). In this respect, already in \textcite{COR-13} it was suggested to use BLS measurements for determining the DMI constant, pointing out that the largest effect can be observed applying a sufficiently large external field H, so that the magnetization lies in-plane ($\phi_M=0$), using the Damon-Eshbach (DE) geometry ($\phi_k=\pi/2$).

It is worth noting that the above dispersion relation is valid up to a certain threshold of the DMI strenght, above which the dispersion relation becomes zero and the ground state is not uniform but  rather chiral and the theory may not be applicable. By numerical calculations it is shown that the asymmetry does not only regard frequency but also amplitude and attenuation length \cite{KOS-14a,KOR-15}. 

Let us also notice that the above description is valid only for a film consisting of a single monolayer. An attempt to take the finite film thickness correctly into account by using mixed exchange boundary conditions at the surface is shown in \textcite{KOS-14a}\footnote{Note: The DMI constant in \textcite{KOS-14a} is defined as the DMI constant in the first atomic layer and therefore is defined as $\widetilde{D}=D d\sqrt{2}/a$, where $a$ is the lattice constant.}. The resulting boundary conditions then are similar, but not equal, to the ones derived for a uniaxial surface anisotropy \cite{SOO-63}. Different surface anisotropies at both interfaces of the magnetic film result in an intrinsic non reciprocity of DE spin waves \cite{HIL-87}. In this case the two contributions are difficult to be distinguished. In fact, already in \textcite{CRE-98}, the presence of DMI can be interpreted as a contribution to the surface anisotropy. In order to analyze the different contributions in a more rigorous way, film thickness dependencies have to be studied, as discussed in \textcite{GLA-16, LUC-20}.

In summary we want to point out that the complete solution in the continuum theory for the bilayer system is not trivial, requiring assumptions concerning the boundary conditions. Especially, care has to be taken if one wants to take all contributions to the spin wave dispersion non-reciprocity (such as surface anisotropies) into account correctly. However, in most cases the theory required for modeling experimental results is simplified by the fact that by subtraction (as explained in the following sections) the DMI related term in the spin wave dispersion relation is directly determined. 

\subsection{Experimental results}

\subsubsection{Brillouin light scattering}

The most popular and widely employed technique for the measurement of the DMI constant from the non-reciprocal spin wave propagation is Brillouin light scattering (BLS) \cite{CAR-99}. As shown in Tables \ref{TableBLS1a}, \ref{TableBLS1b}, \ref{TableBLS2} and \ref{TableBLS3}, more than forty papers have been published during the last five years, reporting the value of the DMI constant in samples where a  FM material (usually Co or CoFeB) is in contact with a HM material (such as Pt, W, Ta, etc.). One of the first BLS experimental works reporting values of the DMI constant determined by BLS was done on Co/Ni multilayers \cite{DI-15,DI-15a, ZHA-15a}. In the same year a series of publications of DMI measurements by BLS on Py, Co and CoFeB followed \cite{VAN-15,STA-15,NEM-15,BEL-15,CHO-15, KIM-15} and the research continues to be very active. 

\begin{figure*}[ht!]
\centering
\includegraphics[width=\textwidth]{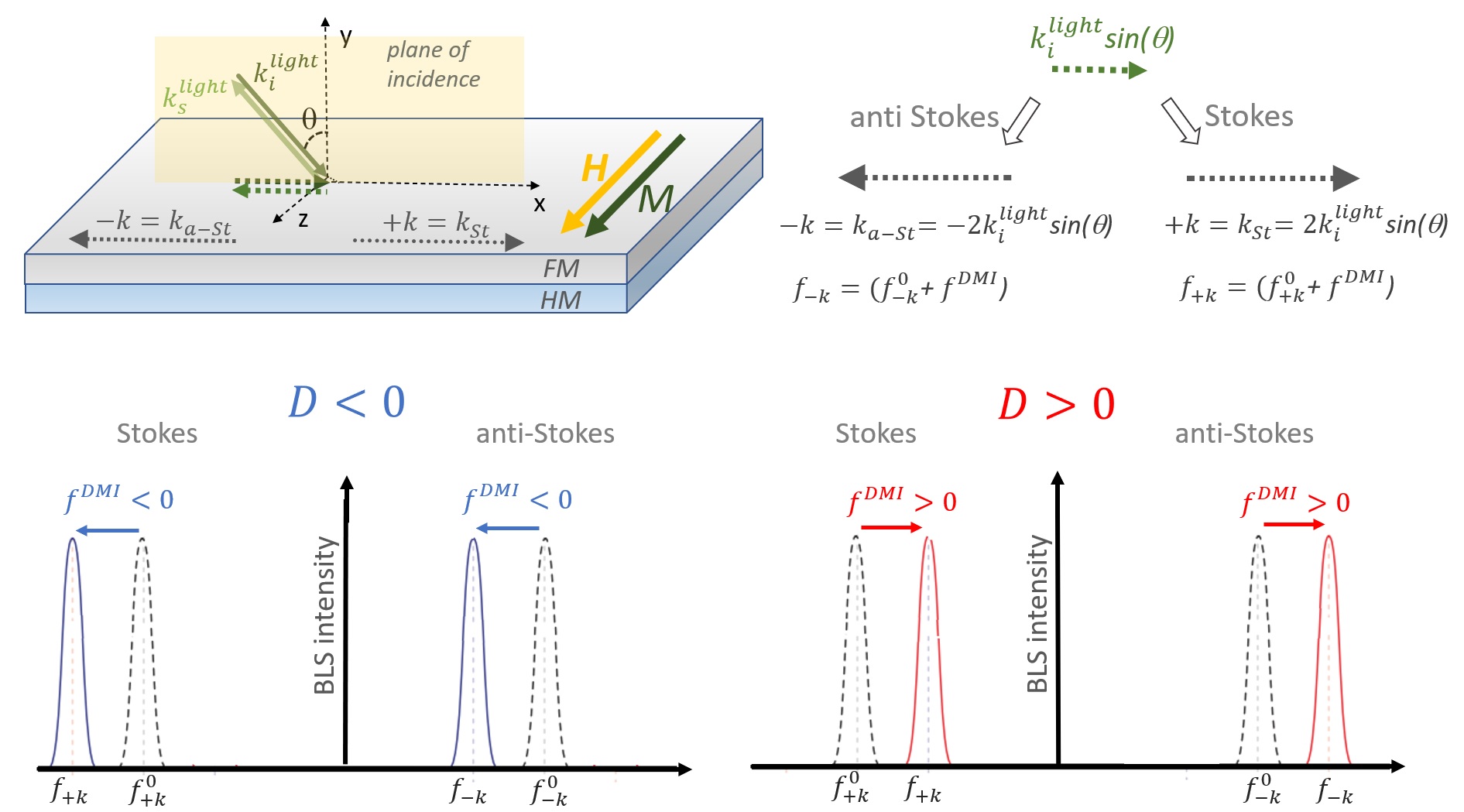}
\caption{Schematic diagram of the BLS interaction geometry. $k_i^{light}$ and $k_s^{light}$ represent the wave vectors of the incoming and of the backscattered light. Due to the conservation of the wave vector component parallel to the surface of the specimen, the length of the wave vector of the SW involved in the scattering process is $2k_i^{light}sin(\theta)$, as shown in the right panel. SWs propagating along the $+k$ $(-k)$ direction correspond to those involved in the Stokes (anti-Stokes) process, i.e in the generation (annihilation) of a magnon.  The bottom panel shows a typical BLS spectrum in the absence of DMI (dashed line) and with DMI (continuous line) for positive or negative values of $D$ i.e. of $f_{DMI}$, respectively.}
\label{fig:SW3}
\end{figure*}

BLS is a classical tool to study spin wave dispersion in general and non-reciprocity in particular. As sketched in Fig.~\ref{fig:SW3} in a BLS experiment a monochromatic light beam is focused on the surface of the specimen under investigation by an objective lens and the light that is back-scattered within a solid angle is collected by the same lens and analysed in frequency by a high-resolution spectrometer, usually a Sandercock-type, multipass tandem Fabry-Perot interferometer \cite{MOC-87}. The physical mechanism of BLS relies on the inelastic scattering of photons by spin waves that are naturally present within the medium under investigation, thanks to either the creation (Stokes process) or the annihilation (anti-Stokes process) of a magnon. This implies that a red-shift or a blue-shift is observed in the scattered light with respect to the incident beam. In wave vector space, magnons experiencing Stokes or anti-Stokes scattering processes correspond to either a positive or a negative wave vector. As a consequence, the non-reciprocity caused by the presence of a DMI interaction leads to an asymmetry in frequency shift of the peaks corresponding to the Stokes or the anti-Stokes process, as anticipated in the previous paragraph (see Fig.~\ref{fig:SW3}).
A rather wide range of $k$ vectors is available, up to about  2.2$\cdot10^5$ rad/cm. The specific value of  $k$ can be varied easily by changing the angle of the incident beam of light, while reversing the applied magnetic field permits to cross-check for the sign of the DMI constant. The frequency resolution is limited by the instrumental characteristics to about 0.1~GHz, because the finesse of the Fabry-Perot interferometer (ratio between the free spectral range and the pass-band window) has an upper limit of about 100 \cite{CAR-99}. This implies a lower limit for the minimum values of the DMI constant that can be measured by this technique. One should also consider that usually several tens of mW of light are focused on the measurement spot, so that the real temperature\footnote{Note: The real temperature depends on the thermal properties of the material under investigation (including the substrate material), as well as on the numerical aperture of the exploited lens (that determines the size of the focused light spot).} of the probed sample region may be lifted above room temperature and this can have an influence on the values of the measured magnetic parameters, including $D$.  
Here we adopt the reference frame presented in Figs.~\ref{fig:SW1} and ~\ref{fig:SW3}, where the $z$-axis is aligned with the static magnetization, the $x$-axis coincides with the propagation direction of SWs and the film normal ($y$- axis) is oriented upward with respect to the free surface of the FM film. With such a choice, the frequency shift of the BLS peaks is
\begin{equation}
f_{DMI}=\omega_{DMI}/2 \pi=(\gamma D k)/(\pi M_s )
\end{equation}
For the case of a FM/HM bilayer, as sketched in Fig.~\ref{fig:SW3}, a negative (positive) value of $f_{DMI}$, i.e. of $D$, indicates that a left (right)-handed chirality is favoured by the DMI. This means that in presence of a positive (negative) value of $D$, in the measured BLS spectra the absolute frequency of the anti-Stokes (Stokes) peak is larger (lower) than that of the Stokes (anti-Stokes) one, and as a consequence $f_{DMI}$ assumes positive (negative) values, as illustrated in the lower panel of Fig.~\ref{fig:SW3}. Please note that if the direction of the applied magnetic field is reversed,  $f_{DMI}$ changes sign due to the reversal of the SW chirality.  

A comparison of the value and the sign of the DMI constant measured using BLS for Co, CoFeB, and a collection of different magnetic materials are reported in Tables~\ref{TableBLS1a}, \ref{TableBLS1b}, \ref{TableBLS2} and \ref{TableBLS3}, respectively. When the values were not explicitly given in the original manuscript, we have calculated or extracted them from the figures and reported them in italic. Note that the signs in the Tables are consistent with the convention of axes, field, $D$, $k$ vector and frequency difference as previously described in Figs.~\ref{fig:SW1} and ~\ref{fig:SW3}. If the authors in the original manuscript have used a different coordinate system and/or method of calculating the frequency asymmetry we recalculated the sign according to our convention.

Many papers discuss the dependence of the DMI constant as a function of the ferromagnetic layer thickness \cite{STA-15,BEL-15,CHA-16,KIM-16} and possible correlation with surface anisotropy \cite{STA-15} or Heisenberg exchange \cite{NEM-15}. Theoretically, a $1/d$ dependence of the DMI constant is predicted \cite{KOS-14a}, as is the case for all interface phenomena, and this makes it difficult to distinguish between the different phenomena. Sometimes a change in slope of the DMI constant versus $1/d$ is observed \cite{CHO-15,KIM-18b,BEL-18,BEL-16a}. \textcite{BEL-18} discuss different origins of this change. The first is a coherent-incoherent growth mechanism transition at a certain thickness accompanied by magnetoelastic anisotropy changes. Others are changes in surface roughness which are causing in-plane demagnetizing magnetic fields. Another is interdiffusion and mixing at the interfaces, which reduces the interface anisotropy. According to the experimental results the authors conclude that the first hypothesis is the most probable. 

Similarly, the dependence of the DMI on the heavy metal thickness has been studied for both Co \cite{KIM-17a} and CoFeB \cite{CHE-18,TAC-17} films in contact with a Pt layer. For both systems the DMI intensity was found to increase with the Pt thickness, $d_{Pt}$, reaching a saturation value when $d_{Pt}$ approaches the Pt spin diffusion length (about 2~nm). This behaviour has been explained by analytical calculations assuming that several Pt atoms, belonging to different layers in the heavy metal, can contribute to the strength of the interfacial DMI \cite{TAC-17}. On the contrary, in \textcite{MA-17}, the DMI strength in the IrMn/CoFeB system was observed to increase even for IrMn layers thicker than the IrMn spin diffusion length (about 0.7~nm). The authors claim that this DMI enhancement can be ascribed to the reduction of the thermal fluctuations of the antiferromagnetic (AFM) spin arrangement when the IrMn thickness increases, suggesting a different microscopic origin with respect to HM/FM systems. 

The sign of the DMI constant was investigated for several material combinations by changing both the stacking order and the multilayer composition \cite{CHO-17}. In systems with a Pt (W) underlayer, negative (positive) values of the DMI constant have been found with a quite overall agreement, indicating that a left-handed (right-handed) chirality is favored. In agreement with the theoretical calculations, it has also been observed that the sign of DMI is reversed when the stacking order is inverted, while its value becomes negligible in symmetric structures where a FM film is sandwiched between two identical HM layers. On the contrary, the DMI sign induced by other materials, such as Ir or IrMn, is still the subject of controversial debate in the literature (see for instance \cite{MA-18,MA-17,BEL-18}). However, most experimental works contradict the theoretical prediction of Ir leading to opposite DMI sign with respect to Pt \cite{YAN-15}. Recently, the influence of electric fields on the DMI strength has been investigated in Ta/FeCoB/TaO$_x$ trilayer \cite{SRI-18}. A strong variation of the $D$ value has been observed and attributed to the dependence of the DMI at the CoFeB/Oxide interface on its oxydation state, that can be electrically tuned.  Among the Co articles, \citealp{KHA-18} investigate the effect of exchange bias with an antiferromagnetic (AFM) overlayer on Co and confirm that DMI is not influenced by exchange bias or the AFM spin order. Finally, the sinusoidal dependence of the frequency non-reciprocity as a function of the angle between the sample magnetization and the SW wave vector, theoretically predicted by \textcite{COR-13}, has been experimentally verified for different multilayer structures \cite{BEL-18,KIM-16,CHO-17,TAC-17,ZHA-15a}.

Let us notice that several studies were able to correlate the presence of a sizeable DMI not only to the frequency position of the BLS peaks but also to their linewidth, i.e. to the damping that affects spin waves. Already \textcite{DI-15} studied the interfacial DMI in an in-plane anisotropic Pt(4)/Co(1.6~nm)/Ni(1.6~nm) film by Brillouin spectroscopy, showing that the measured linewidths of counterpropagating magnons are different, with the difference being more pronounced for larger wave vectors. This could be ascribed to a DMI-induced term that is antisymmetric in the wave vector. Moreover analytical calculations showed that, due to the existence of the DMI, the magnon linewidth is no longer a monotonic function of frequency. In \textcite{CHA-16}, asymmetry in the peak frequency, peak intensity and magnon lifetime were observed in W/CoFeB/SiO$_2$. Also in this case the linewidth for spin-wave propagating in $+ k$ direction is smaller than the same for spin-wave propagating in $-k$ direction, indicating different lifetimes of magnon propagating in opposite directions, induced by DMI. More recently, a clear correlation between the DMI strength and the SW damping were demonstrated by combined FMR and BLS investigation in different kind of systems, such as [Pt(1.5)/Co($d$)/W(1.5)]$_{xN}$ multilayers \cite{BEN-19}, Py(5)/ Cu$_{1-x}$Pt$_x$ bilayers \cite{BOU-19} and He+ irradiated Ta/CoFeB/MgO ﬁlms \cite{DIE-19a}.

Very recently, several works reported that a sizeable interfacial DMI can also arise at the interface between an oxide layer and a ferromagnetic one \cite{Arora2020,NEM-20,LIN-20,KIM-19a}. \textcite{NEM-20} studied the interfacial DMI in Cu(3)/Co$_{90}$Fe$_{10}$(1.5)/Ta($d_{Ta}$)/Oxide/Ta(3) systems, where the oxide layer was prepared by in-situ sample oxidation. Changing the Ta thickness, $d_{Ta}$, an oxide layer of CoFeO$_x$ and/or TaO$_x$  was obtained following the sample exposure to oxygen. The authors found that both the magnitude and sign of the interfacial DMI can be tuned by varying the thickness and the composition of the oxide layer, due to the changes of the electronic structure at the interface, induced by the oxidation process. Moreover, \textcite{LIN-20} showed that in the BaTiO$_3$ (BTO)/CoFeB system the DMI strength is strongly affected by the termination of the oxide layer (TiO$_2$ vs BaO). In particular, a higher value of the DMI constant was found for a TiO$_2$-BTO substrate, and this finding was attributed by first principle calculations to the different electronic states around the Fermi level at the oxide/FM interfaces.

In summary, BLS has rapidly become the most used technique for measuring DMI using spin waves. A reasonable complexity of the experimental setup and a high accuracy in the determination of the spin wave dispersion, together with the fact that no external excitation is needed for this technique,  which relies on thermal excitation, contribute to the success and spreading of this technique.

\begin{table*}
\caption{Overview of DMI measurements for \textbf{Co} thin films via \textbf{BLS} (publication years 2015 to 2018). FM and NM stand for ferromagnetic and non-magnetic layer, respectively. $D$ is the interfacial DMI constant and $D_s = D \cdot d$, with $d$ being the thickness of the ferromagnetic film. Numbers in roman were quoted in the reviewed papers, while numbers in italics were either extracted from figures or calculated using the parameters provided. Signs with $^*$ are according to the convention used in this paper and opposite to that in the original manuscript.} 

\begin{tabular}{c|c|c|c|c|c|c}
\hline 
\hline
\textbf{FM} & \textbf{Bottom NM} & \textbf{Top NM}  & $\mathbf{D}$ & $\mathbf{D_s}$ & \textbf{Sign} &\textbf{Ref}\\ 
(nm) & (nm) & (nm) & (mJ/m$^2$) & (pJ/m) & & \\
\hline 
\hline
Co(0.6-1.2) &	Ta(3)/Pt(3)&	AlO$_x$/(2)Pt(3)&	2.7-1.6	&1.6-1.9& - &\citealp{BEL-15}\\
\hline
Co(1-2)&	Pt(4) & AlO$_x$(2) &\emph{1.2-0.9}	&\emph{1.2-1.8}& n.a. &\citealp{CHO-15}\\
\hline
Co(1)&	Pt(5)&	GdOx(2-5)/Al(2)&\emph{1.6}& 1.6& - &\citealp{VAN-15}\\
\hline
\centering Co(1-1.7)&	Pt(4)&	 \multirow{2}{*}{\centering AlO$_x$(2)}	&	\emph{1.2-0.8} & \emph{1.2-1.4} & -$^*$ &\multirow{2}{*}{\hfil\citealp{KIM-15b}}\\
\centering Co(1.3-1.8)&Ta(4)/Pt(4) &	&	\emph{1.6-1.3} & \emph{2.1-2.3}& -$^*$ & \\\hline
Co(1.06)&	Ta(3)/Pt(3)&	MgO(2)/Ta(2)&	 2.05&2.17& - &\citealp{BOU-16}\\\hline
Co(1.3-2.9)&	Ta(4)/Ir(4)&AlO$_x$(2)	&	\emph{0.7-0.35}&\emph{0.9-1}& n.a. &\citealp{KIM-16c}\\\hline
\multirow{2}{*}{\centering Co(2)}&	Ta(4)& 	Pt(4)	&	0.7&\emph{1.4}& + &\multirow{2}{*}{\hfil\citealp{CHO-17}}\\
&	Pt(4)& 	Ta(4)	&	0.92&\emph{1.84} & - &\\\hline
\multirow{2}{*}{\centering Co(1.6)} &	Ta(4)/Pt(0.8) &	\multirow{2}{*}{\centering AlO$_x$(2)/Pt(1)} &	0.43 &\emph{0.7}&-&	\multirow{2}{*}{\hfil\citealp{KIM-17a}}\\
  &	Ta(4)/Pt(4.8) & &	1.57 &\emph{2.51}&-&	\\
\hline
\multirow{2}{*}{\centering Co(2.5)}&\multirow{2}{*}{\centering Pt(5.4)}& Au (2.5)/Pt(2.6) & 0.60 & 1.51 & -& \multirow{2}{*}{\hfil\citealp{ROW-17}}\\
&& Ir(2.5)/Pt(2.6) & 0.60 & 1.51 & -& \\
\hline
\multirow{5}{*}{\centering Co(1)} & 	Ta(5) &	\multirow{5}{*}{\centering MgO(2)/Ta(2)}&	\emph{0.08} &	0.08	&+& \multirow{5}{*}{\hfil \citealp{MA-18}}\\
&	 	W(5)& &	\emph{0.11}&	0.11	&+&\\
&				Ir(5) &&	\emph{0.34} &0.34&-&	\\
&				Pt(5)& &	\emph{1.59} &1.59&-&	\\
&				 Au(5) &&	\emph{0.22} &0.22&-&	\\\hline
\multirow{2}{*}{\centering Co(3) }&	 \multirow{2}{*}{\centering MgO(2)/Pt(6)} &\multirow{2}{*}{\centering Ru(3)}	&	 0.43\footnote[1]{The sample is a stripe sample with stripe width 300~nm and spacing 100~nm} &1.3\footnotemark[1] & -&\multirow{2}{*}{ \hfil\citealp{BOU-18}}\\
 &		&&	  0.33\footnote[2]{Stripe width: 100~nm, spacing 100~nm} & 1\footnotemark[2] &-& \\\hline
\multirow{4}{*}{\centering Co(1)}&	\multirow{4}{*}{\hfil Ta(5)/Pt(2)} & 	Ir$_{20}$Mn$_{80}$(1.1) &	1.14 &1.14&-	& \multirow{4}{*}{\hfil \citealp{KHA-18}}\\
&	 & 	Ir$_{20}$Mn$_{80}$(1.7) &	1.14 &	1.14	&-& \\
&	 & 	Ir$_{20}$Mn$_{80}$(2.4) &	1.22 &	1.22	&-& \\
&	 & 	Ir$_{20}$Mn$_{80}$(5) &	1.11 &	1.11	&-& \\
\hline
\multirow{2}{*}{\centering Co(0.6)}& \multirow{2}{*}{\hfil Ta(5)/Pt(2)}	 & 	Fe$_{50}$Mn$_{50}$(1) &	1.5 &	1.35\footnote[3]{For $D_s$ calculation, the authors have added a monolayer of Fe from Co/FeMn interface to the effective thickness of the FM layer.}	&-& \multirow{2}{*}{\hfil \citealp{KHA-18}}\\
&	 & 	Fe$_{50}$Mn$_{50}$(2.6) &	1.44 &	1.3\footnotemark[3]	&-& \\
\hline
\multirow{4}{*}{\centering [Co(0.2)/Ni(0.6)]$_2$/Co(0.2)} & Pt(1.2) & \multirow{4}{*}{\hfil\centering Ta(0.8)/TaN(6)}  & \textit{0.49} & \textit{0.88} &-& \multirow{4}{*}{\hfil\centering \citealp{LAU-18}}\\ 
 & Ir(1.2) &  & \textit{0.05} & \textit{0.09}&-& \\ 
 & Pt$_{0.4}$Ir$_{0.6}$(1.2) &  & \textit{0.19} & \textit{0.34} &-& \\ 
 &  Pt$_{0.6}$Ir$_{0.4}$(1.2) &   & \textit{0.24} & \textit{0.43} &-& \\ 
\hline
\hline
\end{tabular}
\label{TableBLS1a}
\end{table*}

\begin{table*}
\caption{Overview of DMI measurements for \textbf{Co} thin films via \textbf{BLS} (publication years 2019 to 2021). FM and NM stand for ferromagnetic and non-magnetic layer, respectively. $D$ is the interfacial DMI constant and $D_s = D \cdot d$, with $d$ being the thickness of the ferromagnetic film. Numbers in roman were quoted in the reviewed papers, while numbers in italics were either extracted from figures or calculated using the parameters provided. Signs with $^*$ are according to the convention used in this paper and opposite to that in the original manuscript.} 

\begin{tabular}{c|c|c|c|c|c|c}
\hline 
\hline
\textbf{FM} & \textbf{Bottom NM} & \textbf{Top NM}  & $\mathbf{D}$ & $\mathbf{D_s}$ & \textbf{Sign} &\textbf{Ref}\\ 
(nm) & (nm) & (nm) & (mJ/m$^2$) & (pJ/m) & & \\
\hline 
\hline
\multirow{4}{*}{\hfil\centering Co(1-2)}&\multirow{4}{*}{\hfil\centering Ta(3)/Pt(3)} & Ir(3) & 0.8 -- 0.3 & 0.8 & - & \multirow{4}{*}{\hfil \citealp{BEL-19a}}\\
& & Cu(3) & 0.9 -- 0.4 & 1.05 & - & \\
& & MgO(1) & 0.9 -- 0.4 & 0.95 & - & \\
& & Pt(3) & 0 & 0 &  & \\
\hline
\multicolumn{3}{c|}{[Pt(1.5)/Co(1-2)/W(1.5)]$_{x1}$} & 0.2 - 0.3 & 0.55 & - & \multirow{3}{*}{\hfil\centering \citealp{BEN-19}} \\
\multicolumn{3}{c|}{[Pt(1.5)/Co(1-2)/W(1.5)]$_{x3}$} & 1.1 - 0.6 & 0.85 & - & \\
\multicolumn{3}{c|}{[Pt(1.5)/Co(1-2)/W(1.5)]$_{x7}$} & 1.2 - 0.6 & 0.90 & - & \\
\hline
Co(0.6) &	Pt(3) &	HfO$_2$(3)&	0.9	& \textit{0.54}& - &\citealp{DIE-19}\\
\hline
\multirow{6}{*}{\hfil\centering Co(0.9)}  &	\multirow{6}{*}{\hfil\centering Ta(5)/Pt(1.5)} &	Ti(2.5)/Pt(2.5) & 1.42& \textit{1.28}& n.a.&\multirow{6}{*}{\hfil\centering\citealp{KIM-19}}\\
 & & Cu(2.5)/Pt(2.5) & 0.87& \textit{0.78}& n.a.&\\
& & W(2.5)/Pt(2.5)& 1.25& \textit{1.13}& n.a.&\\
& & Ta(2.5)/Pt(2.5)& 0.99& \textit{0.89}& n.a.&\\
& & Al(2.5)/Pt(2.5)& 0.92& \textit{0.83}& n.a.&\\
 & & Pt(2.5)/Pt(2.5)& 0.02& \textit{0.02}& n.a.&\\
\hline
\multirow{2}{*}{\hfil\centering Co(1.4)}  &	\multirow{2}{*}{\hfil\centering Ta(5)/Pt(5)} &	MgO(2)/Ta(3) & 1.20& \textit{1.68}& n.a.&\multirow{2}{*}{\hfil\centering\citealp{KIM-19a}}\\
 & & Cu(2)/Ta(3) & 1.05& \textit{1.47}& n.a.&\\
\hline
Co(0.8) & Ta(2)/Pt(2.2) & Ir(0-2)/Ta(4) & \emph{1.8 - 0.88} & \emph{1.64 - 0.7} & -* & { \citealp{SHA-19}}\\
\hline
\multicolumn{3}{c|}{[Pt(3)/Co(1.1)/Ta(4)]$_{x12}$}  	&0.78	&\textit{0.86}& - &\citealp{SAH-19}\\
\hline
Co(1.2) & Ta(2.5)/Pt(0.4-2.2)\footnote[4]{On a glass substrate} &	Pt(2) &	0.27 - 0.42 & \textit{0.32 - 0.5}& + &\citealp{GUS-20}\\
\hline
Co(0.8-2.5) & Ta(3)/Pt(3) &	Ir(3) & \textit{0.95 - 0.3}	 & 0.76 & - &\multirow{7}{*}{\hfil\centering \citealp{BEN-20}}\\
 Co(0.9-2.5)& Ta(3)/Pt(3)\footnote[5]{Annealed at 300$^o$C} &	Ir(3) &	0.6 - 0.2 & 0.53 & - &\\
 Co(1.1-3.1)& Ta(3)/Ir(3)\footnotemark[5] &	Pt(3) &	0.88 - 0.32 & 1.05 & + &\\
Co(1-2.6) & Ta(3)/Ir(3)\footnote[6]{As grown} &	Pt(3) &	0.8 - 0.3 & 0.77 & + &\\
Co(0.8-2.5) & Ta(3)/Pt(3)\footnotemark[5] &	Cu(3) &	\textit{1.31 - 0.42} & 1.05 & - &\\
 Co(1-2.2)& Ta(3)/Pt(3)\footnotemark[5]&	MgO(1) &1 - 0.5	 & 0.95 & - &\\
 Co(1.5-2.1)& Ta(3)/Pt(3)\footnotemark[6] &	MgO(1) &	0.8 - 0.6 & 1.17 & - &\\
\hline
Co(1.2-10) & Pt(3) &	TaO$_x$(0.8)/Al(0.5) &1 - 0.15	 & 1.25\footnote[7]{This value was extracted by the authors from a fit of the experimental data of all samples with various Co thicknesses.} & - &\multirow{3}{*}{\hfil\centering \citealp{BEN-20a}}\\
Co(1.2) & Pt(3) &	TaO$_x$(0.7-0.9)/Al(0.5) &1.2 - 1.5	 & \textit{1.44 - 1.8} & - &\\
Co(1.2) & Pt(3)/Cu(0-2.4) &	TaO$_x$(0.8)/Al(0.5) &1.3 - 0 & \textit{1.56 - 0} & - &\\
\hline
Co(0.7) & Pt(5) &	Ni$_x$O$_y$(15) &1.75\footnote[7]{Sample deposited with Ar pressure of 0.13~Pa}	 & 1.3 & - &\citealp{KOL-21}\\
\hline
\multicolumn{3}{c|}{[W(1)/Co(0.6)/Pt(1)]$_{x10}$} & 2.65 & 1.83 & + & \multirow{2}{*}{\hfil\centering \citealp{JEN-21}} \\
\multicolumn{3}{c|}{[W(1)/Co(0.6)/Pt(1)]$_{x20}$} & 2.49 & 1.72 & + & \\
\hline
\hline
\end{tabular}
\label{TableBLS1b}
\end{table*}

\begin{table*}
\caption{Overview of DMI measurements for \textbf{CoFeB} thin films via \textbf{BLS}. FM and NM stand for ferromagnetic and non-magnetic layer, respectively. $D$ is the interfacial DMI constant and $D_s = D \cdot d$, with $d$ being the thickness of the ferromagnetic film. Numbers in roman were quoted in the reviewed papers, while numbers in italics were either extracted from figures or calculated using the parameters provided. Numbers with $^*$ indicate that a magnetic dead layer was taken into account. Signs with $^*$ are according to the convention used in this paper and opposite to that in the original manuscript.} 
\begin{tabular}{c|c|c|c|c|c|c}
\hline 
\hline
\textbf{FM} & \textbf{Bottom NM} & \textbf{Top NM}  & $\mathbf{D}$ & $\mathbf{D_s}$ &\textbf{Sign}& \textbf{Ref}\\ 
(nm) & (nm) & (nm) & (mJ/m$^2$) & (pJ/m) & & \\
\hline 
\hline
Co$_{48}$Fe$_{32}$B$_{20}$(1.6-3) &	Pt(4) &AlO$_x$(2)&		\emph{0.8-0.4} & \emph{1.2-1.3} & n.a. &\citealp{CHO-15}\\\hline
Co$_{40}$Fe$_{40}$B$_{20}$(0.8) &	MgO(2)/Pt(2) & MgO(2)/SiO$_2$(3)&		1 & \emph{0.8} & -&\citealp{DI-15a}\\\hline
Co$_{20}$Fe$_{60}$B$_{20}$(0.85, 1, 1.5, 2, 3) &		W(2) & SiO$_2$(2)&	\emph{0.25-0.08}& \emph{0.21-0.24}&n.a. & \citealp{CHA-16}\\\hline
\multirow{4}{*}{\centering Co$_{20}$Fe$_{60}$B$_{20}$(1)} &		W(2) & \multirow{4}{*}{\centering MgO(1)/Ta(1)}&	0.25& \emph{0.25} &+*& \multirow{4}{*}{\hfil\citealp{SOU-16}}\\
 &		W(3) & &	0.27& \emph{0.27} & +*&\\
  &		Ta$_{48}$N$_{52}$(1) & &	0.31& \emph{0.31} & +*&\\
 &		Hf(1) & &	0.15& \emph{0.15} &+*& \\\hline
Co$_{20}$Fe$_{60}$B$_{20}$(1) & Ta(5)  & Pt(0.12-0.27)/MgO(2)	& \emph{0.07-0.015} & \emph{0.07-0.015}  & +& \citealp{MA-16}\\\hline
Co$_{20}$Fe$_{60}$B$_{20}$(2) & Ir$_{22}$Mn$_{78}$(1-7.5)  & \multirow{3}{*}{\centering MgO(2)/Ta(2)}	& 0.02-0.13 & \emph{0.04-0.26} & -&\multirow{3}{*}{\hfil\citealp{MA-17}}\\
Co$_{20}$Fe$_{60}$B$_{20}$(0.8-2) & Ir$_{22}$Mn$_{78}$(5)  & 	& \emph{0.15-0.07} & \emph{0.12-0.14} & -&\\
Co$_{20}$Fe$_{60}$B$_{20}$(1.2) & Ir(5)  & 	& \emph{0.17} & \emph{0.2} & +&\\\hline
Co$_{40}$Fe$_{40}$B$_{20}$(2)&	Si/SiO$_2$& Pt(2)/Cu(3) &		0.45\footnote[1]{for $d_{Pt}>$2nm} &	\emph{0.9}&+& \citealp{TAC-17}\\\hline
\multirow{4}{*}{\centering Co$_{40}$Fe$_{40}$B$_{20}$(2) }&	Pt(4) & 	Ta(4) & 0.51 &	\emph{1.02}&-& \multirow{4}{*}{\hfil\citealp{CHO-17}}\\
&	Ta(4) & 	Pt(4) &	0.43&	\emph{0.86}  &+& \\
&	Ta(4) & 	Ta(4) &	0.15&	\emph{0.3} &-& \\
&	Pt(4) & 	Pt(4) &	0.01 &	\emph{0.02}  &-& \\\hline
\multirow{2}{*}{\centering Co$_{20}$Fe$_{60}$B$_{20}$(1.12)} & Ta(3)/Pt(3)  &	Ru(0.8)/Ta(3)	& 0.84 & \emph{0.94} & -&\multirow{2}{*}{\hfil\citealp{BEL-17}}\\
&	Pt(3)/Ru(0.8) &	MgO(1)/Ta(3)	& 0.3 & \emph{0.37} &-&\\\hline
\multirow{5}{*}{\centering Co$_{20}$Fe$_{60}$B$_{20}$(1)} & 	Ta(5) &	\multirow{5}{*}{\centering MgO(2)/Ta(2)} &	\emph{0.04} &	0.04 &  +&\multirow{5}{*}{\hfil\citealp{MA-18}}\\
 & 	W(5) &	 &		\emph{0.07}&0.07 &+&  \\
 & 	Ir(5) &	 &	 	\emph{0.21} &0.21 & +& \\
 & 	Pt(5) &	 &		\emph{0.97}&0.97 &-&  \\
 & 	 Au(5) &	 &	 	\emph{0.17}&0.17 &-&  \\
 \hline
[Pt(5)/Co$_{20}$Fe$_{60}$B$_{20}$(1)/Ti(1)]$_{x1}$ &\multirow{4}{*}{\hfil Ti(5)} &	\multirow{4}{*}{\hfil Pt(5)} &	0.81 &	\emph{0.81} & n.a.&\multirow{4}{*}{\hfil\citealp{KAR-18}} \\
{[Pt(5)/Co$_{20}$Fe$_{60}$B$_{20}$(1)/Ti(1)]$_{x10}$} &	 &	 &	0.79 &	\emph{0.79} &n.a.& \\
{[Pt(5)/Co$_{20}$Fe$_{60}$B$_{20}$(1)/Ti(1)]$_{x15}$} &	 &	 &	0.67 &	\emph{0.67} &n.a.& \\
{[Pt(5)/Co$_{20}$Fe$_{60}$B$_{20}$(1)/Ti(1)]$_{x60}$} &	 &	 &	0.71 &	\emph{0.71} &n.a.& \\
\hline
CoFeB(1)&	Ta(5)/Pt(0.7-4) & MgO(1)/Ta(1) &	\emph{0- 0.87}	&\emph{0- 0.87}&- &\citealp{CHE-18}\\\hline
Co$_{40}$Fe$_{40}$B$_{20}$(0.9) &	W(1-13) &	MgO(1)/Ta(2) &	\emph{0.25-0.7}& \emph{0.23-0.63}&n.a. &\citealp{KIM-18b}\\
\hline
Co$_{20}$Fe$_{60}$B$_{20}$(0.9) & Ta(3)  & TaO$_x$(1)	& \emph{0.08} & \emph{0.05}$^*$  & +& \citealp{SRI-18}\\
\hline
\multirow{3}{*}{\centering Co$_{20}$Fe$_{60}$B$_{20}$(1.5)} & \multirow{4}{*}{\centering Ta(1)}  & Cu(0-2.4)/Pt(4)	& 0.52 - 0.14 & \emph{0.78 - 0.21}  & + & \multirow{4}{*}{\hfil\citealp{BEN-20b}}\\
&  & Ta(0-2.4)/Pt(4)	& 0.52 - 0 & \emph{0.78 - 0}  & + & \\
&  & MgO(0-1.8)/Pt(4)	& 0.55 - 0 & \emph{0.78 - 0}  & + & \\
Co$_{20}$Fe$_{60}$B$_{20}$(1.5-6) &    & Pt(4)	& 0.7 - 0.2 & 1.25\footnote[2]{$D_s$ is obtained from the linear fit of $D$ versus $1/d$ for $d>$1.5~nm ($d$ is the FM thickness) }  & + & \\\hline
\multirow{2}{*}{\centering Co$_{20}$Fe$_{60}$B$_{20}$(2)} & TiO$_2$-BTO  & Pt(4)	& 0.45 & \emph{0.9} & +&\multirow{2}{*}{\hfil\citealp{LIN-20}}\\
&BaO-BTO &	Pt(4)	& 0.56 & \emph{1.12} &+&\\\hline
\hline
\hline
\end{tabular}
\label{TableBLS2}
\end{table*}

\begin{table*}
\caption{Overview of DMI measurements \textbf{various FM materials} via \textbf{BLS}. FM and NM stand for ferromagnetic and non-magnetic layer, respectively. $D$ is the interfacial DMI constant and $D_s = D \cdot d$, with $d$ being the thickness of the ferromagnetic film. Numbers in roman were quoted in the reviewed papers, while numbers in italics were either extracted from figures or calculated using the parameters provided. The value indicated with $^*$ was given for the thickest FM film.} 
\begin{tabular}{c|c|c|c|c|c|c}
\hline 
\hline
\textbf{FM} & \textbf{Bottom NM} & \textbf{Top NM}  & $\mathbf{D}$ & $\mathbf{D_s}$ &\textbf{Sign}& \textbf{Ref}\\ 
(nm) & (nm) & (nm) & (mJ/m$^2$) & (pJ/m) & \\
\hline 
\hline
Co(1.6)/Ni(1.6) &	Pt(4) & MgO(2)/SiO$_2$(3) &	0.44 & \emph{1.4} &-& \citealp{DI-15}\\\hline
Ni$_{80}$Fe$_{20}$(4) &Si substrate & Pt(6) &	\emph{0.056}\footnote{This value was calculated according to the definition of $D$ used in this paper from the original value $\bar{D}$ by $D=\bar{D}a/(\sqrt{2}d)$, where $a$=0.248nm is the lattice constant of Py} & \emph{0.34} &+& \citealp{STA-15}\\\hline
Co(1.6)/Ni(1.6) &	MgO(2)/Pt(4) & MgO(2)/SiO$_2$(3) &	0.44 & \emph{1.4} &-& \citealp{ZHA-15a}\\\hline
Ni$_{80}$Fe$_{20}$(1-13) &		Ta(3)/Pt(6) & SiN &	\emph{0.15-0.025} & \emph{0.15 - 0.33} &-& \citealp{NEM-15}\\\hline
Co$_{2}$FeAl(0.9-1.8) &		Ta(2)/Ir(4) & Ti(2) &	\emph{0.5 - 0.3 }& 0.37$^*$ &-& \citealp{BEL-18}\\\hline
\multirow{2}{*}{Fe(3)} & SiO$_2$ & \multirow{2}{*}{Pt(4)} &	0.22 & \textit{0.67} &+& \multirow{2}{*}{\citealp{ZHA-18}}\\
 & MgO(5) &  &	0.35 & \textit{1.05} &+&\\
\hline
Py(5) &  & Cu$_{1-x}$Pt$_x$(6)\footnote{$x$=0,6.6,28,45,75,100~\%} &	0 - 0.05 & \emph{0 - 0.27} &+& \citealp{BEL-19a}\\
\hline
\multirow{2}{*}{Co$_{90}$Fe$_{10}$(1)/oxide\footnote{Oxidation: 0-1000~s at 99~\% Ar and 1~\% O$_2$}} & Ta(3)/Pt(6) & \multirow{2}{*}{Cu(3)/Ta(2)} &	\textit{1.29-1.65} & \textit{2.52}  &-& \multirow{2}{*}{\citealp{NEM-20}}\\
 & Ta(3)/Cu(6) &  &	\textit{-0.02-0.26}& \textit{0.4}\footnote{The value refers to the longest oxidation of 1000~s} &-&\\
\hline
YIG(10) & GGG substrate &  &	0.01 & \emph{0.1} &n.a.& \citealp{WAN-20}\\
\hline
[Py(1)/Pt(1)]$_{10}$ & Ta(5)/Pt(6) & Au(3) &	0.032 & \textit{0.032} & - & \citealp{AHM-20}\\
\hline
\multirow{6}{*}{Fe(3)} & SiO$_2$(4) & Pt(4) &	0.25 & \textit{0.75} & + & \multirow{6}{*}{\citealp{ZHA-21}}\\
& SiO$_2$(4)/Pt(4) & SiO$_2$(4) &	0.33 & \textit{1} & - & \\
& SiO$_2$(4)/Au(4) & SiO$_2$(4) &	0.07 & \textit{0.21} & + & \\
& SiO$_2$(4) & Au(4) &	0.07 & \textit{0.21} & +& \\
& SiO$_2$(4)/Au(4) & Pt(4) &	0.4 & \textit{1.2} & + & \\
& SiO$_2$(4)/Pt(4) & Au(4) &	0.17 & \textit{0.51} & - & \\
\hline
\hline
\end{tabular}
\label{TableBLS3}
\end{table*}

\subsubsection{Time resolved magneto optical imaging}
\begin{figure}[ht!]
\centering
\includegraphics[width=.45\textwidth]{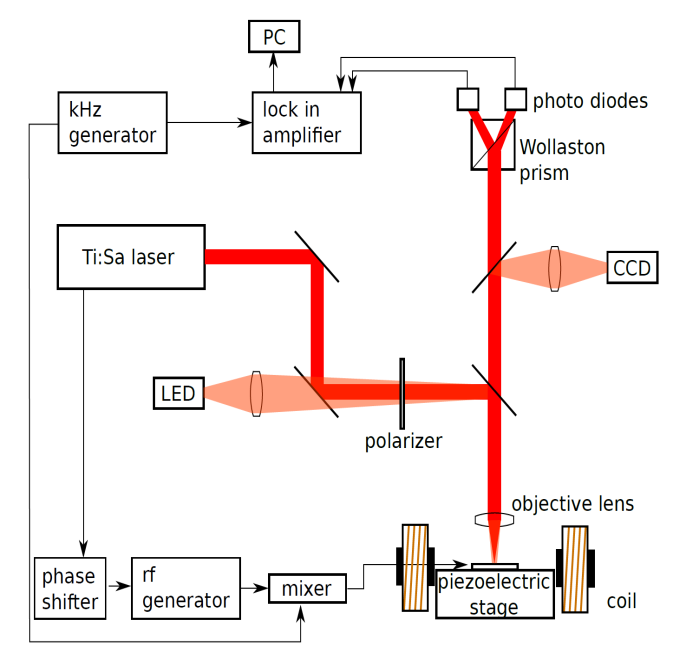}
\caption{Scheme of a time resolved magneto-optic Kerr effect experiment. The pulse train of a femtosecond laser system is synchronized to the sinusoidal excitation field which is generated by sending a microwave current through a coplanar wave guide. The change of polarisation of the reflected light is analysed in an optical bridge and sent to a lock-in analyser synchronized to a microwave mixer/chopper.}
\label{fig:TRMOKE}
\end{figure}

Time resolved magneto-optical Kerr microscopy (TR-MOKE) allows detecting propagating spin waves in real space and as a function of time enabling the determination of the spin wave $k$-vector and decay. TR-MOKE was employed by \textcite{KOR-15} in order to determine DMI in a Pt/Co/Py/MgO multilayer. For imaging spatially and time dependent spin wave signals a focused laser pulse (spot size $\leq$300~nm) of $\lambda=$400~nm and pulse length $\approx$200~fs (repetition rate 80~MHz) is used. The laser pulses are phase locked to a microwave generator connected to a coplanar wave guide (CPW or microwave antenna) deposited on top of the thin film sample as shown in Fig.~\ref{fig:TRMOKE}. The excited $k$-vectors are in the range of 2-10~$\mu$m$^{-1}$. The dynamic out-of-plane magnetization is detected via MOKE along the direction normal to the CPW up to a distance of 5~$\mu$m from the center of the CPW. The spin wave dispersion is determined by using the expression given in \textcite{MOO-15}. The Co film is 0.4~nm thick and has a perpendicular magnetic anisotropy (PMA). During the measurement it is magnetized in plane parallel to the CPW. Due to its low damping, the 5~nm Py film on top of Co facilitates the SW propagation. Damon-Eshbach modes are detected at various distances from the CPW. Therefore it is possible to detect with this technique not only frequency and amplitude non reciprocity, but also a non reciprocity in attenuation length of the spin waves, as predicted by theory. 

Apart from the necessity of a pulsed laser system, the complexity of a TR-MOKE experiment is comparable to BLS, as far as the optical signal acquisition is concerned. However,  SWs need to be excited by a CPW which has to be prepared by lithography directly on top of the sample. The $k$-vectors that can be excited are limited by the CPW geometry. Broadband excitation can be obtained by a single line (as e.g. in \textcite{CIU-16}) with the drawback that the amplitude is reduced with respect to a monochromatic excitation. Since high $k$ values are preferable, the CPW lines have to be rather narrow ($<$1~$\mu$m) which requires electron beam lithography (EBL). Boundary conditions for the generation of plane waves have to be taken into account and it is preferable to have stripes of magnetic material (see supplemental material of \textcite{CHA-14}). The Gilbert damping limits strongly the technique \cite{BAU-14}, since the attenuation length with large damping parameter becomes too small for the SW to be detectable, which is a significant drawback with respect to BLS.

%\begin{figure}[ht!]
%\centering
%\includegraphics[scale=0.35]{fig_LUC_20.jpg}
%\caption{Scheme of an AESWS device. The horizontal stripe is the FM/HM bilayer (purple), on top are two meander shaped CPWs. The measurements are performed by using a Vector Network Analyzer for determination of the mutual inductances L$_{12}$ and L$_{21}$. From \citealp{LUC-20}.}
%\label{fig:SW4}
%\end{figure}

\subsubsection{Propagating spin wave spectroscopy}

\begin{figure}[ht!]
\centering
\includegraphics[scale=0.25]{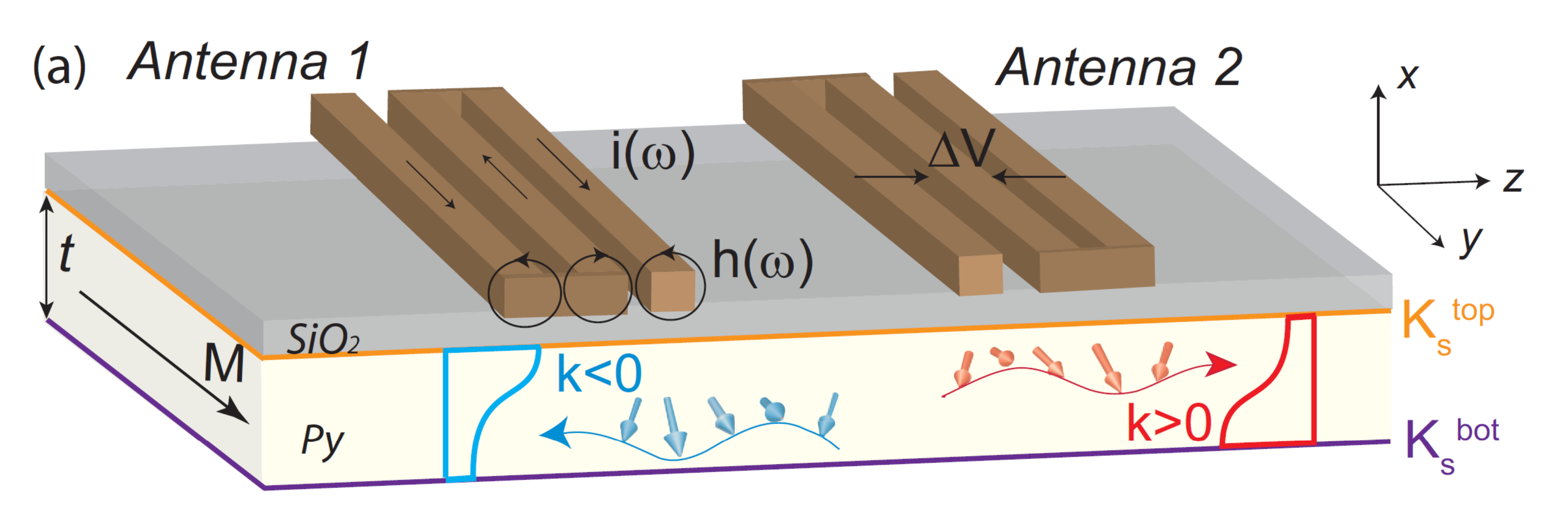}
\caption{Scheme of the PSWS measurement. The two antennas are emitting and receiving HF signals from a VNA, which generates spin waves with an antenna specific $k$-vector distribution. Applying an in-plane magnetic field parallel to the antenna, and focusing on spin waves with $k$ perpendicular to it, an absorption in the signal is observed due to the Damon Eshbach (surface) modes which travel preferentially on one of the film surfaces contributing with different anisotropies ($K_s^{bot}$ and $K_s^{top}$). From \citealp{GLA-16}.}
\label{fig:SW4}
\end{figure}

\begin{figure}[ht!]
\centering
\includegraphics[scale=0.4]{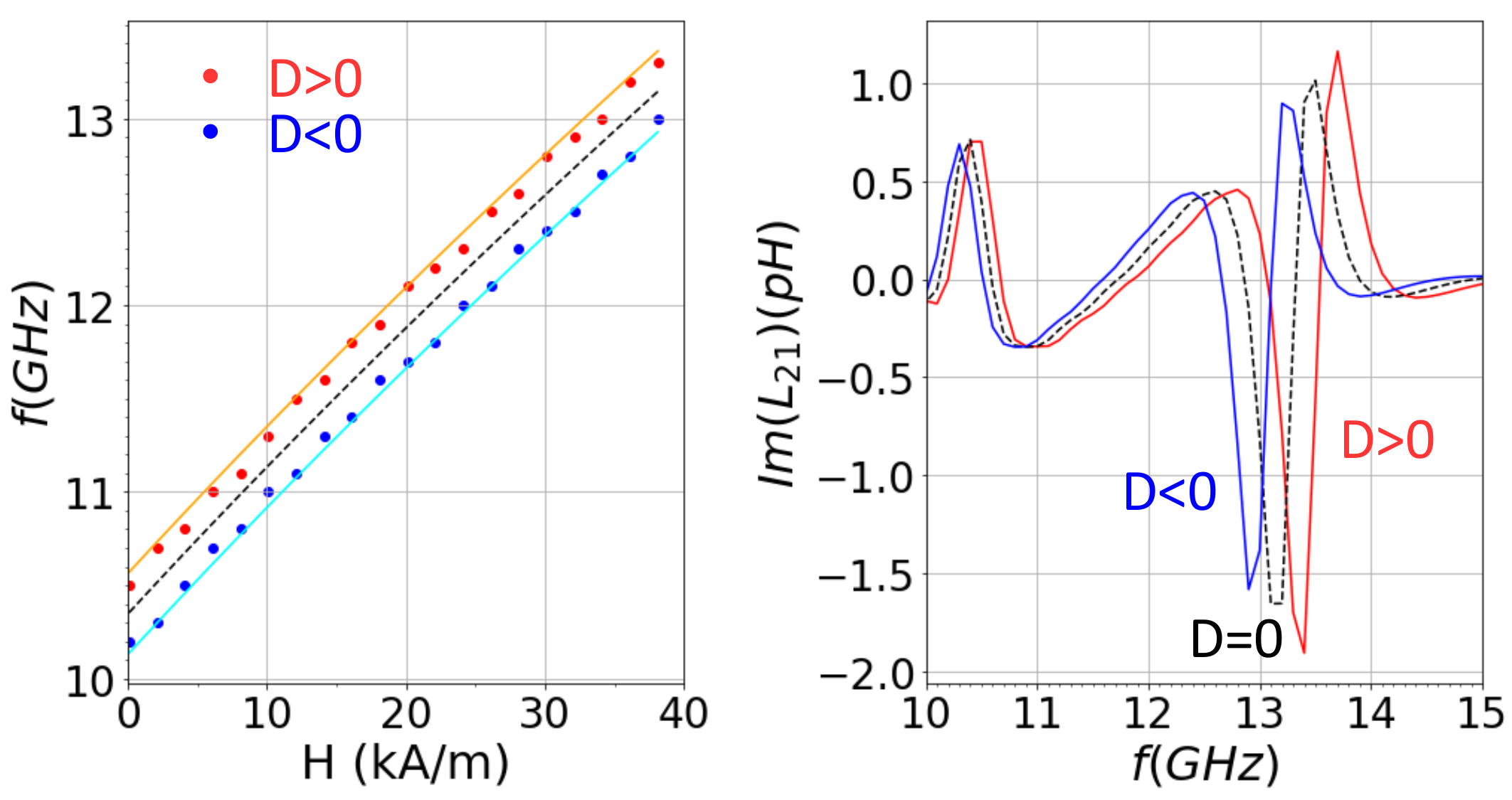}
\caption{Shift of the mutual inductance spectra with $D$. The plot was obtained from numerical calculations based on the LLG approach with use of the surface permeability \cite{EMT-78} on a  magnetic  film (saturation magnetization $\mu_0M_s=$~1.70~T, damping $\alpha=$~0.012, film  thickness $d=$~10~nm, $D=\pm$0.5~mJ/m$^2$) for an antenna with an excitation maximum for a wave vector of $k_{max}=$~10.4~$\mu$m$^{-1}$. Left panel: the frequency shift with respect to the applied magnetic field is shown. The black dashed line is the calculated Damon Eshbach frequency with $D=0$.}
\label{fig:indspec}
\end{figure}

Propagating spin wave spectroscopy (PSWS) (see Fig.~\ref{fig:SW4}) is a tool to characterize spin waves excited in magnetic thin films by measuring the dispersion relation and group velocity all-electrically using a standard high frequency (HF) instrument, a Vector Network Analyzer (VNA) \cite{SCH-73, STA-91, BAI-01}. In the field of magnonics it was widely applied (e.g. \textcite{NEU-10}). As depicted in Fig.\ref{fig:SW4}, the VNA emits HF signals in antenna 1 and receives the transmitted and reflected signals in antenna 2 and 1, respectively (and vice versa). The mutual inductance spectra are obtained from the transmission and reflection coefficients. The inductance spectra show a shift in frequency in presence of DMI (see Fig.\ref{fig:indspec}) from which $D$ is directly obtained through Eq.\ref{eq:omegadmi}. Currently there are only a few experimental works published applying PSWS to measure the DMI constant. 

In~\textcite{LEE-16}, one of the first that applied PSWS, spin waves are excited by a meander shaped CPW (similar to~\textcite{VLA-10}), and detected in a second CPW at a certain distance, thus measuring the spin wave dispersion and group velocity by a VNA \cite{BAI-03}. The measured magnetic multilayers are Pt/Co/MgO, MgO/Co/Pt and MgO/Co/MgO in order to compare samples with positive and negative DMI constant with one with vanishing DMI. The films were patterned into 8~$\mu$m wide stripes and a 40~nm thick AlO$_x$ layer for electric insulation was deposited on top. The CPW consists of Ti(5)/Au(150) and was realized with a primary spatial period of 800~nm ($k$=7.85~$\mu$m$^{-1}$) and subsidiary spatial period of 2250~nm ($k=$~2.8~$\mu$m$^{-1}$). The Co thickness $d$ was changed from 14~nm to 20~nm. The DMI constant is considered here to be a pure interface effect with a characteristic length $\lambda=$~0.25~nm, so that the interfacial DMI $D_i$ is calculated to be $D=D_i \lambda/d$. Assuming that $D_i \lambda$ remains constant the authors obtain a value $D=$~1.8~pJ/m$^2$ for a Co thickness of 2~nm.

In \textcite{KAS-18}, a similar film Ta(2)/Pt(3)/Co(20)/MgO(2) was investigated for electric field effects. Modulations of spin wave frequency were obtained by the gate electric field and the electric field dependence of the DMI was calculated to be 2$\pm$0.93~fJ/Vm. The interfacial DMI was obtained at zero electric field and calculated to be 0.45$\pm$0.15~mJ/m$^2$. The value calculated for $D_s$, as defined in this paper, is very high compared with other techniques. The authors reason that the measured frequency shifts are not caused by the effects of different interfacial anisotropies at the Pt/Co and MgO/Co interface, since no magnetic field dependence was observed. Therefore, they attribute the whole $\Delta f$ to the interfacial DMI. However, it appears that the measured value is overestimated in this way. 

\begin{figure}[ht!]
\centering
\includegraphics[scale=0.2]{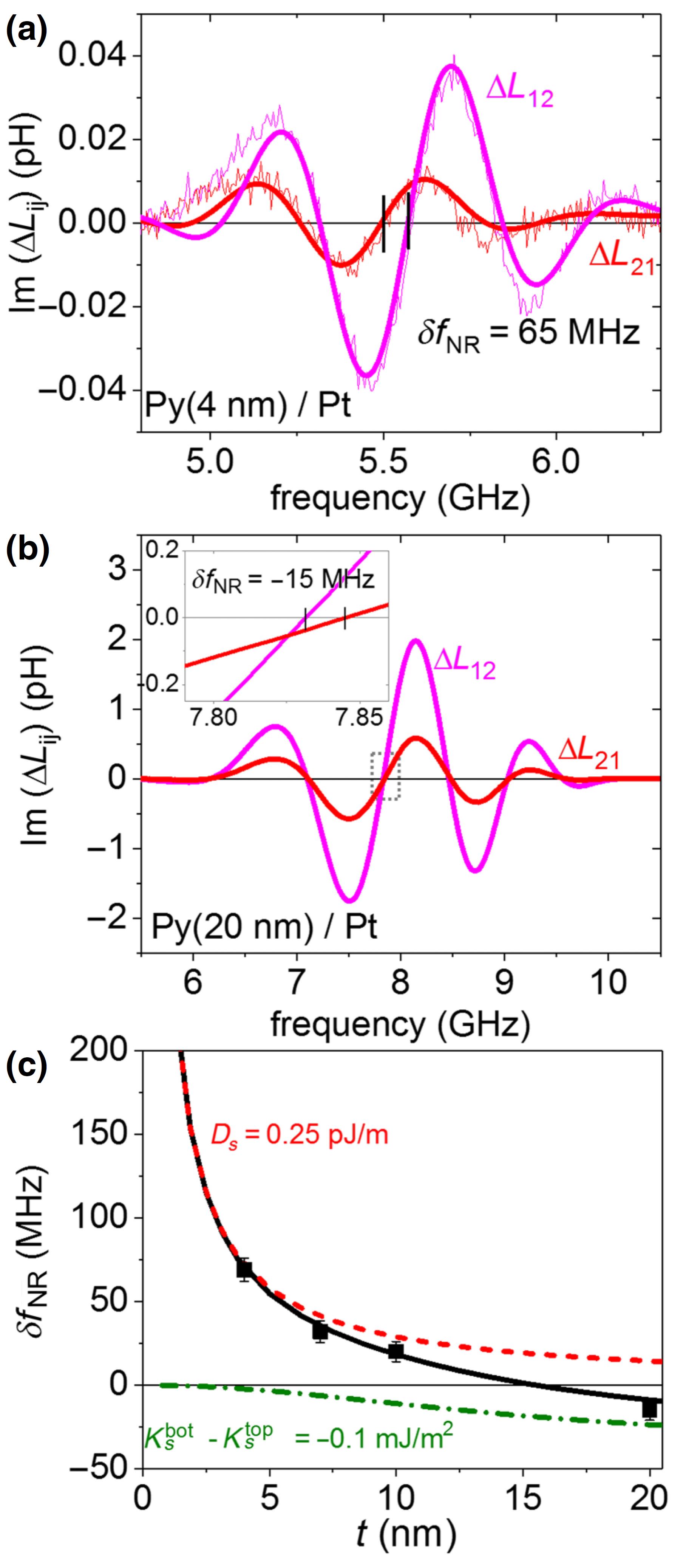}
\caption{(a-b) Mutual inductance spectra of Py/Pt bilayers measured at $\mu_0 H = $~37~mT for spin waves with $k < 0$ and $k > 0$. The frequency of the left-moving spin wave is shifted with respect to the frequency of the right-moving one with a difference $\delta f_{NR}$ (c) Frequency non reciprocity as a function of the Py thickness. Dashed line and dashed-dotted line are the frequency non reciprocity contributions induced by DMI and magnetic anisotropy asymmetry, respectively. From \textcite{GLA-20}.}
\label{fig:SW5}
\end{figure}

\textcite{LUC-20} investigate Ta(4)/X(4)/Co(d)/Y(3)/Pt(2) stacks with $d=$~4-26~nm and (X,Y)=(Pt, Ir),(Ir, Pt) by meander type CPWs with wave vectors $k=$~4, 5.5, 7, 8.5, 10 $\mu$m$^{-1}$. Changing the Co thickness in such a wide range leads to a gradual transition between different Co phases and therefore a change in volume anisotropy in the different samples has to be considered. A detailed analysis of the frequency non-reciprocity due to DMI considering surface anisotropy and changes in volume anisotropy is performed.
Also \textcite{GLA-16,GLA-20} (see Fig. \ref{fig:SW5}) performed an analysis of the influence of a difference in surface (interface) anisotropy on both sides of a Permalloy film on the SW dispersion. In \textcite{GLA-16} the frequency non reciprocity due to different surface anisotropies was calculated revisiting the approach of \textcite{KAL-86}. The important result of the paper is that the frequency non reciprocity due to a difference in surface anisotropies scales as $d^2$, where $d$ is the thickness of the FM film, while the contribution due to DMI scales with $d^{-1}$. In this approach, the exchange spin waves play an important role and a purely magnetostatic approach would lead to unrealistic modal profiles. In \textcite{GLA-20} additionally the spin wave relaxation rate was measured using two CPWs as emitter and detector at different distance. An injected DC current in the Pt layer induced STT via the spin Hall effect and linearly altered the spin wave relaxation time. From this it was possible to obtain additional information such as the spin Hall conductivity.

On the brink of this review is the recent paper \cite{WAN-20} which investigates interfacial DMI at an oxide-oxide interface, here YIG/GGG. This so called Rashba (or iSGE) induced DMI was recently observed in \textcite{AVC-19,DIN-19}. Here YIG films with thicknesses 7,10, 20, 40, 80~nm on a GGG substrate were measured by PSWS and from the asymmetric group velocities the DMI constant was calculated. The values were comparable with the ones obtained from BLS.

\begin{figure}[ht!]
\centering
\includegraphics[scale=0.25]{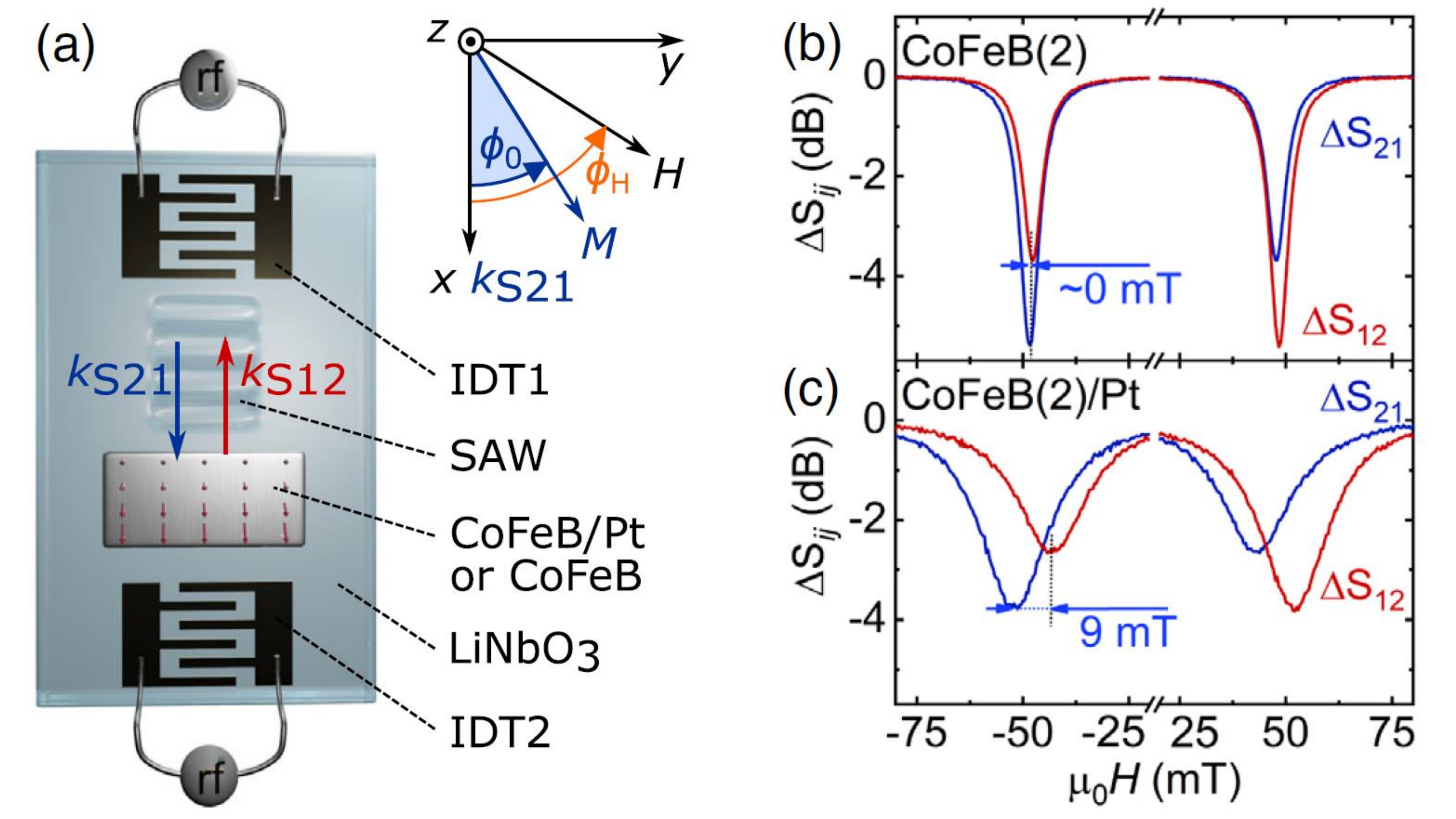}
\caption{(a) Scheme of the experimental setup, sample and
the coordinate system of the magneto acoustic spin wave (MASW) measurement. The transmission coefficients $S_{21}$ and $S_{12}$ for the
oppositely propagating surface accustic waves (SAWs) with wave vectors $k_{S21}$ and $k_{S12}$ are measured by a VNA. The SAW transmission signals of (b) CoFeB(2~nm) and (c) CoFeB(2~nm)/Pt are
shown. CoFeB(2~nm)/Pt shows a nonreciprocity in both transmission
amplitude at resonance and resonant magnetic field. From \textcite{KUE-20}.}
\label{fig:masw}
\end{figure}

A very recent work \cite{KUE-20} employs a technique which, as in PSWS, exploits the SW non-reciprocity in presence of DMI. The magneto acoustic spin wave (MASW) measurement is experimentally  similar to PSWS using a VNA for the detection of the MASW absorption, although it is based on a slightly different physical principle. Via antennas connected to the VNA, surface acoustic waves (SAW) are excited in LiNbO$_3$/CoFeB(1.4-2~nm)/Pt(3~nm) heterostructures which couple to the SWs and generate MASWs with frequencies depending on DMI (see Fig.\ref{fig:masw}). As in a PSWS experiment the $D$ value is obtained from the frequency shift between SWs travelling in opposite directions. The non-reciprocity in the resonant field is caused by the DMI as well as the reduced coupling efficiency due to the SAW-SW helicity mismatch for opposite propagation directions.

In summary, the electric SW measurement is rather straightforward and only commercially available equipment is required. Among the SW techniques it has a very good frequency resolution of a few MHz. As in the TR-MOKE experiments, the excitation of SWs is performed by CPWs, which have to be prepared by EBL in order to obtain high wave numbers, and therefore larger DMI signals. Usually only few wave vectors are available, since each wave vector requires its own CPW geometry and values are limited to about 10~$\mu$m$^{-1}$ using conventional techniques. Only sophisticated techniques allow to reach higher values \cite{LIU-18}. A second CPW is used for detecting the spin wave signal and the signal to noise ratio is in generally low, so that only thick FM films ($>$20~nm) give a clear, well measurable signal. However, the resonance frequency can be extracted well also from a noisy signal. More problematic is the analysis of an amplitude non-reciprocity. Since the FM films are rather thick, using $D_s=D\cdot d$ gives much larger values than other techniques. This may be due to the fact that $D_s$ is not well estimated since large differences of surface properties might occur. In order to avoid the $D_s$ concept, some groups (\textcite{LEE-16}), similar to \textcite{KOS-14a}, attribute the measured DMI value $D_i$ only to a thin layer (monolayer or layer of a characteristic length) close to the interface. This delivers values for an effective $D$ value comparable to other techniques, but depends on the arbitrary choice of the characteristic length. However, while it is shown in several works that the $1/d$ dependence does not hold for thicker samples (see e.g. \textcite{NEM-15}), the PSWS method particularly confirms this, first because ultrathin films are difficult to be measured, second because the surface character of the Damon Eshbach waves might play a role.

\subsubsection{Spin-polarized electron energy loss spectroscopy}
Following the theoretical work of \textcite{UDV-09}, the spin wave dispersion asymmetry due to DMI was firstly measured by spin-polarized electron energy loss spectroscopy (SPEELS) on a double layer of Fe(110) grown on W(110) single crystal~\cite{ZAK-10}. A monochromatized spin-polarized electron beam is scattered from the sample and energy loss or gain as a function of in-plane momentum transfer is measured. The minority electrons create surface SWs by a virtual spin flip scattering process and loose energy, while majority electrons may annihilate thermally excited SWs and gain energy. One measures a peak in the minority spin channel in the energy loss region and a peak in the majority spin channel in the energy gain region (analog to the Stokes and Anti-Stokes peak in BLS) \cite{ZAK-16}. At an electron energy of 4.163~eV (resolution 19~meV) a wave vector of $\pm$0.5-1~\AA$^{-1}$ is employed by changing the angle of the incident beam. A small SW dispersion asymmetry is expected due to Damon-Eshbach surface modes of about 0.1~meV. The energy difference for $\pm k$ is evaluated by inverting the magnetization direction. The modulus of the DMI vector for first neighbors is 0.9~meV and 0.5~meV for second neighbors. With an interatomic distance of $a=$3.16~$\AA$ and a monolayer thickness 1.72~$\AA$ a value of $D=$3.95$\pm$0.6~mJ/m$^2$ is obtained. In a recent work \cite{TSU-20} the technique was taken up again to investigate Co/Fe bilayers with $\textbf{C}_{2v}$ symmetry on W(110) and compared with the previous results. The fundamental role of the crystal symmetry is demonstrated resulting in an extraordinary high DMI. With a thickness of 2$\times$1.72~$\AA$ a value of $D=$6.55$\pm$0.7~mJ/m$^2$ is obtained.

The main advantage of SPEELS is that the wave vector $k=$~500-1200~$\mu$m$^{-1}$ is hundreds of times larger than for the above techniques,  allowing to explore exchange-based spin waves. Since the energy difference of spin waves with opposite $k$ vector grows linearly with $k$ the asymmetry detected is about 100 times larger, so that even very small $D$ constants can be measured. On the other hand, one can recall the necessity of ultra-high-vacuum conditions and the impossibility of applying magnetic fields. Furthermore, spin wave energies calculated usually from a Heisenberg model have to take into account the itinerant character of the spins in ferromagnetic metals by effective exchange constants. A quantitative estimation of the effective exchange constants (isotropic $J$ as well as anisotropic $D$) requires complex ab initio approaches combined with spin dynamic simulations \cite{Bergqvist2013}. This is especially valid for mono- or bilayers of ferromagnets on a heavy metal, as were investigated by SPEELS, where changes in the inter- and intra-layer distances have a non-trivial influence on the exchange parameters \cite{ZAK-16}. 

\begin{table*}
\caption{Overview of DMI measurements of \textbf{various FM materials} via \textbf{spin wave methods different from BLS}. \textcite{KOR-15} use TR-MOKE, \textcite{ZAK-10, TSU-20} use SPEELS, \textcite{KUE-20} use MASW spectroscopy and all the others PSWS. FM and NM stand for ferromagnetic and non-magnetic layer, respectively. $D$ is the interfacial DMI constant and $D_s = D \cdot d$, with $d$ being the thickness of the ferromagnetic film. Numbers in roman were quoted in the reviewed papers, while numbers in italics were either extracted from figures or calculated using the parameters provided. Signs with $^*$ are according to the convention used in this paper and opposite to that in the original manuscript.} 
\begin{tabular}{c|c|c|c|c|c|c}
\hline 
\hline
\textbf{FM} & \textbf{Bottom NM} & \textbf{Top NM}  & $\mathbf{D}$ & $\mathbf{D_s}$ &\textbf{Sign}& \textbf{Ref}\\ 
(nm) & (nm) & (nm) & (mJ/m$^2$) & (pJ/m) & & \\
\hline 
\hline
Fe(2ML)\footnote[1]{ML: atomic monolayer} &		W(110) single crystal &  &	3.95$\pm$0.6\footnote[2]{the value in SI units was taken from the supplemental material of \textcite{TSU-20}, while the original work reports $D$ in meV} & \textit{1.36$\pm$0.2}\footnote[3]{the value was obtained assuming a ML thickness of 1.72$\AA$} &+& \citealp{ZAK-10}\\
\hline
Co(0.4)/Py(5) &	Ta(2)/Pt(2) & MgO(5) &		0.16 & 0.89 &-& \citealp{KOR-15}\\\hline
\multirow{3}{*}{\centering Co(14-20)} &		Ta(3)/Pt(3) & MgO(1.8)/Ta(3) &	\emph{0.19-0.26} & \emph{3.62-4.08} &-$^*$& \multirow{3}{*}{\hfil\citealp{LEE-16}}\\
 &		Ta(3)/MgO(1.8) &  MgO(1.8)/Ta(3) &	\emph{0.03} & \emph{0.54} &-$^*$&\\
&		Ta(3)/MgO(1.8) & Pt(3)/Ta(3) &	\emph{0.11-0.17} & \emph{2.1-2.55} &+$^*$&\\\hline
   Py(6) & Al$_2$O$_3$(21) & Al$_2$O$_3$(5) & \emph{0.007} & 0.04 & - & \citealp{GLA-16}\\\hline
Co(20) &		Ta(2)/Pt(3) & MgO(2) &	0.45 & \emph{9} &-$^*$& \citealp{KAS-18}\\\hline
 \multirow{3}{*}{\centering Co(4-26)} &	Ta(4)/Pt(4) & Ir(3)/Pt(2) &	n.a. & 1.0$\pm$0.2 &-& \multirow{3}{*}{\hfil\citealp{LUC-20}}\\
 &	Ta(4)/Pt(4) & Pt(3)/Pt(2) &	n.a. & 0.1$\pm$0.04 &-& \\
  &	Ta(4)/Ir(4) & Pt(3)/Pt(2) &	n.a. & 1.0$\pm$0.2 &+& \\\hline
  Py(4-20) & Ti(5) & Pt(5,10) &	n.a. & 0.25 & + & \citealp{GLA-20}\\\hline
  YIG(10) & GGG substrate & &	0.0099$\pm$0.0019 & \emph{0.1} &n.a.& \citealp{WAN-20}\\\hline
 Fe/Co (ML)\footnotemark[1] &		W(110) single crystal &	&6.55$\pm$0.7 & 2.25$\pm$0.25\footnotemark[3] & + & \citealp{TSU-20}\\\hline
 Co$_{40}$Fe$_{40}$B$_{20}$(2) & LiNbO$_3$ & Pt(3)	&0.424$\pm$0.001 & \textit{0.85$\pm$0.002} & +$^*$ & \citealp{KUE-20}\\\hline
\hline
\end{tabular}
\label{TableSW1}
\end{table*}

\subsection{Advantages and Limitations}
The main advantage of spin waves methods for determination of the DMI constant is the rather simple expression for the measurable frequency difference that occurs when the sign of $k$ is changed, as seen in Eq.~\ref{eq:dispKost}. This also means that not only the size, but also the sign of the constant can be easily determined, provided that the geometry of the experiment and the polarity of the applied field are specified. However, attention has to be paid to several aspects. 

First of all, as it is the case for DW methods, the saturation magnetization should be determined by external means, such as vibrating sample magnetometry (VSM) or SQUID magnetometry. Then,  the gyromagnetic factor and the anisotropy constants can be obtained by either FMR or BLS measurements as a function of the intensity and the direction of the external field. Furthermore, the thickness of the magnetic active layer has to be evaluated correctly in the derivation of ${D_s}$ from measurement of $\omega_{DMI}$. Not always it coincides with the nominal layer thickness due to interlayer mixing (dead layer) \cite{BEL-18} or proximity effects \cite{ROW-17}. Finally, it is considered that $\omega_0$ in Eq.~\ref{eq:dispKost} is independent of the sign of $k$. This is in general the case, but when comparing different samples (e.g. with different thicknesses), this might be not perfectly true, for example due to different surface/interface anisotropies at both sides of the magnetic film which influence differently the $\pm k$ spin waves and $\Delta \omega_0=\omega_0(k)-\omega_0(-k) \neq 0$. Therefore, such a difference is necessarily reflected in the measured frequency asymmetry. This problem comes from the fact that DE waves travel on opposite surfaces at opposite $k$ directions \cite{CAM-87} and therefore any difference in magnetic properties at the two surfaces, as e.g. surface anisotropy, will reflect in a non-reciprocity of the spin wave spectra. By comparison with theoretical models and the correct measurement procedure their contribution has to estimated as far as possible. For example, the asymmetry due to differences in surface anisotropy or asymmetry in saturation magnetisation $M_s$ is evaluated in \textcite{DI-15} by measuring additionally to a Pt/Co/Ni sample a Co/Ni and a Co/MgO/Ni sample. While in the latter $\Delta \omega_0$ is negligible, meaning that the contribution due to asymmetry in $M_s$ is negligible, the former shows an opposite sign of $\Delta \omega_0$ with respect to the Pt/Co/Ni sample. This means that the two interfaces have different contributions either from anisotropy or DMI, but that a major role is played by the Pt/Co interface with opposite sign with respect to the Co/Ni interface.

Surface effects usually scale as $1/d$ \cite{STA-91,HIL-90} which might open a question on how to distinguish DMI related non-reciprocity from other surface or interface effects. However, the frequency shift due to DMI and surface anisotropy are intrinsically different. As shown in \cite{GLA-20} they have different dependencies on $d$ as the first is proportional to $1/d$ while the latter to $d^2$. Furthermore, the first depends linearly on $k$, while the latter has a more complex dependence, but usually decreases with $k$. The wave vector dependence of the frequency shift due to the surface anisotropy is stronger for thicker films, while the one induced by DMI is larger for thinner films (see e.g. supplementary material of \cite{LUC-20}). This opens the possibility to distinguish the different contributions, depending on the value of $k$, having available films of different thickness. 

It was shown by numeric calculations and experimentally by \textcite{STA-15} that for thin films ($d_{Py}<$~8~nm) the SW reciprocity due to the surface anisotropy is negligible with respect to that induced by DMI,  while for thicker films both contributions have to be considered. On the basis of the above arguments, in the literature it is usually considered that at $k d << 1$ any dependence of $\omega_0$ on the sign of $k$ can be neglected \cite{NEM-15} and that, in absence of DMI, the $\pm k$ spin waves are subjected to an average magnetic energy equal at both sides of the magnetic film \cite{SOO-63}. 

Care has to be taken not only with respect to frequency non-reciprocity, but also to amplitude non-reciprocity. Also here there might be effects due to different boundary conditions at the interfaces of the ferromagnetic film. For example a metallic boundary condition at one side would attenuate strongly the spin wave travelling close to it. Furthermore, methods as PSWS show an amplitude non-reciprocity due to the fact that the CPW couples more to one wave vector than to the other due to the symmetry of the CPW \cite{SER-10}.

Probably the strongest experimental limitation of the methods based on spin waves is that the sample has to be magnetized in-plane in order to be able to detect DE modes. Therefore for samples with strong out-of-plane anisotropy (perpendicular magnetic anisotropy, PMA) it is necessary to apply a sufficiently high in-plane field, which can be experimentally difficult. Furthermore, ultrathin PMA films, give rise to a low signal to noise ratio, not only due to their thickness but often also due to the high damping, so that fitting procedures may have quite large uncertainties. This sets a lower limit to the values of $D$ that may be measured by spin wave methods and therefore these may be considered as reliable for either magnetic film thicknesses above a certain threshold, or with low damping, or high DMI constant.

\section{Spin Orbit Torque method}\label{sec:SOT}

\subsection{Method overview}

\begin{figure*}[ht!]
\centering
\includegraphics[width=.8\textwidth, trim=0 0 0 0,clip]{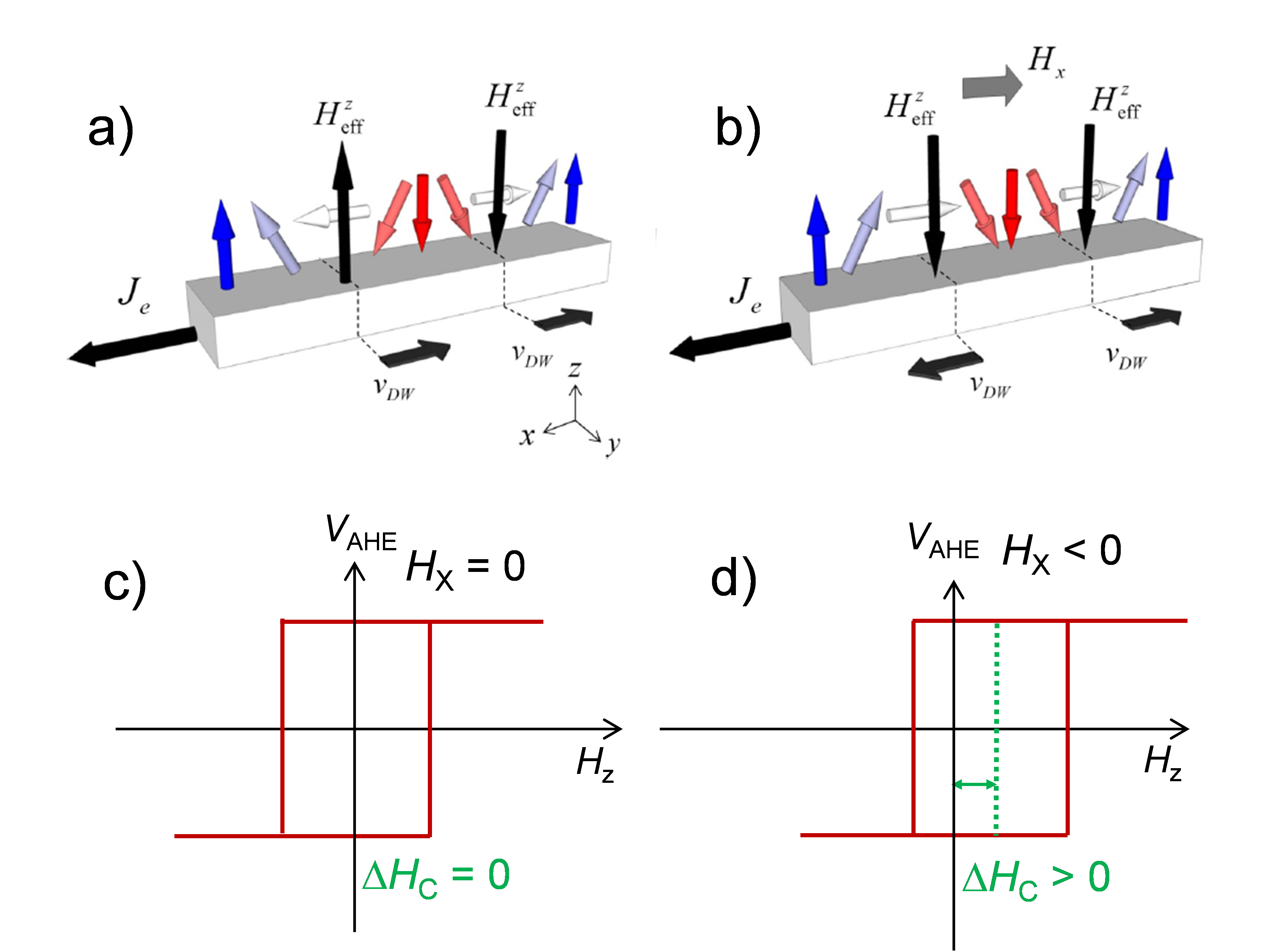}
\caption{(a) Schematics of the current induced effective magnetic field $H_{eff,z}$ at the N\'eel-type chiral domain wall in a HM/FM bilayer with perpendicular magnetic anisotropy. In the absence of an external magnetic field, $H_{eff,z}$ at the DW points along the opposite direction ($H_{eff,z} = 0$) thus the motion of a domain wall $v_{DW}$ is along the –x direction and there is no domain expansion. (b) In the presence of an external magnetic field $H_x$ large enough to align the N\'eel-type DW along $H_x$, the induced $H_{eff,z}$ points along the same direction, which leads to domain expansion due to the domain wall motion towards to the opposite directions. (a) and (b) are taken from \citealp{PAI-16}. (c) In the absence of $H_x$, there is no shift of the anomalous Hall loop. (d) However, a shift of the anomalous Hall loop is expected with the application of $H_x$, which provides a measure of DMI and the spin-torque efficiency.}
\label{fig:SOT1}
\end{figure*}

When a charge current is applied to a HM/FM bilayer, spin torque effects occur due to the strong spin-orbit interaction at the interface. A charge current J$_{e}$ flowing in the HM along the $x$ axis will generate a transverse spin accumulation $\vec{\sigma}$ along the $y$ direction through the spin Hall effect and/or inverse spin Galvanic effect \cite{MAN-19}. The spin accumulation acts on the magnetic moment $\vec{M}$ and results in an effective damping-like magnetic field $\vec{H}_{eff} \approx~\vec{\sigma} \times \vec{M}$. Of particular interest is the out-of-plane effective magnetic field $H_{eff,z}$ induced at the N\'eel-type DW \cite{PAI-16}. As shown in Fig.~\ref{fig:SOT1}, $H_{eff,z}$  depends on the angle $\phi$ between the DW moment and the $x$ axis, and can be quantitatively written as $H_{eff,z}  = \chi_{DL} J_e cos\phi$ (which represents effectivly Eq. \ref{eqSHE}), where the charge-to-spin conversion efficiency (effective damping-like field per unit current density) $\chi_{DL} = (\hbar \xi_{DL}/2e\mu_0 M_S d)$. Here, $\xi_{DL}$, $M_S$, $d$ are the effective damping-like torque efficiency, the saturation magnetization of the FM and thickness of the FM layer.

\subsection{Theory and models}
\textcite{PAI-16} first proposed a method to simultaneously determine the spin-torque efficiency and DMI field $H_{DMI}$ in HM/FM with PMA, based on current assisted DW propagation model. As shown in Fig.~\ref{fig:SOT1}(a), in the case of a homochiral N\'eel-type DW, the net $H_{eff,z}$ which acts on the DW magnetization is expected to be zero due to the opposite signs of $H_{eff,z}$ for up-to-down ($cos\phi = 1$) and down-to-up ($cos\phi = -1$) DWs. In this case, the opposite $H_{eff,z}$ at the DWs moves the domain in the same direction and there is no domain expansion.
However, when an in-plane external magnetic field $H_x$ is applied that is large enough to overcome $H_{DMI}$, the magnetization in the walls will align parallel to $H_x$ as shown in Fig.~\ref{fig:SOT1}(b) and the resulting $H_{eff,z}$ will point along the same direction for both up-to-down and down-to-up DWs. 
This leads to the expansion or contraction of domains. The magnitude and the polarity of $H_{eff,z}$ depends on $H_{x}$ and $J_e$., i.e., $H_{eff,z}(H_{x},J_e)$. Since $H_{eff,z}$ acts as an effective magnetic field for the device, it is thus expected that the out-of-plane hysteresis loop of the bi-layer can be shifted by $H_{eff,z}$.
Therefore, by measuring the shift of the hysteresis loop as a function of $H_x$ and $J_e$, one can determine the magnitude of $H_{DMI}$ as well as $\chi$. 
Specifically, $H_{DMI}$ is determined when the DW is fully aligned by $H_{sat}$, i.e. $H_{DMI}=H_{sat}$, and $\chi_{DL}=H_{eff,z}(H_{sat},J_e)/J_e$.

\begin{figure*}[ht!]
\centering
\includegraphics[width=0.9\textwidth,trim=50 60 50 0,clip]{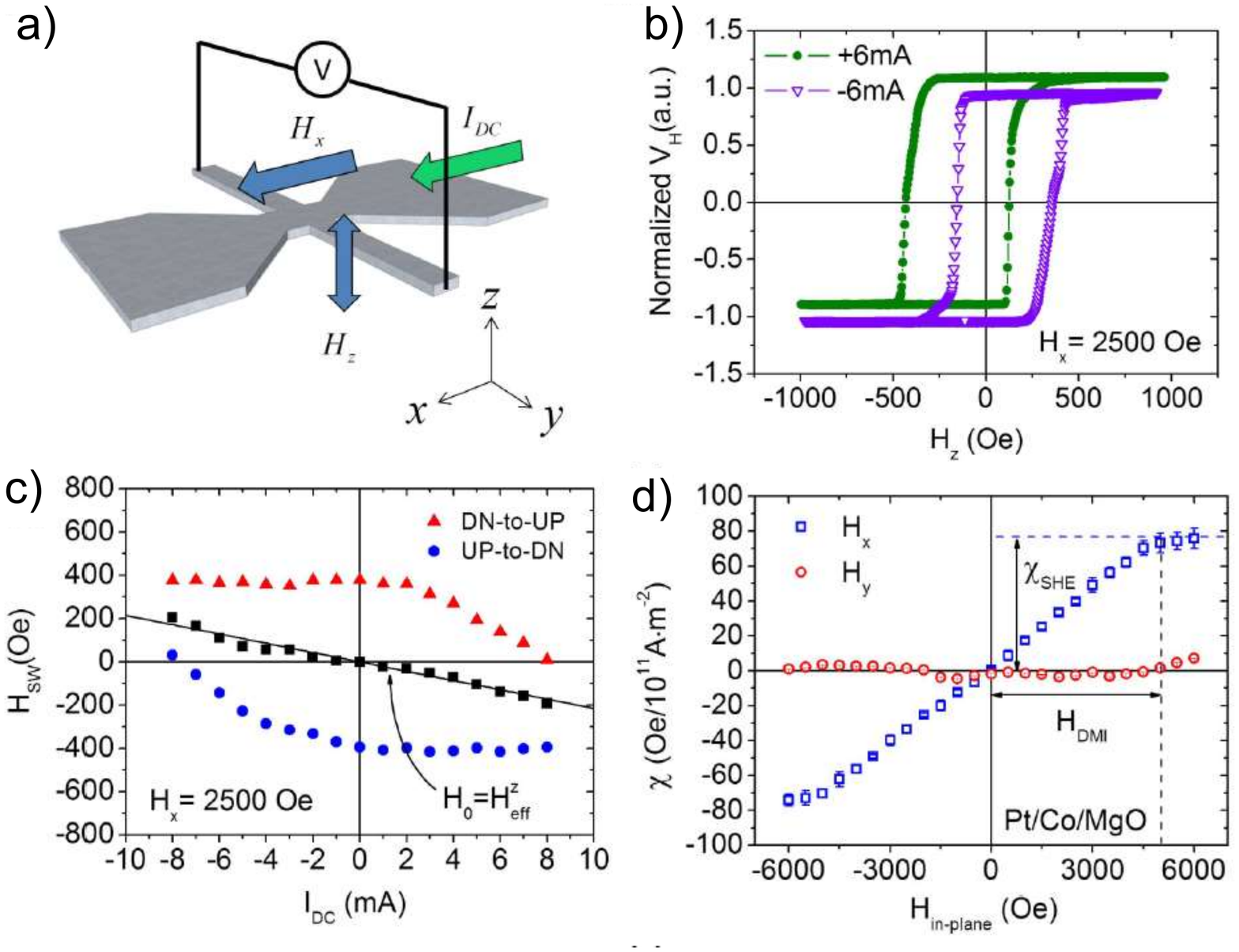}
\caption{(a) Schematics of the measurement of the DMI field utilizing the anomalous Hall effect (AHE). (b) AHE-loops for Pt(4~nm)/Co(1~nm)/MgO(2~nm) with $I_{e} = \pm$6~mA (note that for consistency in the main text, we use $I_e$ ($J_e$) instead of $I_{dc}$ as used in the figure) and an in-plane bias magnetic field $H_x =$~0.25~T. For clarity, slight offsets are introduced for both AHE loops. (c) Switching fields $H_{SW}$ for down-to-up (red triangles) and up-to-down (blue circles) magnetization reversal as a function of $I_{e}$. $H_0 = H_{eff,z}$ represents the average value of the switching fields. (d) Charge to spin conversion efficiency as a function of in-plane bias field for Pt(4~nm)/Co(1~nm)/MgO(2~nm). Blue squares and red circles represent the data obtained with in-plane magnetic field along x and y directions, respectively. Note that we use $\chi_{DL}$ in the text instead of $\chi_{SHE}$. All the figures are taken from \citealp{PAI-16}.}
\label{fig:SOT2}
\end{figure*}

As shown in Fig.~\ref{fig:SOT2}(a), an anomalous Hall effect (AHE) loop, i.e., an anomalous Hall voltage versus out-of-plane magnetic field, is measured by changing both the magnitude and polarity of $I_{e}$ and $H_x$. Under an in-plane bias-field of 0.25~T the AHE-loop shifts to opposite directions at $\pm$6~mA, which indicates the opposite sign of the induced $H_{eff,z}$ due to the SOT. The $I_{e}$ dependence of the up-to-down switching field $H_{SW,up-to-down}$ and down-to-up switching field $H_{SW,down-to-up}$ is summarized in Fig.~\ref{fig:SOT2}(c). Two current-related effects should be considered to explain the switching fields, i.e., the effect of Joule heating and $H_{eff,z}$. Joule heating reduces the coercivity $H_C$ which is proportional to $I_{e}^2$, i.e., $H_C(I_{e}^2)$; while $H_{eff,z}$ is expected to show a linear behaviour with respect to $I_{e}$, i.e., $H_{eff,z}(I_{e})$. Thus, for a fixed $I_{e}$, the switching field can be written as $H_{SW,up-to-down} = H_{eff,z} + H_C$ for up-to-down switching, and $H_{SW,down-to-up} = H_{eff,z} - H_C$ for down-to-up switching. By eliminating the effect of Joule heating, the magnitude of $H_{eff,z}$ can be easily obtained as $H_{eff,z} = (H_{SW,up-to-down} + H_{SW,down-to-up})/2$. As shown in Fig.~\ref{fig:SOT2}(c), $H_{eff,z}$ scales linearly with $I_{e}$, indicating that $H_{eff,z}$ is indeed induced by current-induced spin accumulation due to interfacial spin-orbit interaction. 
One can quantify the conversion efficiency by the slope $\chi(H_x,J_e)= dH_{eff,z}/dJ_e$, and Fig.~\ref{fig:SOT2}(d) shows the magnitude and polarization of $\chi$ as a function of in-plane bias magnetic-field along x and y directions. One can see that $\chi$ remains zero with the application of $H_y$, but $\chi$ increases linearly with $H_x$ and saturates at $\mu_0H_{sat} = $~0.5~T. 
This observation is in agreement with the domain expansion picture, where the DW magnetization changes from an average value of $<cos\phi> = 0$ at $H_x = 0$ to $<cos\phi> = 1$ when $H_x$ fully aligns the DW moment, which provides a measure of $H_{DMI}$ with $H_{DMI}=H_{sat}$. Moreover, the magnitude of $\chi_{DL}$ is obtained when $\chi$ saturates, from which one can determine the charge-to-spin conversion efficiency $\xi_{DL}$.

\subsection{Experimental Results}
Data of DMI constants determined using the SOT method are reported in Tab.~\ref{table:SOT}. Initially, \textcite{PAI-16} quantified the magnitude of $H_{DMI}$ in Pt/Co/MgO, Pt/CoFeB/MgO and Ta/CoFeB/MgO multilayers, and demonstrated opposite signs of the spin Hall angle for Pt and Ta. They also show that a sizeable DMI exists in wedged Pt/Co(wedge)/MgO samples, where an additional spin-orbit torque appears due to the lateral structure  asymmetry and which can be used for in-plane bias magnetic field free switching. By inserting a Pt spacer between a Ta and CoFeB interface, \textcite{CHE-18} found that both the spin-torque efficiency and the DMI constant $D$ gradually decrease as the Pt thickness increases to 1~nm. Note that we did not include ultra thin layers in our summary since the layer formation may not be complete in this case. To generate larger interfacial DMI to support a skyrmion phase, a non centrosymmetric superlattice structure [HM1/FM/HM2]$_{N}$ with PMA is sometimes adopted because of the enhanced interlayer coupling \textcite{MOR-16}. \textcite{ISH-19} demonstrated that the DMI at Ir/Co and Pt/Co interfaces shows comparable magnitude with the same sign, leading to a reduced DMI in Pt/Co/Ir trilayers in agreement with several measurements using BLS and DW methods. \textcite{KHA-18a} showed that the DMI in Pt/Co/Ru trilayers is comparable to Pt/Co/Ir. 
\textcite{Yan-21} measured the spin-orbit torque and DMI in [Ni-Co]/Ir heterostructures. By alloying Cr into Pt, \textcite{Qua-20} recently found that both the spin-torque efficiency and the DMI can be modulated, reaching a maximum at a Cr concentration of about 25\%. 

\textcite{PAI-16} found that $H_{DMI}$ in a Pt(4~nm)/Co(1~nm)/MgO(2~nm) stack and a Pt(4~nm)/CoFeB(1~nm)/MgO(2~nm) stack differ by a factor of two. 
Since DMI is an interfacial effect, the magnitude of $H_{DMI}$ is expected to be inversely proportional to the thickness of FM. However, no clear experimental results with the SOT method show this trend indicating that DMI can have a complicated relation to the FM thickness. Interestingly, it has been shown in W/FM/MgO structures (FM = CoFeB, FeB) that $H_{DMI}$ changes sign upon increasing the thickness of FM \cite{DOH-19}, indicating that there could be competing mechanisms contributing to DMI as well as to the interfacial spin-orbit interaction \cite{YUN-18,CHE-18a,CAO-18a}.

Recently, measurements have demonstrated the presence of DMI in heavy metal/magnetic insulator heterostrucures, where the DMI is strong enough to stabilize skyrmions \cite{DIN-19,SHA-19a} and for fast domain motion \cite{AVC-19}. Ferrimagnetic insulators are attractive due to their lower Gilbert damping in comparison with ultrathin ferromagnetic metals. Using SOT, \textcite{DIN-19} quantified the DMI in thulium iron garnet (TmIG)/Pt bi-layers, and have shown that the magnitude of the DMI constant is about 1-2 orders smaller than in metallic heterostructures. They also showed that the magnitude of the DMI constant is inversely proportional to the TmIG thickness, indicating that the DMI in TmIG/Pt bilayer is indeed an interfacial effect. In \textcite{DIN-20a} the origin of the DMI in TmIG/HM bilayers on Gd Garnet (GGG) substrates could be attributed to the GGG/TmIG interface by investigating the FM and HM thickness dependent DMI and SOT.

\begin{table*}
\caption{Overview of DMI measurements of \textbf{various FM materials} by \textbf{spin-orbit torque induced effective field} under an in-plane bias magnetic field. FM and NM stand for ferromagnetic and non-magnetic layer, respectively. $D$ is the interfacial DMI constant and $D_S = D \cdot d$, where $d$ is the thickness of the FM. Numbers in roman were quoted in the reviewed papers, while numbers in italics were either extracted from figures or calculated using the parameters provided.} 
\begin{tabular}{c|c|c|c|c|c|c}
\hline 
\hline
\textbf{FM} & \textbf{Bottom NM} & \textbf{Top NM}  & $\mid$\textbf{$\mu_0 $}\textbf{H$_\mathbf{{DMI}}$}$\mid$ & $\mid$\textbf{$\mathbf{D}$}$\mid$ & $\mid$\textbf{$\mathbf{D_s}$}$\mid$ & \textbf{Ref}\\ 
(nm) & (nm) & (nm) & (mT) & (mJ/m$^2$) & (pJ/m) & \\
\hline 
\hline
Co(1) & Pt(4) & \multirow{10}{*}{\centering MgO(2)} & 500 & 3.0 & \textit{3.0} & \multirow{10}{*}{\hfil \citealp{PAI-16}}\\
CoFeB(1)& Pt(4) & & 250 & 1.8 & \textit{1.8} & \\
CoFeB(1)& Ta(6) & & 25 & 0.6 & \textit{0.6} & \\
Co(0.65)& Pt(4) & & 110 & 1.45 & \textit{0.94} & \\
Co(0.80)& Pt(4) & & 200 & 1.99 & \textit{1.59} & \\
Co(0.92)& Pt(4) & & 290 & 2.12 & \textit{1.95} & \\
Co(1.00)& Pt(4) & & 400 & 2.64 & \textit{2.64} & \\
Co(1.10)& Pt(4) & & 450 & 2.91 & \textit{3.20} & \\
Co(1.23)& Pt(4) & & 310 & 2.49 & \textit{3.06} & \\
Co(1.43)& Pt(4) & & 160 & 1.62 & \textit{2.32} & \\
Co(1.52)& Pt(4) & & 100 & 1.37 & \textit{2.08} & \\
\hline
\multirow{4}{*}{\centering Co(0.5)}& \multirow{4}{*}{\centering Pt(4)} & Ta(2) & 110$\pm$10 & 1.01$\pm$0.12 & \textit{0.51$\pm$0.06} & \multirow{4}{*}{\hfil\citealp{YUN-18}} \\
& & Ta(4) & 110$\pm$10 & 0.75$\pm$0.09 & \textit{0.38$\pm$0.05} & \\
& & Ta(6) & 80$\pm$10 & 0.70$\pm$0.09 & \textit{0.35$\pm$0.05} & \\
& & Ta(8) & 190$\pm$10 & 1.4$\pm$0.12 & \textit{0.7$\pm$0.06} & \\
\hline
Co(1.2) & Pt(4) & Ir(1) & 65$\pm$6 & 1.09$\pm$0.14 & \textit{1.31$\pm$0.17} & \multirow{8}{*}{\hfil\citealp{KHA-18a}}\\
Co(1.2) & Pt(4) & Ir(2) & 81$\pm$8 & 1.42$\pm$0.18 & \textit{1.7$\pm$0.22} &  \\
Co(1) & Pt(4) & Ir(1) & 130$\pm$13 & 2.06$\pm$0.26 & \textit{2.06$\pm$0.26} &  \\
Co(1) & Pt(4) & Ru(2) & 140.4$\pm$14 & 2.66$\pm$0.33 & \textit{2.66$\pm$0.33} &  \\
Co(0.8) & Pt(4) & Ru(2) & 212$\pm$21 & 2.40$\pm$0.3 & \textit{1.92$\pm$0.24} &  \\
Co(0.8) & Pt(4) & Ru(3) & 218$\pm$22 & 2.30$\pm$0.29 & \textit{1.84$\pm$0.23} &  \\
Co(0.8) & Pt(4) & Ru(4) & 238.6$\pm$24 & 2.56$\pm$0.32 & \textit{2.05$\pm$0.26} &  \\
{[Pt(1)/Co(0.8)/Ru(1.3)]$_{x2}$}& Pt(3) &  & 220$\pm$22 & 2.07$\pm$0.26 & \textit{1.66$\pm$0.21} &  \\
\hline
CoFeB(1.3)& Ta(3) & MgO(1) & 30 & 0.22 & \textit{0.29} & \citealp{CHE-18a} \\
\hline
CoFeB(1.4)& Mo(4) & MgO(2) & 20 & 0.35 & \textit{0.49} & \citealp{CHE-18}\\
\hline
FeB(0.96) & \multirow{4}{*}{\centering $\alpha$-W(4)} & \multirow{4}{*}{\centering MgO(1.6)} & 25 & n.a. & n.a. & \multirow{4}{*}{\hfil\citealp{DOH-19}} \\
FeB(0.86) & & & 30 & n.a. & n.a. &  \\
FeB(0.76) & & & 18 & n.a. & n.a. &  \\
FeB(0.56) & & & 50 & n.a. & n.a. &  \\
\hline 
{[Pt(0.6)/Co(0.9)/Ir(0.6)]$_{x3}$}& Pt(2) &  MgO(2) & 80 & 0.4 & \textit{0.36} & \multirow{5}{*}{\hfil\citealp{ISH-19}} \\
{[Pt(0.6)/Co(0.9)/Cu(0.6)]$_{x3}$} & Pt(2) &  MgO(2) & 180 & 1.8 & \textit{1.62} & \\
{[Pt(0.6)/Co(0.9)/Ir(0.6)]$_{x1}$} & Pt(2) &  MgO(2) & 110 & 0.7 & \textit{0.63} & \\
Co(0.9) & Ir(7) & MgO(2) & 220 & 1.6 & \textit{1.44} & \\
Co(0.9) & Pt(2)/Ir(1) & Ru(1) & 150 & 1.1 & \textit{0.99} & \\
\hline 
TmIG\footnote[1]{Thulium Iron garnet}(2.9)& Pt(7) & n.a. & 20$\pm$1 & 0.036$\pm$0.02 & \textit{0.1$\pm$0.06} & \citealp{DIN-19} \\
\hline 
TmIG\footnotemark[1](2.9-16)& \multirow{4}{*}{\centering GGG\footnote[2]{Gd$_3$Ga$_5$O$_{12}$} substrate} & Pt(7) & \textit{22 - 1.5}  & \textit{0.035 - 0.003} & \textit{0.13}\footnote[3]{the value was obtained from a linear fit over all TmIG thicknesses}  & \multirow{4}{*}{\centering \citealp{DIN-20}} \\
\multirow{3}{*}{\centering TmIG\footnotemark[1](5.4)}& & Pt(0.5-7) & n.a. & \textit{0.024 - 0.029} & \textit{0.13}\footnote[4]{the value was obtained from averaging over all Pt thicknesses} &  \\
&  & W(5)/Cu(3) & n.a. & \textit{0.025} & \textit{0.135} &  \\
&  & Cu(3)/Pt(1.5) & n.a. & \textit{0.025} & \textit{0.135} &  \\
\hline
Co(0.8)& Cr$_{0.25}$Pt$_{0.75}$(8)/Pt(0.4)& AlO$_x$(2)& 100& 0.87& \textit{0.70}& \citealp{Qua-20}\\
\hline
{[Ni(0.34)/Co(0.16)]$_{2}$}& Cu(2.5)& Ir(4)& 17.4& 0.82& \textit{0.82}& \citealp{Yan-21}\\
\hline
\end{tabular}
\label{table:SOT}
\end{table*}

\subsection{Advantages and Limitations}
Being different from the methods based on DW motion and asymmetric spin-wave propagation, the method based on current-induced shifts of the anomalous Hall loops is a straightforward way for determining the DMI field without involving complicated mathematical models. However, this method requires perpendicular magnetic anisotropy of the ferromagnetic material, which typically limits this method to ultrathin ferromagnetic materials. Non-square shapes of the AHE-loops also lead to inaccuracies in the determination of $H_{DMI}$. To determine the DMI constant, one requires a numerical value for the exchange stiffness; usually an assumed value is used. Moreover, this method is not capable to determine the sign of DMI.

\section{Comparison of the methods and discussion of problems in the determination of the DMI constant}

\subsection{Comparing the DMI constants obtained by the different methods}

\begin{figure*}[ht!]
\centering
\includegraphics[width=.75\textwidth, trim=10 10 60 10, clip]{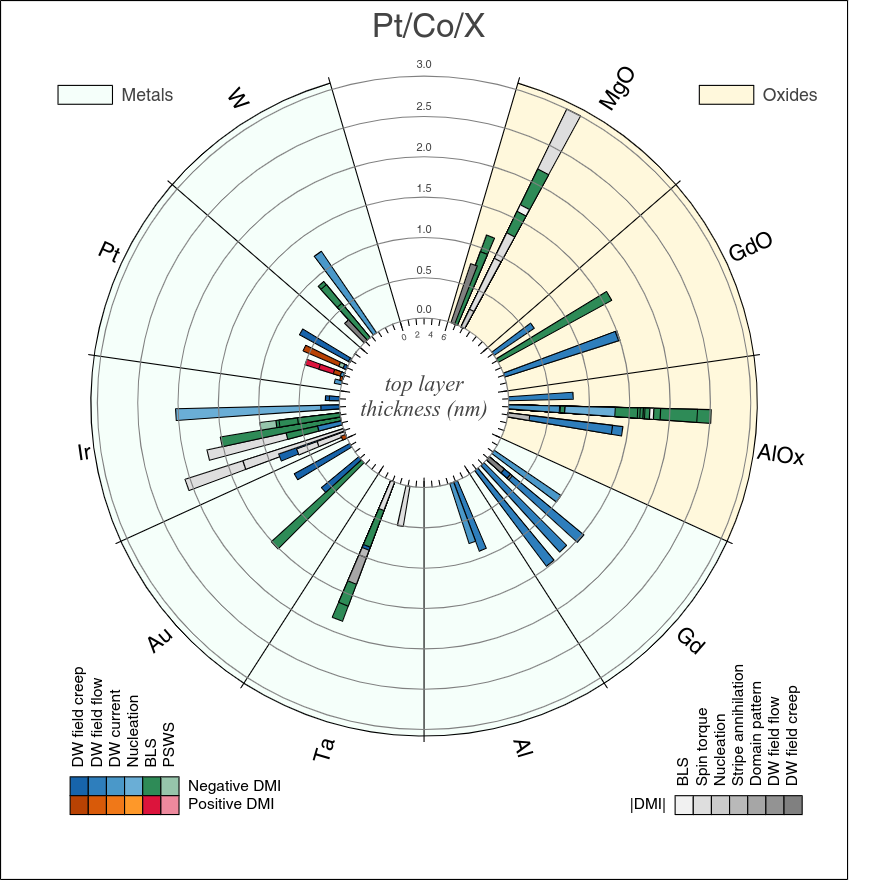}
%\caption{Literature data of the DMI constant $D_\mathrm{s}$ for the Pt/Co/X thin films, where the top layers X = (Pt, Ir, Au, Ta, Al, Gd, Mg, AlO$_x$, GdO, MgO). The shapes of the symbols refer to different experimental methods, while their size and color reflect the amplitude and the sign of $D_\mathrm{s}$, respectively, as specified in the legends on the right. Data for methods not able to determine the sign of $D_\mathrm{s}$ (i.e, domain pattern, stripe annihilation, nucleation, and spin torque) are in grey. The thickness of the bottom layer of Pt varies between 0.8~nm and 30~nm, with more than 75\% of the data for a thickness larger that 3~nm. See \href{run:./figs27_29.html}{here} for an electronic version of the figure.}
\caption{Literature data of the DMI constant $D_\mathrm{s}$ for the Pt/Co/X thin films, where the top layers X = (MgO, GdO, AlO$_x$, Al, Ta, Au, Ir, Pt, W) can be metals (light-green) or oxides (light-yellow). The blue/green and orange/red palettes are for experimental methods where the sign is known (negative and positive, respectively), while the grey palette is for data where only the magnitude of $D_\mathrm{s}$ is available. The thickness of the bottom layer of Pt varies between 0.8~nm and 30~nm, with more than 75\% of the data for a thickness larger that 3~nm. The top layer thickness (in nm) is shown by the internal ticks (up to 6 nm). The large circles are the scale of $D_\mathrm{s}$ (up to 3 pJ/m). See \href{run:./figs28_29.html}{here} for an electronic version of the figure, where it is possible to get the full description of each data (stack composition, DMI value, reference) by hovering the mouse over the bars.}
\label{fig:Ds_Pt_Co_X}
\end{figure*}

\begin{figure*}[ht!]
\centering
\includegraphics[width=.75\textwidth, trim=10 10 60 10, clip]{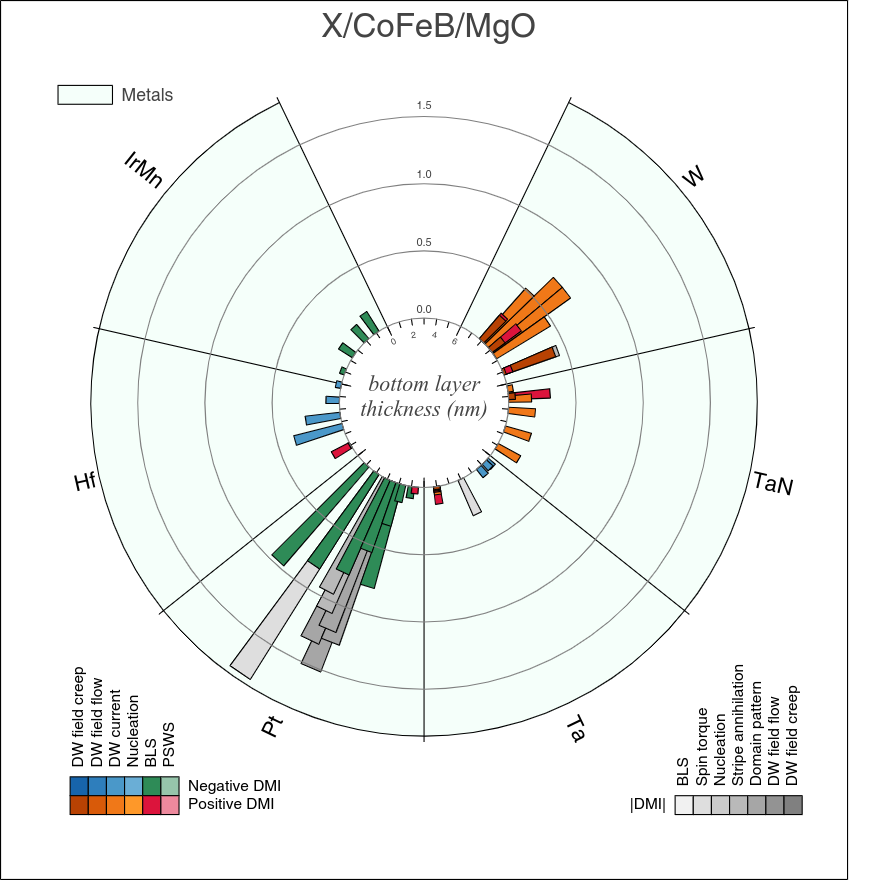}
\caption{Literature data of the DMI constant $D_\mathrm{s}$ for the X/CoFeB/MgO thin films, where the bottom layers X = (IrMn, Hf, Pt, Ta, TaN, W) are the most popular metals in the literature.  The blue/green and orange/red palettes are for experimental methods where the sign is known (negative and positive, respectively), while the grey palette is for data where only the magnitude of $D_\mathrm{s}$ is available. Most of the data ($> 70\%$) are for a MgO thickness of 2~nm, with nearly $20\%$ for 1~nm. The bottom layer thickness (in nm) is shown by the internal ticks (up to 6 nm). The large circles are the scale of $D_\mathrm{s}$ (up to 1.75 pJ/m). See \href{run:./figs28_29.html}{here} for an electronic version of the figure, where it is possible to get the full description of each data (stack composition, DMI value, reference) by hovering the mouse over the bars.}
%The shapes of the symbols refer to different experimental methods, while their size and color reflect the amplitude and the sign of $D_\mathrm{s}$, respectively, as specified in the legends on the right. Data for methods not able to determine the sign of $D_\mathrm{s}$ (i.e, domain pattern, stripe annihilation and spin torque) are in grey. Most of the data ($> 70\%$) are for a MgO thickness of 2~nm, with nearly $20\%$ for 1~nm. See \href{run:./figs27_29.html}{here} for an electronic version of the figure.
\label{fig:Ds_X_CoFeB_MgO}
\end{figure*}

When comparing the DMI constant $D$ measured by different methods, it becomes clear that the values sometimes do not coincide within the uncertainties of different measurements, or are even vastly different (as shown in Figs.~\ref{fig:Ds_Pt_Co_X}-\ref{fig:Ds_X_CoFeB_MgO}), even if the measured stacks are nominally the same or at least very similar. 
As examples shown in Fig.~\ref{fig:Ds_Pt_Co_X}, disagreement on the values are found for Pt/Co/Ir stacks (measured by nucleation, DW creep and BLS), Pt/Co/Ta stacks (measured by BLS, domain pattern and SOT) or Pt/Co/AlO$_x$ and Pt/Co/MgO (measured by stripe annihilation, BLS and DW flow). Therefore, it is important not only to consider the advantages and limitations of each individual method, which influences their applicability, but also their intrinsic differences, which might influence the measurement result. The aim of this section is to discuss the intrinsic differences and make some hypotheses about their influence on the measurement result, although currently systematic studies comparing the various methods are unfortunately rare. Furthermore, in most of the available studies, samples are investigated where the different techniques agree well, while the disagreement is rarely discussed. 
%One important point is to distinguish static or quasi-static methods (as the methods discussed in sections \ref{sec:domainwidthexperiments} to section \ref{sec:NVmagnet}) from dynamic methods as current- (section \ref{sec:experimentDWcurrent}) or field-driven (section \ref{sec:experimentDWfield}) DW motion, spin wave-based methods (section \ref{sec:spin waves}) or SOT-based methods (section \ref{sec:SOT}). Within the dynamic methods one still has to differentiate different dynamic regimes, such as the creep and flow regime for DW motion, and the spin wave regime that is concerned with completely different frequency regimes.
We suppose that the presence of defects or disorder evidences some of the intrinsic differences between the techniques. In general, methods based on domain wall motion are affected by the fact that pinning influences DW dynamics. Pinning is a local phenomenon that - in the case of domain wall motion - may nevertheless have a global influence on the average domain wall movement and thus may complicate the quantitative determination of $D$. Furthermore, quasistatic methods based on imaging usually rely on observing the DW or a number of domains in a certain position, and therefore close to pinning sites. They depend on the local energy landscape at a pinning site which may be different from the one far from the pinning site and the result of the measurement may vary even from spot to spot in a single sample due to the distribution of inhomogeneities and defects. Analytical models and simulations usually do not take these imperfections into account and use average parameters, which may lead to intrinsic errors. Considering a larger sample area may reduce such an effect. Also nucleation of domains, as it is needed also for measurements in creep regime, occurs at pinning sites and single measurements may not be representative. In addition, strain in the thin films might play a role and, in fact, several recent works investigate defect- or strain-induced DMI \cite{FER-19, MIC-19, DEG-20}. Therefore, a statistic assessment across a wafer, measuring at many different locations, could help to understand more about defect related differences in the measurements, but is currently lacking in the literature. 

Dynamic methods often average out these effects. Note however, that an average $D$ value obtained across a large film area in presence of inhomogeneities might not represent the intrinsic value. DW creep motion can be considered an intermediate regime, where DWs move due to thermal activation from one pinning site to the next. These measurements show generally good agreement with static methods, such as the stray field measurements by NV center magnetometry \cite{SOU-16,GRO-16}. However, measurements in the creep regime require simplified models of the domain wall dynamics, as detailed in section \ref{sec:theoryDW}. In fact, concerning measurements of DW motion in the creep regime by asymmetric bubble expansion, there are currently various models proposed which evaluate the DMI from the domain velocity profile and fit the data to extract the DMI field. In addition, the evaluation of the DMI constant depends on the DW width or the DW energy, parameters known only up to a certain accuracy. They are often calculated from the exchange stiffness and the anisotropy constant. Both parameters should be therefore determined experimentally, which requires a substantial effort especially for the exchange stiffness in thin films. At that point direct measurements of the DW width by imaging techniques may become a valuable alternative, even if pretty much more complicated. 

SW methods have the advantage of a more straightforward evaluation of the DMI constant by measuring a frequency shift, which is possible with high accuracy. Concerning the model, only the gyromagnetic ratio $\gamma$ and the saturation magnetization $M_s$ have to be known (the exchange constant is not needed for long wavelength SWs), parameters more easily to be accessed experimentally. Pinning sites usually do not influence the dispersion relation of spin waves directly, but only affect the measured spin wave decay length which is reflected in the line width in frequency domain methods. The most widely employed, direct and effective method is BLS, as it does not require any external spin wave transduction or any kind of patterning, contrary to TR-MOKE and PSWS. Influence of defects, grain boundaries, inclusions and surface roughness play a marginal role, as is generally the case for the spin wave methods. To this respect, it has also to be considered that the revealed SW  have wavelength ranging from  half a $\mu$m to a few microns and that measurements average over a rather large sample area. In the case of BLS and TR-MOKE  the typical spot size is of the order of 10~$\mu$m and in the case of PSWS a typical CPW is also about 10~$\mu$m long, while DWs have a width of the order of 10~nm.  Moreover, the propagation distance of SW, even for thermal SW detected by BLS, is of the order of several microns and the influence of nanometric inhomogeneities is averaged out (for instance they may result in a broadening of the BLS peaks, with a minor influence on the mean frequency) so that SW methods are rather insensitive to them due to the different length scales involved. Another simplification of spin wave methods might be that the measurement does in general not require measuring (and evaluating by a model) the spin wave dispersion $\omega (k)$ itself but only the difference $\omega (k) - \omega (-k)$, and is therefore a kind of differential method.

The differences in applicability of the various methods explain why a direct comparison on the same sample is rarely reported in the literature. Only in few papers \cite{VAN-15, SOU-16, KIM-19, SHA-19, PHA-16, LAU-18}, a quantitative comparison of the values of $D$ is reported, measured by both BLS and DW-based methods. In general, a reasonable agreement is attained for the different systems if BLS results are compared to those extracted from analysis of DW motion in the flow-regime\footnote{Note, however, that the DW methods use additional parameters with respect to the spin wave methods, as e.g. the exchange stiffness, which may be adapted in order to fit the BLS results.}. Instead, when DW motion is analysed in the creep regime, the values obtained by DW-based methods are generally significantly higher or lower than those obtained by BLS. As mentioned, this may depend critically on the accurate analysis of the measured creep data. However, while in the creep regime disorder plays a crucial role, spin wave propagation might be rather insensitive to it. This could be owing to the different magnetic state necessary for the measurement: in the creep regime (nucleating and expanding a bubble domain) the measurement is performed at small applied fields, so at a state where the energy landscape starts to vary locally. Instead the spin wave methods are applied close to saturation, where disorder in the magnetization state becomes negligible. The level of discrepancy may be also strongly influenced by the material properties and layer combinations, as well as by the specific thickness and chosen substrate \cite{SOU-16, SHA-19}. In fact, since in the thin film systems DMI is an interface effect, details concerning the structural and chemical quality of the materials, in particular of the interfaces, play a crucial role and should be stated in the papers. Furthermore, in order to judge the relative accuracy and reliability of different methods it is indispensable to compare measurements performed on identical samples and a more intense research effort should respond to this need. 

Concerning methods that determine the DMI field under the influence of a current (such as current induced DW motion and SOT field measurements), the main difficulty is the interpretation of the results, taking into account the different contributions of spin orbit torques. Damping and field like torques may have different origins as the SHE and the iSGE, and both contribute to the observed dynamics. Furthermore, in metallic stacks, the current flow in the plane is not well defined as it is partly in the FM and partly in the HM material, depending on the relative resistances and the interface scattering and transparency \cite{STE-20}. 

An attempt to classify the applicability of three methods (BLS, SOT efficiency, and DW velocity)  according to DMI strength and FM layer thickness was performed by \textcite{KIM-19}. According to their scheme,  BLS applies better to larger $D$ values and FM film thicknesses, while DW methods are suitable for smaller values, instead SOT applies to small values of $D$, but higher FM layer thicknesses. In an intermediate range all three methods are applicable and a direct comparison of the methods on the same sample is possible. In fact for the samples Pt(2.5)/Co(0.9)/X(2.5) (X~=~Ti, Ta, Al, Pt) they find an excellent agreement within a 5\% difference. While in the case of Pt/Co/W, a 10\% difference between DW-based and SW-based methods occurs, and Pt/Co/Cu presents a much larger 40\% difference. They argue that the limitation of BLS for measuring low DMI values is the interferometer resolution, while in film thickness, supposing films with PMA, it is due to to the high fields necessary to overcome the out-of-plane anistropy. DW velocity and SOT methods are limited in measuring high DMI values by the maximum in-plane field that can be applied. The DW velocity is limited to smaller film thicknesses by chiral damping \cite{Kim2018} with respect to the SOT method where spin transfer torque contributes, instead at higher thicknesses both are limited due to dendrite domain formation.
The above considerations are corroborated by a synoptic view of the Tables of results and Figs.~\ref{fig:Ds_Pt_Co_X}, and \ref{fig:Ds_X_CoFeB_MgO} discussed later: one may find that in several examples of nominally similar systems, different groups obtained relatively different results. In this context, even if BLS seems at the moment the most popular and efficient method to determine DMI in layered systems, more systematic multi-technique investigation of the DMI on the same samples, for different materials combination and layer thicknesses (including a cross-check concerning the mutual consistency of results relative to different regions on the same sample) would be highly desirable.

%\begin{figure*}[ht!]
%\centering
%\includegraphics[width=.45\textwidth]{Ds_Pt_Co_Pt.png}
%\includegraphics[width=.45\textwidth]{Ds_Pt_Co_Ir.png}
%\caption{Literature data of the DMI constant $D_\mathrm{s}$ for Pt/Co/Pt, and Pt/Co/Ir/X thin films, where the top layer X is Pt or Ta (not specified). The shapes of the symbols refer to different experimental methods, while their size and color reflect the amplitude and the sign of $D_\mathrm{s}$, respectively, as specified in the legends on the right of each panel. Data for $D_\mathrm{s} = 0$ are plotted as black dots. See \href{run:./figs27_29.html}{here} for an electronic version of the figures.}
%\label{fig:Ds_Pt_Co_Pt_Ir}
%\end{figure*}

\subsection{Comparing the DMI constants in different materials}

The two most common material combinations investigated in the literature are heterostructures of a) HM/CoFeB/MgO and b) Co with Pt as a top or bottom layer.
We begin by focusing on Pt/Co/X heterostructures, with the top layer X = W, Pt, Ir, Au, Ta, Al, Gd, AlO$_x$, GdO, and MgO, as shown in Fig.~\ref{fig:Ds_Pt_Co_X}. When the sign of $D_\mathrm{s}$ is not available a gray scale is used, otherwise a positive sign is indicated by orange/red palette and negative by a blue/green one. The sign is mostly negative because it is dominated by the Pt/Co interface which gives a negative sign (according to the conventions used in this paper). The oxides, AlO$_x$, GdO and MgO, as top layers show the largest (negative) DMI values \cite{BOU-16}, with the record (excluding PSWS measurements) of $|D_{s}| =$~3~pJ/m in Pt(4)/Co(1)/MgO(2) thin films measured using the spin-orbit torque method \cite{PAI-16}. One could therefore assume that all the top layers contribute with a negative $D_\mathrm{s}$, adding up to the total negative $D_\mathrm{s}$ measured. In \textcite{CHA-19} it was shown by investigating oxide wedge samples that the Co/$M$O$_x$ interface (with $M$ = Al or Gd) gives a contribution to the DMI caused by the Pt underlayer which depends on the oxidation state and that by tuning the Co oxidation an optimization of DMI can be reached. However, a positive sign for $D_\mathrm{s}$ (not shown in the figure) was reported measured by PSWS in Pt/Co/MgO films with a thick Co layer (up to 20~nm) \cite{LEE-16,KAS-18}. 
The other elements as top layers show more or less contradicting results, as for Ta, Pt, and Ir. The stacks  Pt/Co/Ir/X stacks show a large variety in magnitude and sign, for instance. Note that the introduction of an Ir thin layer in Pt/Co/Ir/Pt reduces the magnitude of $D_\mathrm{s}$ and changes the sign for thicker Ir \cite{HRA-14}. In Ir/Co/MgO a negative value for $D_\mathrm{s}$ was found by BLS, concluding that Ir contributes as a bottom layer with a negative DMI and positive as a top layer, so in the same way as Pt. This fact is in contrast with the theoretical prediction that the contribution of the interface Ir/Co was opposite to that of Pt/Co \cite{YAN-15,YAM-16}. Opposite contributions would make it possible to increase the DMI by fabricating Pt/Co/Ir trilayers. As discussed in Sections \ref{sec:experimentDWfield} and \ref{sec:SOT}, the combination Pt/Co/Ir seems to give smaller values than the addition of the $D_\mathrm{s}$ values. Furthermore, Ir as a bottom layer gives a positive $D_\mathrm{s}$ measured by BLS in Ir/CoFeB/MgO and a negative value for Ir/Co/MgO, so the sign does also depend on the ferromagnetic material. A simple addition of $D_\mathrm{s}$ values of top and bottom interfaces seems not to be feasible, as is also shown for Pt/Co/W. Even though W contributes with opposite sign to the DMI with respect to Pt and its contribution to the DMI strength in general is higher than for FM/oxide interfaces, the total DMI value of the stack is smaller than the one obtained by Pt/Co/oxide stacks. The addition of $D_\mathrm{s}$ values of the top and bottom interface can be discussed at the Pt/Co/Pt stack. For instance, films with the same thickness of the two Pt layers should, in principle, have a zero DMI, as for epitaxial films. However, a perfect or nearly-perfect compensation of the top/bottom contributions is found in a few cases \cite{PHA-16,KIM-18,AGR-19,KIM-19,BEL-19a,GEH-20} (not shown in figure), and more generally small values of DMI are found, both positive and negative, as shown in figure.  Additionally, current-induced motion experiments in Pt/Co/Pt samples show negative values independent of the thickness of top or bottom layer \cite{FRA-14}, while creep regime experiments give mostly positive values. In summary, Fig.\ref{fig:Ds_Pt_Co_X} points out the strong influence of film preparation, differences in interface properties of top and bottom layers, etc. on the DMI, which explain the high variability in magnitude and sign of the $D_\mathrm{s}$. 

The interpretation of the data for the X/CoFeB/MgO heterostructures (with X = IrMn, Hf, Pt, Ta, TaN, W as shown in Fig.~\ref{fig:Ds_X_CoFeB_MgO}), is a bit more clear\footnote{Note: The plot contains both compositions of CoFeB, the Fe rich Co$_{20}$Fe$_{60}$B$_{20}$ and the Co rich Co$_{40}$Fe$_{40}$B$_{20}$.}. Similarly to the previous figure, methods not able to determine the sign are shown with gray palette, while positive signs are indicated by a orange/red palette and negative signs by the blue/green one. In general, the different methods give values compatible in magnitude, with the exception of Pt/CoFeB/MgO, where BLS and domain pattern/spin torque methods diverge. The CoFeB/MgO bilayer sputtered directly on the substrate (so without bottom layer X) has a small positive $D_\mathrm{s} = 0.13$~pJ/m \cite{CHE-18}. This value is compensated in the case of Pt and IrMn bottom layers showing a negative $D_\mathrm{s}$ for the largest thickness (up the maximum value of about -1.75~pJ/m for the Pt layer, measured by spin torque methods). On the contrary, W and TaN increase the positive value of $D_\mathrm{s}$, up to about 0.7~pJ/m for the W layer. Less clear is the contribution of Hf and Ta. Hf leads to a negative $D$ obtained by current-induced domain wall motion measurements for a film thickness larger than 2~nm, while a positive $D_\mathrm{s}$ is obtained in the creep regime and by BLS for a 1~nm thick layer. It is interesting to note that the negative value is decreasing as the Hf becomes thicker, while for Pt or IrMn $D_\mathrm{s}$ increases for thicker films. Ta gives a variation of $D_\mathrm{s}$ from negative to positive when increasing the thickness, measured by current-induced domain wall motion \cite{TOR-14}, which is in contradiction with the positive value for the stack without bottom layer. Furthermore, \textcite{KAR-18a} reported different signs when measuring by creep regime and by current-induced domain wall motion. This contradiction may originate from the small value of $D_\mathrm{s}$ in those Ta samples.

Despite all these difficulties, a few conclusions can be reached concerning the sign and strength of the DMI. If the order of the stack is reversed, then one finds a reversal of the sign by symmetry. The strength of the DMI may be different due to the different growth conditions of a reversed stack, but the sign reversal is effectively verified as in Pt/Co/Ta and Ta/Co/Pt trilayers measured by BLS \cite{CHO-17}. All the BLS experiments are consistent with a negative (positive) value of $D_\mathrm{s}$ for a Pt bottom (top) layer regardless the FM composition, as veriﬁed for CoFeB, Co, Co/Ni, Ni$_{80}$Fe$_{20}$, and independent of whether the top layer is an oxide or another heavy metal. This result is also consistent with the sign detected for current-induced domain wall motion and field driven motion in the flow regime. Concerning the absolute value of $D_\mathrm{s}$, the highest values are obtained for Pt/FM interfaces in combination with oxide top layers. In general, Co seems to be more sensitive to the interface quality and more contradictions are found in the literature. Clear trends are observed for CoFeB compounds concerning a) the sign, as Pt as a bottom layer gives negative sign, while Ta and W give positive ones; and b) the trend in HM thickness dependence, increasing with a saturation value for Pt, decreasing for Hf and non monotonic for W with a maximum value for an intermediate thickness.

\subsection{Influence of growth conditions on the DMI constant}

Due to the large spread of materials used, the investigation of the influence of growth condition on DMI is not conclusive so far. \textcite{WEL-17} studied the effect of sputter-deposition conditions on DMI in Pt/Co/Pt structures. They found that the growth temperature modifies the interfacial roughness. The different quality of the top Pt/Co interface and the lower Co/Pt one introduces a structural inversion asymmetry, which results in a net DMI field in this nominally symmetric structure. This explains the discrepancies of the sign of $D$ for Pt/Co/Pt stacks as mentioned in \ref{sec:experimentDWfield}. Regarding the effect of post annealing, \textcite{KHA-16} studied the influence of annealing on DMI in Ta/CoFeB/MgO. Here the DMI field $H_\mathrm{DMI}$ is determined by magnetic field driven domain motion in the creep regime. They found that both $H_\mathrm{DMI}$ and the DMI constant $D$ vary with annealing temperature, reaching a peak at 230~$^\circ$C and then decreasing as the temperature is further increased. They also found that the dependence of interfacial perpendicular magnetic anisotropy field $H_\mathrm{K}$ on annealing temperature follows a similar trend as DMI, suggesting a connection between these parameters. They suggested that the increase of $H_\mathrm{DMI}$ and $H_\mathrm{K}$ is due to an improved ordering of atoms at the Ta/CoFeB interface. Higher annealing temperature leads to diffusion of B atoms out of CoFeB as well as intermixing at the interface, which significantly reduces $H_\mathrm{DMI}$ and $H_\mathrm{K}$. \textcite{CAO-20} also reported a similar trend in annealing temperature dependence of $H_\mathrm{DMI}$ in Pt/Co/x/MgO structures (x = Mg or Ta) investigated by magnetic field driven domain motion. A maximum $H_\mathrm{DMI}$ is obtained at an annealing temperature of 300~$^\circ$C, which is independent of the MgO thickness. Similar to \textcite{KHA-16}, they also propose that the enhanced $H_\mathrm{DMI}$ is due to the improved crystalline quality upon annealing. However, \textcite{FUR-17} used current-driven domain wall motion to study the effect of annealing on $H_\mathrm{DMI}$ in Pt/[Co/Ni] structures. They show that annealing causes a significant reduction of $H_\mathrm{DMI}$, domain wall velocity, perpendicular magnetic anisotropy, as well as spin-orbit torques, which is attributed to the diffusion of Co atoms across the Pt/Co interface.

\subsection{Outlook and conclusion}

In this review nearly than one hundred and fifty articles measuring interfacial DMI have been analyzed. The interest in this topic is still growing, as illustrated by the rapidly increasing number of publications and citations. It may be surprising that only in recent years the topic of chiral magnetism gained such a relevance in the research community, as parity breaking and chirality is intrinsic to magnetic systems.

%\begin{figure*}[ht!]
%\centering
%\includegraphics[width=0.8\textwidth]{citations.png}
%\caption{Web of Science (WOS) search result in "Topic": "Dzyaloshinskii" AND "Moriya"; a) number of articles published per year; b) number of total citations per year}
%\label{fig:citations}
%\end{figure*}

New opportunities have been foreseen recently moving the interest away from the traditional HM/FM bilayers towards synthetic antiferromagnets (SAF) and oxides.   SAFs exploit the RKKY exchange interaction between two ultrathin FM layers separated by a non magnetic spacer layer, which can be tuned by the spacer layer thickness \cite{DUI-18}. In these systems, with a HM as spacer layer, it was shown that the DMI is enhanced by the dipolar field between the FM layers \cite{FER-19, MEI-20}. Besides having the advantages of antiferromagnets, such as negligible stray fields and stability against magnetic fields \cite{BAL-18}, these SAFs exhibit asymmetric DWs and spin wave dynamics. The possibility of tailoring DMI in SAFs makes them extremely interesting for applications of chiral magnetism and topological spin structures \cite{LEG-20, VED-20}. Since this topic goes slightly beyond the review, it is not included in the tables, however it is worth mentioning the upcoming interest in SAFs \cite{HAN-19, BOL-20, TAN-20, TSU-20, MEI-20, FER-19}.

A second topic that has come up recently, and which brings us to the microscopic origin of DMI, is DMI in FM oxides \cite{WAN-20} and oxidized metallic FM \cite{NEM-20} which is not fully understood, yet, but might be related to local charge transfer at the interfaces. In \cite{NEM-20} the enhanced DMI is suggested to be caused by an electric dipole moment induced by hybridization and charge transfer at the oxygen/FM metal interface \cite{BEL-16b}. This iSGE-induced DMI \cite{KIM-13} was also observed for a Co/graphene interface \cite{YAN-18}. In both cases density functional theory (DFT) calculations were fundamental to interpret the results and the calculations showed that the DMI originates from the FM layer, instead of the HM layer \cite{YAN-18a,YAN-18}. However, the fact that hybridization at the interface plays an essential role makes the analysis complex and categorizing materials or stacks according to their DMI strength becomes almost impossible. Interface intermixing, interface roughness, dead layers and proximity effects all were known to affect the DMI. In fact, a detailed interface characterization should be performed in order to obtain a complete picture. Although clearly certain materials give higher DMI than others, the interfacial DMI may depend on the interface properties more than on material properties themselves. For example, annealing may change the DMI value by about 20-30\% \cite{BEN-20} or exchanging top and bottom layer in \textcite{CHO-17} leads to changes of about 30\%. The iSGE-induced DMI is closely related to materials for spin-to-charge conversion which recently attracted attention \cite{CHE-16, ROJ-19, DIN-20}. A key feature is again the mechanism of SOC, but not necessarily a material with strong SOC, such as HM, is required, but SOC can also be induced at the interface by the iSGE. Similar to other SOC related phenomena (such as spin-orbit torques), the origin of DMI, either from bulk SOC or SOC at the interface, is impossible to disentangle, and have to be treated in a common theoretical approach. The DMI was firstly described by a phenomenological thermodynamic theory, but it appears now fundamental, for developing materials designed for future applications, that the microscopic origin of DMI has to be understood in more detail.

A third new emerging topic investigates the possibility to stabilize chiral spin textures in centrosymmetric magnetic insulators, where high DW speeds (up to 400 ms$^{-1}$ are reached with minimal currents of $10^6$ A/cm${-2}$ \cite{VEL-19}. In systems like Tm$_3$Fe$_5$O$_{12}$ (TmIG) a small DMI of the order of a few $\upmu$J/m$^2$ has been detected in DW racetracks, and confirmed by scanning nitrogen-vacancy magnetometry. In particular, TmIG thin films grown on Gd$_3$Sc$_2$Ga$_3$O$_{12}$ exhibit left-handed N\'eel chirality, changing to an intermediate N\'eel-Bloch configuration upon Pt deposition. Similarly, \citealp{LEE-20} show that Pt, W, and Au induce strong interfacial DMI and topological Hall effect, while Ta and Ti cannot, providing insights into the mechanism of electrical detection of spin textures in these magnetic insulator heterostructures.

A fourth and very important research field is that related to the search for systems with large and tailored DMI,  paving the way towards the stabilization and exploitation of topological spin structures, such as spirals, helices, merons, skyrmions and anti-skyrmions at temperatures well above room temperature and supporting high-frequency dynamics \cite{FIN-16}. This would represent a clear advantage to move towards applications, where non-equilibrium operation is needed to achieve fast switching and information processing.  In this respect, the stabilization of topological magnetic objects and the control of their dynamical properties can be made more efficient and reliable by patterning periodic arrays of two-dimensional dots or anti-dots. These systems are also the battleground of the rising and exciting field of magnonics, where information is carried and manipulated by spin waves propagating in periodically modulated media, i.e. artificial magnonic crystals. Integration of  DMI in magnonic crystals is expected to open unexplored possibilities, due to the appearance of a sizeable asymmetry in the magnonic band structure, enabling the possibility of unidirectionally energy transfer and magnetic damping tuning \cite{GAL-19a}.  In this context, the concept of topologically protected chiral edge spin waves existing in the bandgap of a topological magnetic material and propagating in a certain direction with respect to the bulk magnetization direction appears to be very promising \cite{WAN-18}. Based on this idea, reconfigurable topological spin-wave diodes, beam splitters, and interferometers can be in principle designed and realized for new devices for information and communication technology.

Finally, even if this review was focused on experimental methods and results, it is clear that new challenges and advances for this research field, also in terms of functionalization/engineering of materials,  are connected with achieving a better understanding of the microscopic origin of DMI, relying on quantum mechanical and  atomistic calculations \cite{YAN-15} and integrating them into a multiscale modelling chain, whose uppermost level are the micromagnetic simulations able to mimic the behavior of real devices. One necessary step for this goal will be to shortcut the gap existing between research communities that use either classical or quantum physics, so as to setup efficient computer codes, operating in a multiscale framework, capable of providing quantitative predictions and recipes for the next generation of materials and devices. In terms of experiments, characterising the atomic scale details of real interfaces is a huge but important challenge that must be met in order to tease out the ultimate causes of the variation in DMI observed among the disparate reports to date. 

\section*{Acknowledgment}
The project 17FUN08-TOPS has received funding from the EMPIR programme co-financed by the Participating States and from the European Union’s Horizon 2020 research and innovation programme. F.G.S. acknowledges the support from Project No. SA299P18 from the Consejer\'ia de Educaci\'on of Junta de Castilla y Le\'on. L.C. and C.B. acknowledge the support from DFG through SFB 1277. C.H.M. acknowledges support from the EPSRC, grant number EP/T006803/1. G.D. acknowledges the support from the European Union H2020 Program (MSCA ITN 860060). We kindly thank G. Chen, A. Schmid and T.P. Ma for the unit conversion of the $D$ values in Table \ref{table:imaging}. M.K. thanks H. T. Nembach, J. M. Shaw and A. Magni for fruitful discussions. A special thank to P. Landeros for critical reading of the manuscript and useful suggestions.

%\bibliography{DMI_biblio}
%\bibliography{dmi_paper}
%apsrmp4-2.bst 2018-12-27 (MD) hand-edited version of apsrmp4-1.bst
%Control: author (3) reversed first dotless

\end{document}